\documentclass[aps,twocolumn,superscriptaddress,longbibliography]{revtex4-1}
\pdfoutput=1 %Flag for ArXiv to compile document with PDFLaTex
\usepackage{natbib}
\usepackage{comment}
\usepackage{bm}% bold math with $\boldsymbol{...}$
\usepackage{csquotes} %[babel,english=british]% % key=babel tilpasser quotationmarks til 
\usepackage{dsfont} % Matematik symboler: Kun til store bogstaver samt tallet 1 . Syntax \mathds
\usepackage{amssymb}    % need for more unusual mathsymbols
\usepackage{tensor} 
\usepackage{siunitx} % to write  \si{\angstrom} unit angstrom 
\usepackage{amsmath}    % need for subequations 
\usepackage{graphicx}   % need for figures
\usepackage{subfigure}  % use for side-by-side figures
\usepackage{mathrsfs}
\usepackage{hyperref}   % use for hypertext links, including those to external documents and URLs
\usepackage{color}      % use if color is used in text
\usepackage{comment} % to comment out
\usepackage{tikz}  % A drawing tool for Latex and Beamer
\newcommand{\Ab}{\mathbf{A}}
\newcommand{\ab}{\mathbf{a}}           % bold face lower case a
\newcommand{\abh}{\hat{\mathbf{a}}}    % bold face lower case a with hat
\newcommand{\B}{\text{B}}              % upright capital letter B in math mode
\newcommand{\Bb}{\mathbf{B}}           % bold face capital lettter B
\newcommand{\bb}{\mathbf{b}}           % bold face lower case b
\newcommand{\F}{\text{F}}              % upright capital letter F in math mode
\newcommand{\Fb}{\mathbf{F}}           % bold face capital lettter F
\def\k{{\text{k}}}              % upright lover case letter j in math mode 
\newcommand{\kb}{\mathbf{k}}           % boldface lower case k in math mode
\newcommand{\M}{\text{M}}              % upright capital letter P in math mode
\newcommand{\Mb}{\mathbf{M}}
\newcommand{\p}{\text{p}}              % upright lower case  p in math mode 
\newcommand{\pt}{\tilde{p}}
\newcommand{\pb}{\mathbf{p}}
\newcommand{\pbt}{\tilde{\mathbf{p}}}
\newcommand{\pbht}{\hat{\tilde{\mathbf{p}}}}
\def\P{{\text{P}}}              % upright capital letter P in math mode 
\newcommand{\Pb}{\mathbf{P}}           % bold face capital lettter P
\newcommand{\q}{\text{q}}              % upright capital letter P in math mode
\newcommand{\qb}{\mathbf{q}}
\newcommand{\qbt}{\tilde{\mathbf{q}}}
\newcommand{\Q}{\text{Q}}              % upright capital letter P in math mode
\newcommand{\Qb}{\mathbf{Q}}
\newcommand{\rb}{\mathbf{r}}
\newcommand{\Sb}{\mathbf{S}}           % bold face capital lettter S
\newcommand{\vb}{\mathbf{v}}
\newcommand{\g}{\gamma}
\newcommand{\D}{\Delta}
\newcommand{\Db}{\bm{\Delta}}  % boldface
\newcommand{\Dt}{\tilde{\Delta}}
\def\Dtb{{\tilde{\bm{\Delta}}}}
\def\d{{\delta}}
\newcommand{\m}{\mu}
\newcommand{\mb}{\bm{\mu}}
\newcommand{\s}{\sigma}        % un-primed
\def\sb{{\bm{\sigma}}}  % boldface 
\newcommand{\tauh}{\hat{\tau}}    % hat & boldface
\newcommand{\taub}{\bm{\tau}}    % boldface
\newcommand{\dd}{\textup{d}}
\begin{document}

%\preprint{}

\title{Neutron scattering off Weyl semimetals}
%\title{{\color{blue} QUESTION: What is a good title ?}  Detecting type-$1$ Weyl nodes by inelastic neutron scattering.}
%{\bf title 1:} Probing type-$1$ Weyl nodes by inelastic neutron scattering.\\
%{\bf title 2:} Detecting type-$1$ Weyl semimetals by inelastic neutron scattering.\\ 
%{\bf title 3:} Detecting type-$1$ Weyl nodes by inelastic neutron scattering.? }
\author{Michael Bjerngaard}
%\email[]{mbjerng1@jhu.edu}
\affiliation{Department of Physics and Astronomy, Johns Hopkins University, Baltimore 21218, United States}
\author{Bogdan Galilo}
\affiliation{Cavendish Laboratory,  University of Cambridge, Cambridge  CB3 0HE, United Kingdom}
%\affiliation{T.C.M. Group,  Cavendish Laboratory,  University of Cambridge,  JJ Thomson Avenue, Cambridge, CB3 0HE, United Kingdom}
\author{Ari M. Turner}
%\email[]{turner.ari@physics.technion.ac.il}
\affiliation{Department of Physics, Technion, Haifa 3200003, Israel}


\date{\today}

\begin{abstract}
We present how to detect type-$1$ Weyl nodes in a material by inelastic neutron scattering. Such an experiment first of all allows one to determine the dispersion of the Weyl fermions.  We extend the reasoning to produce a quantitative test of the Weyl equation taking into account realistic anisotropic properties. These anisotropies are mostly contained in
the form of the emergent magnetic moment of the excitations, which determines how they couple to the neutrons. Although there are many material parameters, we find several quantitative predictions that
are universal and demonstrate that the excitations are described by solutions to the Weyl equation.  The anisotropic coupling between electrons and neutrons implies that even fully unpolarized neutrons can reveal the spin-momentum locking of the Weyl fermions because the neutrons
will couple to some components of the Weyl fermion pseudospin more strongly. 
On the other hand, in an experiment with polarized neutrons, the scattered neutron beam remains fully polarized in a direction that varies as a function of momentum transfer (within the range of validity of the Weyl equation).
This allows measurement of the chirality of Weyl fermions for inversion symmetric nodes.  Furthermore, we estimate that the scattering rate may be large enough for such experiments to be practical; in particular, the magnetic moment may be larger than the ordinary Bohr magneton, compensating for a small density of states.
\end{abstract}

%\pacs{???}
%\keywords{?}

\maketitle


%section: Introduction 
\section{Introduction}The Weyl equation, first applied in high-energy physics to describe neutrinos, has recently been connected to condensed matter physics, where it describes materials whose electronic excitations have a strong coupling between spin and orbital degrees of freedom.  In experiments\cite{Xu2015aa,Yang2015a,Lv2015a} guided by band structure calculations\cite{Wan2011a,Weng2015a,Huang2015a}, Weyl fermions have recently been realized in the context of Weyl semimetals (WSM) in crystalline solids, photonic crystals\cite{Lu2015a}, and magnon bands\cite{Mook2016a,Li2016a,Li:2017aa}. 
%{\bf TSM stuff?}The WSM being the most elementary in this new class, host topological features\cite{Nielsen1981a,Nielsen1983a,Volovik2003a,Wan2011a} e.g. Fermi arcs, anomaly, and phase stability all due to its bulk band topology of Weyl nodes separated in the Brillouin zone.\\

Except for establishing magnetic structure\cite{Liu2017a,Guo:2014aa,Nakajima:2015aa}, spin dynamics\cite{Itoh2016a}, and probing magnon excitations\cite{Yao2017a,McClarty2017a}, neutron scattering has by and large been absent in revealing the physics in topological semimetals\cite{Armitage2018a,Bernevig2018a}. WSMs, however, are characterized by the property that their excitations are spin-momentum locked. This indicates that inelastic neutron scattering (INS) could measure these as it is a probe well-suited for measuring magnetic properties of excitations. However, it has long been known, that INS is a technique that has severe difficulties probing electronic excitations due to kinematic restrictions, form factor and low density of states at the Fermi level. For normal metallic systems, the cross-section intensity was predicted\cite{Silver1984aa} to be as low as $10^{-4} -10^{-3}\;{\rm mb/ meV\, sr\, f.u.}$. 
At first glance, the prospects of probing excitations in WSMs seem worse, since the cross-section should be limited by the small density of states at a Weyl point.  However, the coupling of the neutron to Weyl fermions has a contribution from orbital currents in addition to the usual form factor that determines the rate of neutron scattering. This can be large enough to compensate for the small density of states.  As a proof of concept, we employ a toy model to estimate the cross-section with this coupling included; with some optimistic assumptions, the cross-section can be as large as $10^{-2}\;{\rm mb/ meV\, sr\, f.u.}$, which is similar to the rates of scattering associated with other spin$-\frac12$ related phenomena, that \emph{have} been observed\cite{Goremychkin2018aa,Vignolle:2007aa,Walters:2009aa,Fujita2012aa,Janoschek2015aa}.

The Weyl equation (when applied to fundamental particles) describes a particle 
which is massless and therefore always moves at the speed of light in some 
direction, and which also has a handedness--the spin is aligned to the velocity. 
This is described mathematically by a two-component spinor wave-function. 
In a Weyl semimetal the two components correspond to two different Bloch states that happen to be degenerate at a specific crystal momentum, and the fact
that they are described by the same equation as relativistic particles nearby
is an emergent effect.  In particular, qualitative properties 
of a Weyl semimetal that agree with relativistic Weyl fermions are the 
correlation between the velocity and the orientation of the pseudo-spinor (degree of freedom that transform as spin) on the Bloch sphere and the existence of handedness.
The chirality is especially important because it alone determines the magnitude of
the ``chiral anomaly," which leads to macroscopic phenomena such as a strong
magnetoresistance.

This article models the coupling of Weyl fermions to neutrons and calculates the INS cross-section in detail. 
We show that although a Weyl semimetal may not have any  permanent magnetic ordering, neutrons will still become polarized when they are
scattered. When a neutron scatters from the system, it excites an electron from some state below the Fermi energy to one above. 
The chance of the electron's velocity being deflected in a given direction depends on the 
angle between this direction and the initial and final spins of the neutron 
(which in principle can be controlled experimentally). 
If this can be seen in an  experiment, it would be a sign of spin-momentum locking.  INS would provide information that other experimental
techniques cannot obtain. For example, it would go beyond
ARPES in being able to resolve all three components of momenta and so would
be able to probe spin-momentum locking more cleanly. INS would correctly
distinguish a Weyl semimetal from a narrow gap semiconductor because the
spin-momentum locking does not occur in a narrow gap semiconductor (at least not
at low energies). Besides the specific problem discussed in this paper,
of how to deduce the properties of Weyl excitations from neutron scattering, the detailed analysis of the scattering cross-section suggests that highly unusual types of particle-hole excitations could be generated by a scattering event.

There are two difficulties with using neutron scattering to understand
Weyl semimetals in this way. Neutron scattering creates
a continuum of particle-hole pairs.  Only the momentum transfer from the
neutron is known, and this can result from many different combinations of 
momenta of the excited particle and hole, each of which corresponds
to a different change in the neutron spin.  However, at the maximum momentum 
transfer (for a given energy transfer) the electron
velocity must switch sign.  This determines the \emph{direction} of the initial
and final velocity, and the \emph{magnitudes} are not
needed to detect spin-momentum locking.  
The other difficulty is that although the excitations are essentially 
described by Weyl equation, the coupling of the neutrons to the electrons is not
simply proportional to the emergent magnetic moment and depends on many material
dependent parameters.  The differential cross-section is thus given
by a relativistic expression that is distorted in a complex way.
Nevertheless, we show that there is remarkably
a pattern hidden in this function that has a stable character reflecting
the topological chirality of the nodes.

After presenting the results on the differential cross-section,
this article focuses on finding good ways to interpret the neutron scattering
as a function of spin and momentum, 
especially given that there are many unknown parameters.
The article proceeds as follows:
The scattering process (under circumstances we discuss in Sec. \ref{Kinematics_Appendix}) can be mapped to a relativistic process. %, like the creation of an electron and a neutrino from a given amount of energy and momentum. 
The cross-section can thus be determined by using Lorentz invariance (with details of the calculation given in Appendix \ref{Susceptibility_Appendix}). The scattering rate for neutrons is equivalent to the rate of excitation of relativistic Weyl fermions with an applied field of a certain polarization determined by g-factors (see Sec. \ref{MagnetizationOperator_Appendix}) of the WSM-neutron coupling. In particular, we discuss the size of these -- in materials in which the two Weyl nodes have very close momenta. Here the g-factors can be very large, so that the effective magnetic moment is much greater than that of an ordinary electron. The cross-section (see Sec. \ref{Interpretation_Appendix}), while affected
by the material-dependent g-factors, still has properties
that  capture Weyl fermion physics solely. 
%Due to the Lorentz invariance, we have found predictions (see Sec. \ref{Interpretation_Appendix}) that are "universal", i.e. that are independent of these material dependent g-factors, which capture Weyl fermion physics solely. 
\newpage
Our main findings are: 
\paragraph*{$(1)$}
\indent By varying the energy and looking at the corresponding range of the nonzero cross-section, one can indirectly measure the dispersion of the Weyl excitations, their velocity and principal directions (see Sec. \ref{Interpretation_Appendix:Dispersion}).
\paragraph*{$(2)$}
\indent The spin-momentum locking manifests itself as dependence of the cross-section on the angle of momentum transfer. It is readily observable in a fully unpolarized experiment (see Sec. \ref{DiffCrossSectionMagnetic_Appendix:ConventionalMagneticScattering}), because an unpolarized beam acts as if it is polarized thanks to the anisotropy of the neutron coupling parameters.  Furthermore,
one can obtain quantitative identities that are ``universal" in that
they are satisfied by the cross-section independently of the coupling constants.
\paragraph*{$(3)$}
\indent %One can control the momentum of the particle-hole pair by focusing on the maximum momentum transfer for a given energy. 
%Focusing on the maximum momentum transfer for a given energy, one can see the spin-momentum locking especially clearly. 
If the initial neutron beam is perfectly polarized (see Sec. \ref{PolarizedMeasurement}) with maximum momentum transfer, then the scattered beam is rotated in a definite direction by the interaction with the spins of the Weyl fermions, so the neutrons deflected by any given amount remain perfectly polarized. 
\paragraph*{$(4)$}
\indent
With both beam (initially) and detector polarized, one can measure the chirality for inversion symmetric nodes.
% That is, one can tell whether the electronic
%excitations' momenta are parallel or antiparallel to their pseudospin, even though one can measure the pseudospin only through its interaction with the neutron,
%an interaction which has an unknown arbitrary form.

%\newpage
%section: kinematics
\section{Kinematics and spin-momentum locking\label{Kinematics_Appendix}}
%--------------- Begin: Intra vs. Inter Weyl node scattering ----------------------------------------------------------------------------
Let us consider scattering between two Weyl nodes, at momenta $\kb_{0,1}$ and $\kb_{0,2}$.  
Suppose that the Hamiltonians near these can be put into the idealized form 
\begin{equation}
H_{0,i}(\kb) = \chi_{i} v_{\rm F} \sb \cdot \left(\kb - \kb_{0,i}\right),  \label{Kinematics_Appendix_Eq1:HamiltoniannoLin}
\end{equation}
by changing coordinates if necessary.
Here $v_{\rm F}$ is the velocity of Weyl particles and $\chi_i=\pm 1$ their handedness that we will be interested in measuring. The vector of pseudospin Pauli matrices is $\sb$. 
The Weyl equation has two solutions corresponding to the conduction and valence band, labelled by $\eta=\pm1$. These solutions have the form $\psi_{i,\eta}(\rb) =  e^{i\kb \cdot \rb/\hbar} \vert \pb; \chi_{i}\eta \rangle$, where it is 
convenient to introduce $\pb=\kb-\kb_{0,i}$, the momentum measured relative to the Weyl point. Here, $\vert\pb;\chi_i\eta\rangle$ represents the $2$-component spinor pointing either parallel or antiparallel to the momentum, according
to $\chi_i\eta=\pm 1$. 

In general, the Hamiltonians may have a more complicated form (described below);
however, as we show at the end of this section, most of the asymmetries of the Hamiltonian may be eliminated under
assumptions about inversion or time-reversal symmetry.  There is
just one Lorentz-violating term that cannot be eliminated, which causes certain characteristics of our results to break down.  But the conceptual
picture of how neutron scattering reflects spin-momentum locking does not change.

\begin{figure}
\includegraphics[width=0.85\columnwidth]{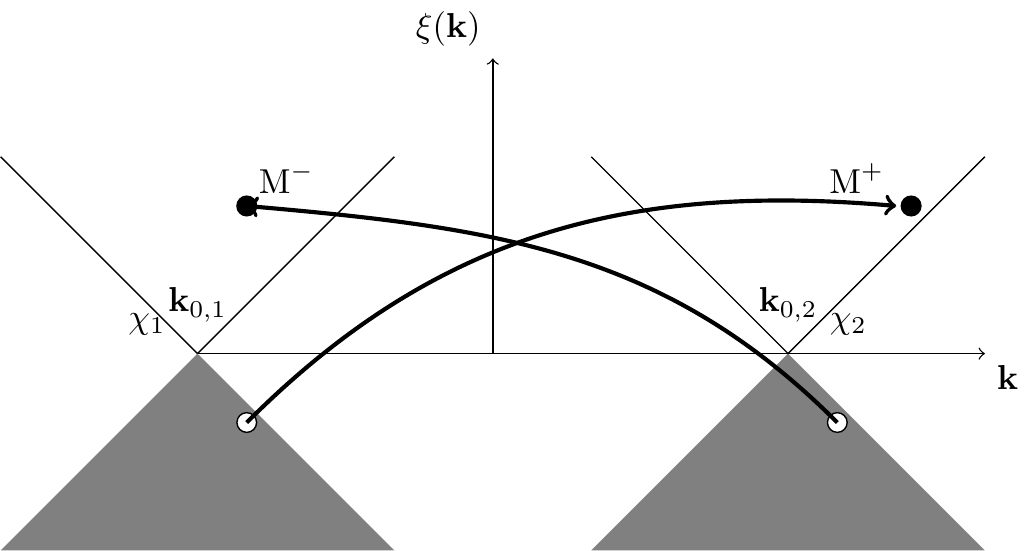}
\caption{Low energy region of two isotropic Weyl nodes located at $\kb_{0,2} = -\kb_{0,1}$ with chirality $\chi_{2}$ and $\chi_{1}$, respectively. At zero temperature the filled Fermi sea (grey) is half-filled.\label{Kinematics_Appendix_fig1:WeylSubspaces}}
\end{figure}

If the material is initially in the ground state, a neutron with initial momentum $\qb_{i}$ can scatter an electron from one Weyl node to another, exciting a Weyl fermion with momentum $\kb_{f}$, and creating a hole below the Fermi energy with momentum $\kb_{i}$ near the other Weyl point, see Fig.
\ref{Kinematics_Appendix_fig1:WeylSubspaces}. As a result of this scattering process the neutron loses energy and its momentum is changed to $\qb_{f}$. For a neutron momentum transfer $\qb = \qb_{i} - \qb_{f}$ and 
change in Weyl momentum $\D\kb = \kb_{f} - \kb_{i}$, the momentum conservation is represented by a factor $\d^{3}(\qb - \D \kb) = \d^{3}(\pb + \Db)$, where it is convenient to  introduce new variables $\Db$ and $\pb$.
The first is defined by $\Db = \Delta\kb_{0} - \qb$, i.e., the deviation
between the transferred momentum and the vector connecting the exact positions of the nodes $\Delta\kb_{0}$. The second is defined by $\pb=\pb_f-\pb_i$
where the variables $\pb_i,\pb_f$ are
the parts of the momenta that appear in the Weyl equation, i.e., the deviation
of each momentum from the corresponding Weyl point. These momenta may be 
regarded as a sort of ``kinetic momentum" because they determine the direction the particle moves and the spin state, while $\kb_{0,1}$ and $\kb_{0,2}$ are just constant offsets. In this article, we consider only absorption processes, where neutrons transfer energy $\hbar \omega = (|\qb_{i}|^2-|\qb_{f}|^2)/2 m_{\rm n}$ to the WSM  with 
accordingly a change in energy $\Delta \xi^{\rm w}$ of the electrons.

\begin{figure}
\includegraphics{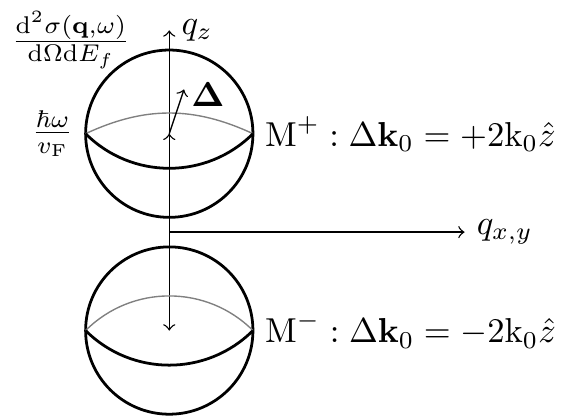}
\caption{Region of nonzero scattering between two nodes at $\kb_{0,2} = -\kb_{0,1} = \k_{0}\hat{z}$. The cross-section as a function of momentum transfer $\qb$ varies within spheres for $0 \leq \lvert \bm{\Delta} \rvert \leq \hbar \omega/v_{\rm F}$.\label{Kinematics_Appendix_fig1:Cross-section_Transitions}}
%\caption{Illustration of the cross-section distribution in coordinates of momentum transfer $\qb$ for nodes located at $\kb_{0,2} = -\kb_{0,1} = \k_{0}\hat{z}$. Intensity varies 
%for $0 \leq \lvert \bm{\Delta} \rvert \leq \hbar \omega/v_{\rm F}$.\label{Kinematics_Appendix_fig1:Cross-section_Transitions}}
\end{figure}

The most basic thing one can measure using neutron scattering is the region of $\qb,\omega$-space in which the cross-section is nonzero. Because the neutron scattering produces two excitations, there is a range of $\omega$’s for each $\qb$ rather 
than a sharp dispersion, similar to the two-particle part of the structure factor in a magnon system, for example. 
The change in energy of the electron, due to scattering from a negative energy state at the first node to a positive energy state at the second node, is $\Delta \xi^{\rm w} = v_{\rm F}|\pb_f|-(-v_{\rm F})|\pb_i|$ so energy 
conservation is  described by $\delta[\hbar\omega-v_{\rm F}(|\pb_f|+|\pb_i|)]$. Graphically, the transferred ``kinetic momentum" $-\mathbf{\Delta}$ is represented by a vector connecting the end-points of $\pb_f$ and $\pb_i$ and the energy is proportional to the sum of their lengths. 
Thus, by the triangle inequality %$\hbar\omega\geq v_{\rm F}|\Db|$
%\begin{equation}
$\hbar\omega\geq v_{\rm F}|\Db|$. %\label{Kinematics_Appendix_Eq1:NonzeroSphere}
%\end{equation}
Suppose one plots the scattering cross-section at a fixed energy transfer.  Then the inequality
%Eq. (\ref{Kinematics_Appendix_Eq1:NonzeroSphere}) 
says that the scattering cross-section is nonzero only inside of a sphere; the sphere is expected to appear with a strong relief as the cross-section jumps sharply from zero at its 
surface. In an actual experiment, if one plots 
the cross-section at a fixed $\hbar\omega$ as a function of the momentum transfer $\qb$, one will see two spheres of radii $\hbar\omega /v_{\rm F}$ centered at $\pm \Delta\kb_0$ as in Fig. \ref{Kinematics_Appendix_fig1:Cross-section_Transitions}, which 
corresponds to transitions (see Fig. \ref{Kinematics_Appendix_fig1:WeylSubspaces}) from the first Weyl node to the second, or vice versa, which we call $\M^{\pm}$ transitions. The $\M^{\pm}$ transitions are displaced in momentum because the physical momentum differs from $\pb$ by offsets $\pm \Delta\kb_0$.  %We will be assuming belowing that the two Weyl points are at opposite momenta, so that $\kb_{0,2}=-\kb_{0,1}=\kb_0$ and $\Delta\kb_0=2\kb_0$. 
The way the cross-section varies within these spheres is interesting to understand in detail, because it is connected to spin-momentum locking (see Sec. \ref{DiffCrossSectionMagnetic_Appendix:ConventionalMagneticScattering}).  

\subsection{Conditions for Lorentz invariance and its consequences\label{LorentzInvariance:Kinematics_Appendix}}
We will see below that Lorentz invariance leads to some special properties of the cross-section. First, there is a discontinuity of the cross-section
at the surface of the spherical regions in momentum space where the cross-section
is nonzero.  
Second, the variation of the cross-section as a function of momentum can be found using Lorentz transformations.\\

In contrast to a relativistic description of Weyl fermions, a condensed matter WSM manifestly breaks\cite{Grushin2012a} Lorentz invariance, because nodes are separated in momentum space and the $i^{th}$ Weyl node expanded to linear order in the 
momentum has the general anisotropic form
\begin{eqnarray}
H_{0,i}(\kb) &=&   \s_{0} \vb^{(i)}_{0} \cdot  \pb  + \s_{l} v_{\rm F}\lambda^{(i)}_{l\,m}\p_{m} , \label{Kinematics_Appendix_Eq2:HamiltonianAisotropic}
\end{eqnarray}
where  $\s_0$ is the identity matrix and $\lambda_{lm}$ is a matrix of parameters (we use Einstein's summation convention)\footnote{Note that the factorization
of the coefficients of the second term as $v_{\rm F}\lambda^{(i)}$ is arbitrary;
$v_{\rm F}$ can be chosen in a convenient way, and the remaining factors which describe the anisotropy are placed in $\lambda^{(i)}$}.

Now, we will focus on scattering between a pair of nodes that are related by either time-reversal or inversion symmetry. By this symmetry, we may
assume the nodes are at $\kb_{0,1}=-\kb_0$ and $\kb_{0,2}=\kb_0$.
By a linear transformation $\pbt  = T\pb$ (see Appendix 
\ref{PrincipalAxisTransformation_Appendix}),  the Hamiltonian of the $i^{\mathrm{th}}$ ($i=1,2$) low energy region can be turned into
\begin{equation}
H_{0,i}(\tilde{\kb}) =  \s_{0} \vb^{(i)}_{0} \cdot  \pbt + \chi_{i} v_{\rm F} \sb \cdot \pbt \quad , \quad \pbt = \tilde{\kb} - \tilde{\kb}_{0,i} . \label{Kinematics_Appendix_Eq1:Hamiltonian}
\end{equation}
The type of symmetry connecting the Weyl nodes determines their relative chirality; for time-reversal and inversion symmetry, the chiralities are equal and negative of one another, respectively. 

The transformation $T$ was chosen such that the second term in Eq. (\ref{Kinematics_Appendix_Eq2:HamiltonianAisotropic}) transforms into the standard isotropic form of Eq. (\ref{Kinematics_Appendix_Eq1:Hamiltonian}).
 If the term $\vb^{(i)}_{0}$ is negligible, then the Hamiltonian is clearly isotropic and even has a relativistic symmetry. Importantly, because of the time-reversal or inversion symmetry, the transformation $T$ is the same for both nodes; i.e. the nodes have their \emph{principal axes aligned} and are \emph{isotropic} in a single coordinate system.  This is crucial for our calculation of
the cross-section; without it we would not be able to use Lorentz symmetry, and the contour of constant energy would not have the simple ellipsoidal shape that is found in Section \ref{Interpretation_Appendix}.  As a consequence, the regions of nonzero scattering would not end
sharply. In order to compare experimental results to this theory, it will be necessary to determine the transformation. We show in Section \ref{Interpretation_Appendix:Dispersion}  that it is easy to see the form of $T$ experimentally from a plot of the structure factor at fixed energy. The transformation must be chosen to have a determinant of $1$ to ensure that the density of states for exciting Weyl fermions does not change. Thus $v_{\rm F}$ will be the geometric mean of the three principal velocities of the original anisotropic dispersion.\\

The following conditions are the precise conditions under which
Lorentz invariance can be assumed:
\begin{enumerate}
\item  The nodes involved in the scattering are aligned (or nearly aligned) with the chemical potential. This requires
careful doping for the materials discovered so far, but in a material where all the Weyl nodes are at the same energy, due
to symmetry, it can be an automatic property of a compound with an even
number of electrons per unit cell.
\item Scattering is between two nodes connected by either time-reversal or inversion symmetry.
\item  The three components of $\vb^{(i)}_{0}$ in Eq. (\ref{Kinematics_Appendix_Eq1:Hamiltonian}) vanish.  Although \emph{this} condition would not usually
be satisfied exactly, we will assume it to be, in order to be able to use Lorentz invariance.
A small nonzero $\mathbf{v}_0$ does not change the predictions too much and, in fact, any type-I WSM $|\mathbf{v}_0| \leq v_{\rm F}$
is analytically tractable as will be discussed in Section \ref{Interpretation_Appendix:Dispersion}.
%{\color{blue} Also, the more realistic  Hamiltonian in which
%$\mathbf{v}_0\neq 0$ is a nice Hamiltonian to study since
%it is valid at low energy in a wide range of cases and includes 
%only one parameter that has an effect on anything, 
%the ratio of $|\mathbf{v}_0|$ to $v_{\rm F}$}.
\end{enumerate}
Under these conditions, the dynamics of the excitations of
the material are entirely Lorentz invariant, but their interaction
with neutrons is not. Thus the cross-section will not be Lorentz invariant,
but it can be predicted using Lorentz symmetry.  It turns out
that the cross-section for a given initial and final neutron polarization
is a certain component of a relativistic tensor (see Sec. \ref{CrossSectionMagneticFormula_Appendix}); the tensor for any net momentum $\Dtb$
can be obtained by applying a Lorentz transformation to that in the rest frame.
The cross-section is not Lorentz invariant for the same reason that the
life-time of a particle depends on its velocity--namely, the lifetime
is only one component of a 4-vector while the cross-section is one component
of a 4-tensor.
%The cross-section for a given initial and final neutron polarization then turns out to be a certain component of a relativistic tensor as described in Section \ref{CrossSectionMagneticFormula_Appendix}. One can start with the case where the electrons are created at rest, $|\Dtb|=0$, and then do a Lorentz transformation to find
%the cross-section of a moving pair of electrons, $|\Dtb| \neq 0$. The cross-section changes for the same reason that objects undergo Lorentz contractions because their length is only one component of a four-vector. Measuring the rate of
%scattering by neutrons is therefore like measuring only one component of a Lorentz invariant tensor--if one measures the lifetime of a particle, it changes as the particle moves faster because the lifetime is only the fourth component of a four-vector.  
In the case of a moving particle, the Lorentz invariance can be 
proven by using a detector that is moving at the same speed as the particle, in which case the lifetime is the same as the rest-lifetime of the particle. In our case, the neutrons are \emph{not} Lorentz invariant, so there is  no way to accelerate the
\enquote{detector}; we can only measure  certain components of the scattering tensor in one reference frame.

\subsection{Kinetic limitations on scattering between nodes at the same momentum}

Consider now the case of intranode scattering, i.e., a transition within a single Weyl
node.  In this case,  $\D \kb_{0} = \bm{0}$.
The conservation of energy and momentum give the same conditions
on the transferred momentum and energy as above.  
However, in contrast to the case of distinct nodes where 
$\qb_i-\qb_f=\Delta \kb_0-\Db$, there is no offset to the momentum,
and this makes it much
more difficult to see anything using neutron scattering. The same
conclusion will apply to scattering between two Weyl nodes at the same
point (e.g., in a Dirac material).  First,
it is clearly impossible to access the center of the spherical region
described above, because $|\Db|=0$ implies that no momentum is transferred;
therefore, the neutron's momentum is unchanged, and so no energy is transferred either. For internode scattering, $|\Db| = 0$ only implies that the transferred
momentum is $\Delta\kb_0$, and so the neutron's energy can change, allowing
it to create excitations in the material.

%. The inequality Eq. (\ref{Kinematics_Appendix_Eq1:NonzeroSphere}) set a threshold value $|\qb_c|$ of the incident neutron momentum, above which neutrons can scatter Weyl fermions
%$|\qb_f| \geq |\qb_c| = m_{\rm n}v_{\rm F}$. The lowest bound is reached when $\qb_f = \qb_i$.

Second, there are no possible scattering events
at all (with any transferred momentum) if the neutron has too small an energy. We initially assume an isotropic system, so that transformed
and untransformed coordinates are the same, e.g. $\qb_i=\tilde{\qb}_i,\qb_f=\tilde{\qb}_f$.  Using %Eq. (\ref{Kinematics_Appendix_Eq1:NonzeroSphere})
 $\hbar \omega \geq |\Dtb| v_{\rm F}$, the triangle inequality, $|\Dtb|\geq |\tilde{\qb}_i|-|\tilde{\qb}_f|$, and conservation of energy, $\hbar\omega=\frac{\hbar^2}{2m_N}(\qb_i^2-\qb_f^2)$, % the energy transfer is constrained to $\hbar \omega \geq v_{\rm F}(|\qb_{i}|-|\qb_{f}|)$, 
we obtain $|\qb_{i}|+|\qb_{f}| \geq 2 m_{\rm n} v_{\rm F}$, a restriction on the neutron momenta. Since the neutron loses energy and momentum, this relation constrains the velocity of the incident neutron $v_{\rm n} = |\qb_{i}|/m_{\rm n}$ to
%$v_{\rm n} \geq v_{\rm F}$.
\begin{equation} 
v_{\rm n} \geq v_{\rm F}. \label{Kinematics_Appendix_Eq1:IntranodeConstraint}
\end{equation}
In the more general case where the electron's speed is direction-dependent, the neutron's speed must exceed the maximum possible speed of the electron if one is to see the full region of scattering $|\Dtb|\leq \hbar\omega/v_{\rm F}$.

Hence, the Fermi velocity of the node determines a characteristic velocity scale for the neutrons\cite{Galilo2013}, implying that only neutrons moving faster than $v_{\rm F}$ can scatter on a single Weyl node. For example, ARPES measurements of tantalum phosphide\cite{Xu:2016aa} indicates a velocity of about $v_{\rm F}\approx 1.5 \times 10^{5}\; {\rm m\, s^{-1}}$, which greatly exceeds the speed\cite{Carron2007a} of a thermal neutron $v_{\rm n}^{\rm thermal} = 2 \times 10^3\;{\rm m\,s^{-1}}$. In order to reach a speed of $10^5\;{\rm m\,s^{-1}}$ a neutron has to be rather hot, carrying an energy of the order of $10^2\;{\rm eV}$, which is far beyond what thermal neutron sources can offer and belongs within the resonance energy range. However, with the advent of ever-new WSMs, ones that allow observation of intranode scattering may be found\cite{Guo2018aa}. Hence, although we focus in this paper on scattering
between separated Weyl nodes, Appendix \ref{IntraNode_Appendix} points out some differences
that appear for intranode scattering.
%--------------- End: Intra vs. Inter Weyl node scattering -----------------------------------------------------------------------------

%-------- Notice: Declare Manuscript to be Masterfile from which compilation is done  ----------------------------------
%%% Local Variables:
%%% mode: latex
%%% TeX-master: "Manuscript"
%%% End:

%section: Magnetization Operator
\section{Operators for Neutron-Weyl Fermion Interaction\label{MagnetizationOperator_Appendix}}
%--------------- Begin: MagnetizationOperator_Appendix ----------------------------------------------------------------------------

%Neutron scattering depends on the matrix elements of the electronic current
%operator $\mathbf{J}(\mathbf{r})=-\deltaH/\delta\mathbf{A}(\mathbf{r})$. With the help of a full band structure of the material, these matrix
%elements can be calculated as a function of the momenta of the initial
%and final electron states.  Near a Weyl point, one can focus on a few parameters from this calculationThese enter into an effective transition current density\cite{Steinsvoll1967a,Trammell1953a,Hirst1997a}. \ref{DifferentialCrossSection_Appendix_Eq6b:DifferentialCrossSectionMagnetic_Polarized_I}

A Weyl fermion has two internal states, similar to a spin, but these do not necessarily correspond to spin--we call them pseudospin
instead.  The two states could, for example, be two orbitals of atoms with positive and negative \emph{orbital} angular momentum $L_z$,
or could differ in both spin and orbital degrees of freedom, or they could differ in some other way (they do not have to correspond to atomic orbitals of single atoms in fact). Because of this, the operator that interacts with the magnetic field of the neutron is not simply proportional to $\bm{\sigma}$. In this section, we will derive the most general form that this operator takes.  It differs from the ordinary magnetic moment in an additional way--namely, it induces
transitions between two different Weyl nodes.

\subsection{Magnetic Moments of Weyl Fermions}
The interaction\cite{Balcar1989a,Hirst1997a} of a neutron with the WSM is treated in the Born approximation, where the vector potential\footnote{The vector potential $\Ab(\rb-\rb_{\rm n}) = (\mu_{0}/4\pi) \mb_{\rm n}\times (\rb-\rb_{\rm n}/\lvert \rb-\rb_{\rm n} \rvert^{3})$ at spacepoint $\rb$ induced by a neutron magnetic moment operator $\mb_{\rm n} = \gamma \frac{\hbar}{2}\taub$ at $\rb_{\rm n}$, where $\gamma = \frac{g \mu_{\rm n}}{\hbar}$ with nuclear magneton magnetic moment $\m_{\rm n} = \frac{e\hbar}{2 m_{\rm n}}$, neutron g-factor $g$. The permeability of free space is $\mu_{0}$.} operator $\Ab(\rb)$ of the neutron's magnetic moment interacts with the currents of the electronic system. If a full band structure
is available, a direct way to calculate the structure factor would be to 
evaluate the matrix elements of the exact current operator (including
spin and orbital parts) between the Bloch states.
%use the expression for the local electronic current density $\mathbf{J}(\rb) = -\delta H/\delta \Ab(\rb)$ 
%Near a Weyl point, one can focus on a few parameters
%from this calculation, which yields an effective transition current density, see Appendix \ref{IntraNode_Appendix} for discussion. 
Near a Weyl point, one can focus on a few parameters from this calculation, which can be represented as an effective anomalous magnetic moment operator. See Appendix D for a discussion of why the interaction cannot be found by the minimal
substitution in this case.

The basic idea is that the Weyl
Hamiltonian in the vicinity of $\kb_{0,i}$ can be developed just from information about
the degenerate states exactly at these points.
The Hamiltonian at a nearby point $H_{0,i}(\kb_{0,i}+\pb)$ can be understood
by treating $\pb$ as a perturbation.  We project it into
the $2-$fold degenerate subspace $D_{i} = \{\lvert s; \kb_{0,i} \rangle\}$ \emph{exactly at} the nearby Weyl point, enumerated by arbitrary pseudospin label $s = \pm$. These are not necessarily different spin states; they are just any two degenerate states, and could differ in orbital structure instead of spin for example. For momenta $\pb \neq 0$ away from the node, the projected Hamiltonian can be expanded to first order as  
$\mathbf{w}^{(i)} \cdot \pb$ which removes the degeneracy, where $\mathbf{w}^{(i)} = \partial H_{0,i}(\pb+\kb_{0,i})/\partial \pb|_{\pb = \bm{0}}$ is
a vector of $2 \times 2$ matrices. % Taylor expanding $H_{0,i}(\pb + \kb_{0,i})$ around $\kb_{0,i}$ to first order in $|\pb|$ and projecting into subspace $D_{i}$ yields 
 Expanding in terms of Pauli matrices gives the effective low energy Weyl Hamiltonian Eq. (\ref{Kinematics_Appendix_Eq2:HamiltonianAisotropic}), under the assumption
that the nodes are aligned at the chemical potential.
Note that the states $|s,\kb_{0i}\rangle$ are not 
eigenstates at a nonzero $\mathbf{p}$; the energy eigenstates 
take the form $\sum_s c_s(\mathbf{p})|s;\kb_{0i}\rangle$ where the
$c_s$'s form an eigenvector of Eq. (\ref{Kinematics_Appendix_Eq2:HamiltonianAisotropic}).
%\begin{equation}
% \lvert \kb; \e;i \rangle = \sum_{s \in D_{i}} c^{(\e)}_{s;i}(\pb)\lvert s; \kb_{0,i} \rangle  \label{section:MagnetizationOperator_Appendix_Eq1:WeylSpinorStates}
%\end{equation}
%as the expansion coefficients $c^{(\e)}_{s;i}$ diagonalize Eq. (\ref{Kinematics_Appendix_Eq2:HamiltonianAisotropic}).

As mentioned above,
neutron scattering depends on the matrix elements of the electronic
current operator.  These matrix elements have a complicated dependence on the ``kinetic momenta" $\pb$ of the states involved.
However, this dependence can be derived from a simple effective description.
There is an effective operator, a simple $2\times 2$ matrix that
describes the electronic current within the low-energy subspaces.  This
matrix has no momentum dependence (to a good accuracy).  
However, it is defined with
respect to the basis $|s;\kb_{0i}\rangle$ which are not energy eigenstates;
the momentum-dependence of the matrix elements appears
because these eigenstates depend on momentum as
$\sum_s c_s(\mathbf{p})|s;\kb_{0,i}\rangle$. 

For an $\M^{+}$ transition, we need only the current's overlaps between states of the degenerate subspaces $D_{1}$ and $D_{2}$. 
 %Each component of $\mathbf{J}$
%thus gives a $2\times 2$ matrix, and all three components form a vector  $\bm{\mathcal{J}}(2\kb_{0})$ of $2\times 2$ matrices. 
The current $\mathbf{J}$ forms a vector $\bm{\mathcal{J}}(2\kb_{0})$ of $2\times 2$ matrices. 
The dependence on $\mathbf{p}$ can
be neglected since
the basis states are constant within the first order approximation aside from multiplication by $e^{i\pb\cdot\rb}$ to change the crystal momentum.
(The basis states are nearly constant by the perturbation theory approach
discussed above; the \emph{eigenstates} vary strongly because
$\pb$ acts as a perturbation to a degenerate Hamiltonian).
 Within the effective
Weyl fermion description, $\bm{\mathcal{J}}(2\kb_0)$ is the $1^\mathrm{st}$
quantized operator
corresponding to the current; it has the same matrix elements for corresponding
states in the effective and more realistic descriptions.
Conservation of momentum gives
\begin{equation} 
\langle s; \kb_{0,2} +\pb_2\rvert  \mathbf{J}(\qb) \lvert \kb_{0,1}+\pb_1; s' \rangle  = \d^{3}(\qb - 2\kb_{0}+\bm{\Delta}) \bm{\mathcal{J}}(2\kb_{0})_{ss'} ,\label{section:MagnetizationOperator_Appendix_Eq1:CurrentOperator}
\end{equation} 
without any dependence of the matrix elements on $\Dtb$, which is valid for $|\Dtb| \ll |2\kb_{0}|$ as is considered in this article\footnote{A Bloch band model of the nodes would be a more accurate treatment and higher orders in $|\pb|$ could be included. If doing so, the overlap would be \unexpanded{$\langle s;\kb_{2} \vert  \bm{\mathcal{J}}(\qb) \vert \kb_{1};s' \rangle  \approx \d^{3}\left(\qb + \Db - 2\kb_{0} \right)  \bm{\mathcal{J}}(\qb)_{ss'}$, and $\bm{\mathcal{J}}(\qb)$} would approximately be a constant matrix in the Bloch band states. Consequently, the coupling $\Fb^{\mu}$ to be introduced below would become dependent on $\qb = \D \kb_{0} - \Db$.}. The electron-neutron coupling can now be reduced to
\begin{equation}
H_{\Ab} =  -\int_{V} \dd \rb\, \mathbf{J}_{eff}(\rb)\cdot \Ab(\rb - \rb_{n})
\label{section:MagnetizationOperator_Appendix_Interaction:HA}
\end{equation}      
where the Weyl-fermion current is given by
\begin{equation}
\mathbf{J}_{eff}(\rb,t)=\Psi_2^\dagger(\rb,t)\bm{\mathcal{J}}(2\kb_0)\Psi_1(\rb,t)+h.c.
\label{section:MagnetizationOperator_Appendix_current}
\end{equation}
and $\mathbf{A}(\rb-\rb_{\rm n})$ is the vector potential of a neutron at $\rb_{\rm n}$.%; since the momentum dependence is weak, we replace the current operator in momentum
%space by a constant which produces in real space this local term. 

This current can be interpreted as a magnetic moment.  We first need
a crucial fact that can be obtained by using conservation of charge $\partial \rho/\partial t + \bm{\nabla}\cdot \bm{\mathcal{J}} = 0$ and Heisenberg's equation of motion $\partial \rho/\partial t = (i/\hbar)\left[H,\rho \right] $ for the local electronic particle density operator. 
One finds that $2\kb_0\cdot\bm{\mathcal{J}}(2\kb_0)=0$ since
the matrix elements of $[H,\rho]$ are 0 between degenerate states.  Hence the transition current density is purely transverse with respect to $2\kb_{0}$ and can therefore be expressed as $\bm{\mathcal{J}}(2\kb_{0}) =- i 2\kb_{0}/\hbar \times \bm{\mathcal{M}}(2\kb_{0})$. This operator has the interpretation of a magnetization operator. Substituting
for $\bm{\mathcal{J}}$ in 
Eq. (\ref{section:MagnetizationOperator_Appendix_current})
in terms of $\bm{\mathcal{M}}$ and replacing $i2\mathbf{k}_0/\hbar$ by the gradient (which is valid for momenta
near the nodes), we find $\mathbf{J}=\mathrm{curl}\ \mathbf{M}$ where
\begin{equation}
\mathbf{M}(\rb,t) = \Psi_{1}^{\dagger}(\rb,t) \bm{\mathcal{M}}^{\dagger} \Psi_{2}(\rb,t) + \Psi_{2}^{\dagger}(\rb,t) \bm{\mathcal{M}} \Psi_{1}(\rb,t) .\label{section:CurrentOperator_Appendix_Eq.1:MagnetizationOperator2}
\end{equation}  This allows one to express
the interaction between the neutron and the electrons, Eq. (\ref{section:MagnetizationOperator_Appendix_Interaction:HA}), as the standard form for
the energy of a dipole in a magnetic field:
\begin{equation}
H_{\Bb} = - \int_{V} \dd \rb\, \mathbf{M}(\rb)\cdot \Bb(\rb-\rb_{n}). \label{section:MagnetizationOperator_Appendix_Interaction:HB}
\end{equation}      
%and, including the conjugate $\tau^{-}$ transition, is given by 
%\begin{equation}
%\mathbf{M}(\rb,t) = \Psi_{1}^{\dagger}(\rb,t) \bm{\mathcal{M}}^{\dagger} \Psi_{2}(\rb,t) + \Psi_{2}^{\dagger}(\rb,t) \bm{\mathcal{M}} \Psi_{1}(\rb,t) ,\label{section:CurrentOperator_Appendix_Eq.1:MagnetizationOperator2}
%\end{equation}
Furthermore, the magnetization
$\bm{\mathcal{M}}$, being a $2 \times 2$ matrix,
can be expanded as:
\begin{equation}
\bm{\mathcal{M}} = \mu_{\rm B}\s_{\mu}\Fb^{\mu}  \quad, \quad \mu = 0,1,2,3 .\label{section:CurrentOperator_Appendix_Eq.1:MagnetizationOperator1}
\end{equation}
%This is somewhat similar to the expression $-\frac{g}{2}\mu_B\bm{\sigma}$ for
%the moment of an electron, but with an arbitrary matrix of coefficients.
Defining the $j^{th}$ component of $\Fb^{\mu} \in \mathds{C}^3$  to be $\Fb^{\mu}$, % = \F^{\mu}_{\hspace*{0.15cm}j} = \F_{\mu j}$, 
a $4\times 3$ matrix which describes the coupling between the \enquote{magnetic} degree of freedom $j$ and the \enquote{pseudospin} degree of freedom $\mu$. 
Since these indices transform differently (one with
spatial rotations and one with redefinition of the pseudospin basis),  $\F^{\mu}_{\hspace*{0.15cm} j}$ is not a geometrical object. It is merely a collection of complex coupling coefficients which relate the magnetic moment
to the spin, similar to the factor $\frac{g}{2}$ for an electron spin-$\frac{1}{2}$ magnetic moment $(g/2)\mu_{\rm B}\sb$ which Eq. (\ref{section:CurrentOperator_Appendix_Eq.1:MagnetizationOperator1}) is a generalization of. 
Roughly, $\Fb^{\mu}$ can be interpreted as the \enquote{anomalous} components of a \enquote{Weyl magnetic moment}. %, whose interaction with the magnetic field $\mathbf{B}$ of a neutron is Eq. (\ref{section:MagnetizationOperator_Appendix_Interaction:HB}). 
However, it is not completely right to use this analogy. The reason is that the interaction involves a transition between states of two \emph{different} nodes. Hence, the presence of the \enquote{anomalous magnetic moment} coupling $\Fb^{\mu}$ is a \emph{quantum effect from the bands}, which acts like a force on the pseudospin.

%One can relate the coupling and  transition current density to each other through Eq. (\ref{section:MagnetizationOperator_Appendix_Interaction:HB}) and
%\begin{equation}
%H_{\Ab} =  -\int_{V} \dd \rb\, \bm{\mathcal{J}}(\rb)\cdot \Ab(\rb - \rb_{n}). \label{section:MagnetizationOperator_Appendix_Interaction:HA}
%\end{equation}

The parameters $\mathbf{F}^\mu$ can be determined numerically if one has
developed a realistic band structure model. 
%The two ``spin" states  at one of the Weyl points are not simply spin-up and spin-down
%versions of the same wave-function, but are just some
%pair of degenerate wave-functions.  
Evaluating the current operator (including
the currents associated with the spin) between the pair of degenerate wavefunctions gives a $2\times2$
matrix from which one can obtain the $\mathbf{F}$'s.
With respect to $\kb_{0}$ these can be divided into longitudinal and transverse parts $\mathbf{F}^\mu=\mathbf{F}^\mu_\parallel+\mathbf{F}^\mu_\perp$.  
%These can be divided into longitudinal and transverse parts wrt. $\kb_{0}$, viz.,  $\mathbf{F}^\mu=\mathbf{F}^\mu_\parallel+\mathbf{F}^\mu_\perp$.  
We have the freedom to set $\mathbf{F}^\mu_\parallel=\bm{0}$ and by Eq. (\ref{section:CurrentOperator_Appendix_Eq.1:MagnetizationOperator1}) and the relation between $\bm{\mathcal{M}}$ and $\bm{\mathcal{J}}$,
$\mathbf{F}_\perp$ can be found in two stages as:
%With the resulting expression Eq. (\ref{section:MagnetizationOperator_Appendix_AnomalousCoupling}) one can calculate the coupling numerically if employing a realistic band structure model by using the Schr{\"o}dinger current in Bloch band states. The coupling $\Fb^{\mu} = \Fb^{\mu}_{\parallel} + \Fb^{\mu}_{\perp}$ is defined for a $\tau^{+}$ process as above (in the original coordinate system) and is decomposed into a parallel component, for which we have the freedom to set $\Fb^{\mu}_{\parallel} = \bm{0}$, and two transverse components\footnote{Notice that $\Fb^{\mu}$ defined in Eq. (\ref{section:MagnetizationOperator_Appendix_AnomalousCoupling}) applies to nodes Eq. (\ref{Kinematics_Appendix_Eq2:HamiltonianAisotropic}) \emph{not} necessarily isotropic. In addition, if the nodes were not aligned at the chemical potential the current would have a longitudinal contribution $\bm{\mathcal{J}}_{\hspace*{-0.10cm}\parallel}$. But the cross-section is always independent of $\bm{\mathcal{J}}_{\hspace*{-0.10cm}\parallel}$, so such an offset in node positions will not add any terms.}   
\begin{subequations} 
\label{section:MagnetizationOperator_Appendix_AnomalousCoupling}
\begin{eqnarray}
\Fb^{\mu}_{\top} &=& \widehat{\kb}_{0} \times \Fb^{\mu} =  \frac{i\hbar}{2 \lvert 2 \kb_{0} \rvert \mu_{\rm B}}\text{Tr}\left[\bm{\mathcal{J}}(2\kb_{0}) \s^{\mu} \right],  \\ 
\Fb^{\mu}_{\perp} &=& \Fb^{\mu}_{\top} \times \widehat{\kb}_{0}. 
\end{eqnarray}
\end{subequations}
Contrary to conventional purely magnetic scattering, the coupling Eq. (\ref{section:MagnetizationOperator_Appendix_AnomalousCoupling}) is determined by sixteen real numbers without invoking constraints from symmetry. These contain information from bands solely, so without a specific band model these are unknown. Thus the coupling is structurally much more complicated than the bare coupling of neutrons with matter, which is just a single number with magnitude $g/2 = 1$. That is, $\Fb_\perp^{0} \neq \bm{0}$ \emph{generally} and $\Fb_\perp^{i}\cdot \hat{j} \neq \d_{ij}$ \emph{always}, (by the constraint $2\kb_0\cdot\Fb_\perp^i=0$), and can even be very asymmetric with either a larger or smaller value than the bare coupling. Furthermore,  $\Fb^{\mu}_{\perp}$ may become divergent upon approaching $|2\kb_{0}| \to 0$ a topological phase transition. An example of these features is illustrated in Section \ref{4bandWSM_Appendix} for a toy model.

%--------------- End: MagnetizationOperator_Appendix -----------------------------------------------------------------------------

%-------- Notice: Declare Manuscript to be Masterfile from which compilation is done  ----------------------------------
%%% Local Variables:
%%% mode: latex
%%% TeX-master: "Manuscript"
%%% End:

%section: Magnetization Operator
%subsection: 4bandWSM_Appendix
\subsection{Example: minimal $4$-band toy model of inversion invariant WSM\label{4bandWSM_Appendix}}
\begin{figure}
\includegraphics[width=0.85\columnwidth]{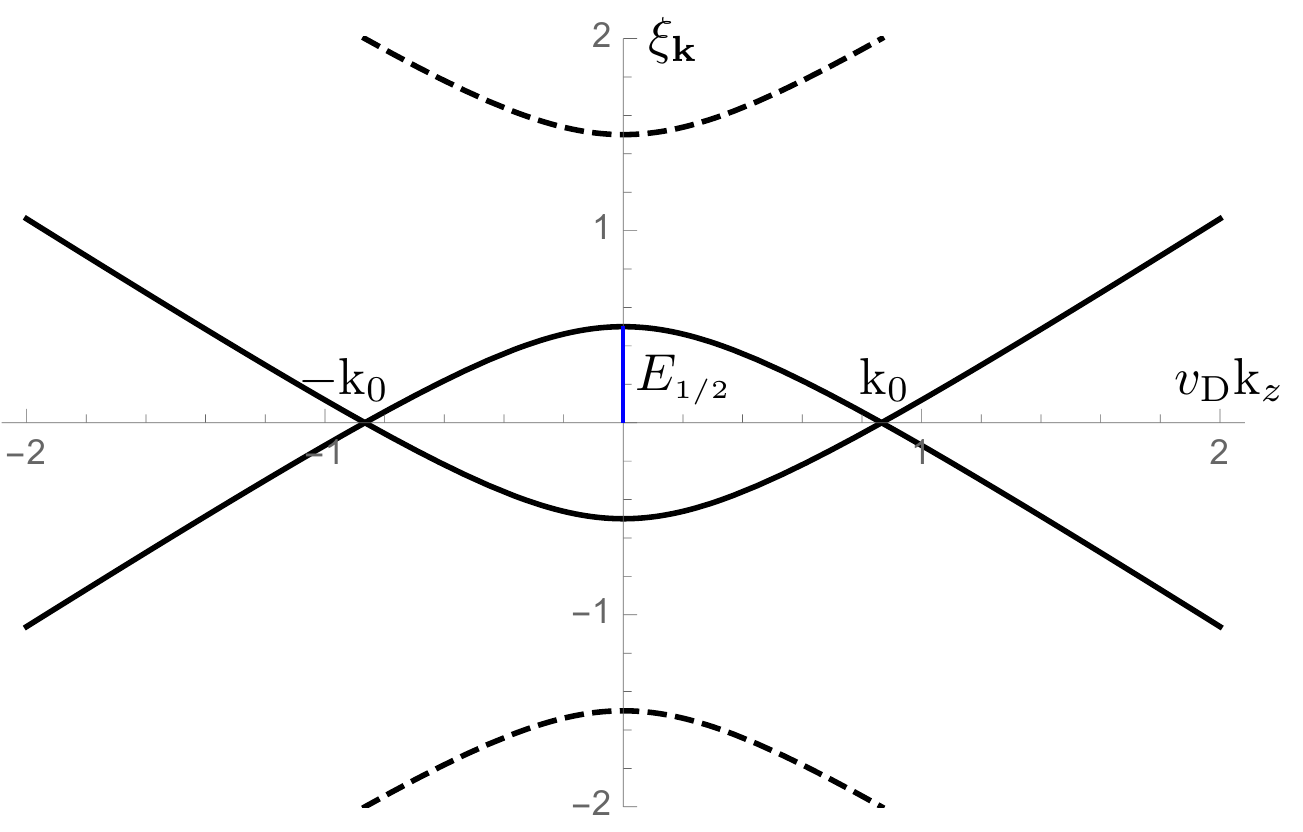}%[scale=1]
\caption{Energy spectrum of $4$-band model for $|\delta|/m = 0.5$. The half-energy gap (blue line) $E_{\scriptscriptstyle 1/2} = m -|\d|$ is indicated.\label{4BandModel_d04m05}}
\end{figure}

Analogous to Ref. \onlinecite{Burkov2011a,*Burkov2011b}, a minimal time-reversal breaking and inversion invariant WSM can be obtained by starting
with a material that is tuned to the transition between a topological
and normal insulator and introducing magnetic impurities.  
In a time-reversal symmetric material that is tuned to the transition point, 
the gap is closed producing
3D Dirac points, which we suppose to be at momentum 0.  
The Dirac points are described by
a Hamiltonian $H_{3\,\rm{D}} =  v_{\rm D} \kb \cdot \sb \tau^{z}$.
%, writtenin terms of two sets of Pauli operators $\bm{\sigma},\bm{\tau}$. 
These may be regarded as two Weyl nodes, 
labelled by $\tau^{z} = \pm 1$, and they have opposite chiralities,
also given by $\tau_z$. The $\bm{\sigma}$s correspond to the spin
of the state, while $\tau$ labels different bands.  As one moves
away from the topological transition,  a hybridization term appears  $H_{\delta} = \delta \s^{0} \tau^{x}$ that couples the nodes with strength $\delta$ and produces a gap.
%, which can also be interpreted as a mass term in the Dirac equation. 
Returning to the transition point and introducing magnetic impurities $H_{\rm Z} = -m \s^{z} \tau^{0}$ 
that are assumed to order ferromagnetically along the z-direction and interact
equally with both orbitals breaks time-reversal symmetry and separates the nodes in momentum space.  If the hybridization term is
present as well and not too large then it will not open a gap and
the Weyl points will remain stable as long as $m>|\delta|$ assuming 
that $m>0$.
This yields a basic minimal $4$-band toy model whose Hamiltonian $H^{0}_{4} =  H_{3\,\rm D} + H_{\delta} + H_{\rm Z}$ has nodes at $\kb_{0,2} = -\kb_{0,1} = \k_{0}\hat{z}$, where $v_{\rm D} \k_{0} = \sqrt{m^2-\delta^2}$, and its energy spectrum is plotted in Fig. \ref{4BandModel_d04m05}.
%
%\begin{figure}
%\includegraphics[width=0.95\columnwidth]{./Fig1_4bandWSM_Apendix}%[scale=1]
%\caption{Energy spectrum of $H^{0}_{4}$ for $|\delta|/m = 0.5$. The half-energy gap (blue line) $E_{\scriptscriptstyle 1/2} = m -|\d|$ is indicated\label{4BandModel_d04m05}}
%\end{figure}
%
\begin{comment}
\begin{figure}
\begin{minipage}[c]{0.45\columnwidth}
\centering
\begin{tikzpicture}
\node[anchor=south west, inner sep=0] at (0,0) {\includegraphics[width=0.75\textwidth]{./4BandModelInversionWSM_Spectrum_m1d05_Axes}};
\node at (6.5,2.05) {$v_{\rm F} \k_{z}$};
\node at (3.05,4.15) {$\xi_{\kb}$};
\draw[very thick, blue] (3.025,1.975) -- (3.025,2.475);
\node at (3.45,2.15) {\tiny $E_{\scriptscriptstyle 1/2}$};
\end{tikzpicture}
\end{minipage}\hfill
\begin{minipage}{1\columnwidth} % right
\caption{Energy spectrum of $H^{0}_{4}$ for $|\delta|/m = 0.5$. The half-energy gap (blue line) $E_{\scriptscriptstyle 1/2} = m -|\d|$ is indicated.\label{4BandModel_d04m05}} 
\end{minipage}
\end{figure}
\end{comment}
Each node $i = 1, 2$ has a degenerate subspace $D_{i} = \left\{ \lvert s;\kb_{0,i} \rangle \right\} $ enumerated by pseudospin $s = \pm$. The Hamiltonian is inversion symmetric, i.e. $P H^{0}_{4}(\kb) P^{-1} = H^{0}_{4}(-\kb)$, where
inversion is $P=\sigma^0\tau^x$.  As explained in Appendix \ref{PrincipalAxisTransformation_Appendix} and \ref{Interaction_Appendix}, in order to be
sure that the effective Hamiltonian can be transformed into an
isotropic form, the inversion symmetry must act as the identity--this
\emph{is} true within the space of degenerate states
%\footnote{{\color{green}The states expressed in the basis \unexpanded{$|\sigma_z\tau_z\rangle=\lvert\nolinebreak\uparrow\nolinebreak\uparrow\nolinebreak\rangle,\lvert\nolinebreak\uparrow\nolinebreak\downarrow\nolinebreak\rangle$}, \unexpanded{$\sloppy{\lvert\nolinebreak\downarrow\nolinebreak\uparrow\nolinebreak\rangle},\lvert\downarrow\downarrow\rangle$} are \unexpanded{$\lvert +;\kb_{0,1} \rangle = \left(u_{-}, u_{+}, 0, 0 \right)^{\rm T}$} and \unexpanded{$\lvert -;\kb_{0,1} \rangle = \left(0, 0, u_{-},-u_{+}\right)^{\rm T}$} are normalized such that $u_{-}^{2} + u_{+}^{2} = 1$ with $u_{\pm} = (1/\sqrt{2}) \sqrt{1\pm \sqrt{1-(\lvert\delta\rvert/m)^2}}$ for a given $ 0\leq \delta\rvert<m $. If $\delta<0$, the sign of $u_-$ is also flipped.}}, 
since $P\lvert s, \kb_{1}\rangle = \lvert s, \kb_{2}\rangle$. As expected the effective low energy Hamiltonians at the
two Weyl points %are negatives of one another; they 
have the form of Eq. (\ref{Kinematics_Appendix_Eq2:HamiltonianAisotropic}) with $\lambda^{(1)} = - \lambda^{(2)} = diag(+1,+1,-\sqrt{1- (\delta/m)^2})$, and $\vb_{0}^{(i)} = \bm{0}$. 

As we consider only scattering within the low energy sector of the nodes, the coupling Eq. (\ref{section:CurrentOperator_Appendix_Eq.1:MagnetizationOperator1}) is determined by evaluating the matrix elements of the current exactly at the Weyl node positions, i.e. evaluating the left-hand side of Eq. (\ref{section:MagnetizationOperator_Appendix_Eq1:CurrentOperator}) for the eigenfunctions
of our model with $\pb_1=\pb_2=\bm{0}$,
and comparing to the right-hand side evaluated using the
effective description, Eq. \eqref{section:CurrentOperator_Appendix_Eq.1:MagnetizationOperator1}. Note that in the effective model,
the spin operators $\sigma^i$ are redefined to act on the two-dimensional
subspace, e.g., $\sigma^z|s;\kb_{0,i}\rangle=s|s;\kb_{0,i}\rangle$, whereas
the eigenstates are not eigenfunctions of the original $\sigma_z$.
The $\Fb^\mu$ can then be solved for [giving Eq. \eqref{section:MagnetizationOperator_Appendix_AnomalousCoupling}].
The current operator in this model is $\bm{\mathcal{J}} = e v_{\rm D} \sb \tau_{z}$; this is obtained by introducing a coupling to the vector potential into $H^{0}_{4}$ by a minimal substitution (see Appendix \ref{IntraNode_Appendix} for
justification) and then comparing the term linear
in $\mathbf{A}$ with Eq. \eqref{section:MagnetizationOperator_Appendix_current}.
Consequently $\Fb_{\perp}^{\mu}$ has nonzero components $\F^{x}_{\perp,x}$ and $\F^{y}_{\perp,y}$, which both have the same magnitude,
\begin{equation}
\F^{\perp}_{x\, x}  =  \frac{e v_{\rm D}}{\mu_{\rm B}} \frac{\delta}{m} \frac{\hbar}{\lvert 2\kb_{0} \rvert} =\pm \frac{m_{\rm e} v_{\rm D}^{2}}{v_{\rm D} \k_{0}} \frac{(v_{\rm D} \k_{0})^2 - E_{\scriptscriptstyle 1/2}^2}{(v_{\rm D} \k_{0})^2 + E_{\scriptscriptstyle 1/2}^2} , \label{4bandWSM_Appendix_Eq1:Fxx}
\end{equation}
where $E_{\scriptscriptstyle 1/2} = m -|\d| \leq v_{\rm D} \k_{0}$ is the half-energy gap at $\kb = \bm{0}$ indicated in Fig. \ref{4BandModel_d04m05}. 
The second expression
is written in terms of parameters of the bands' dispersion; the sign just depends on the sign of $\delta$ which cannot be seen from the dispersion. 

\begin{figure}
\includegraphics[width=0.85\columnwidth]{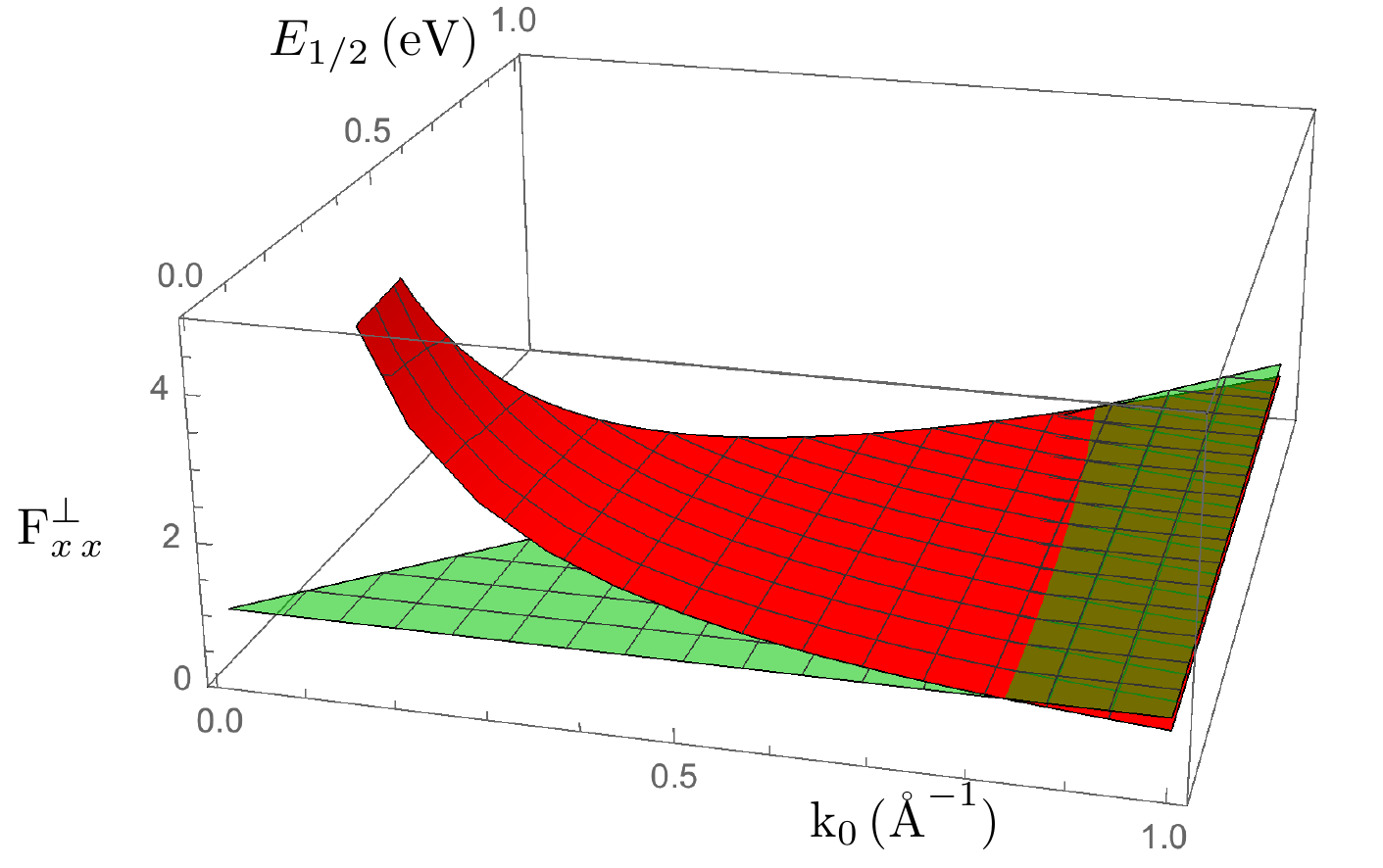}
%\caption{Plot of Eq. (\ref{4bandWSM_Appendix_Eq1:Fxx}) for $v_{\rm D} = c / 300$ in red as a function of $\k_{0}$ (units of $\si{\angstrom}^{-1}$) and $E_{\scriptscriptstyle 1/2}$ (units of ${\rm eV}$). The green plane corresponds to $\mathcal{M}/\mu_{\rm B} = g/2 = 1$ of the electron.\label{Coupling}}
\caption{Coupling of neutron to Weyl fermions. The coupling, Eq. (\ref{4bandWSM_Appendix_Eq1:Fxx}), for $v_{\rm D} = c / 300$ is plotted (red) as a function of node position $\k_{0}$ and half-energy gap $E_{\scriptscriptstyle 1/2}$ of spectrum in Fig. \ref{4BandModel_d04m05}. The bare coupling of neutron to electrons, i.e. $g/2 = 1$, is the (green) plane.\label{Coupling}}
\end{figure}

For example, for a Fermi velocity of order $v_{\rm D} = c / 300$, 
the magnetic moment per Bohr magneton for the internode coupling, i.e. its $g/2$-factor, is plotted in Fig. \ref{Coupling} as a function of node position and half-energy gap. Hence the coupling of a neutron to nodes is comparable to, smaller or even much larger than that of the electron and may diverge upon approaching the topological phase transition.  The cross-section will be estimated in the Section \ref{Interpretation_Appendix:Dispersion}.
%{\color{green} The cross-section will be estimated in the next section, and is of magnitude $\sim (\F^{\perp}_{x\,x})^2 (g r_{0}/2)^2 D(\Dtb,\omega)$ per unit volume, where $g$, $r_{0}$, $D(\Dtb,\omega) \sim (\hbar\omega/v_{\rm F})^2$ are the neutron's g-factor, the classical electron radius, and density of states of the scattering process [see Eq. (\ref{Interpretation_Appendix:Eq1_DOScalculation})] respectively. I THINK THE FOLLOWING ISN'T REALLY RIGHT: Conventional wisdom would suppose $\F^{\perp}_{x\,x}$ to equal $1$. But from Fig. \ref{Coupling} one sees that  $\F^{\perp}_{x\,x}$ is larger than $1$  for a wide range of node distances and half-energy sizes. For these, the cross-section is amplified much more than what intuitively would be expected.} 
The above features hold, at least for this toy model which does not represent a realistic model.  However, these features could be more generic in nature and 
hence present in real WSMs, but this question is left unanswered here.  Alternatively some Weyl materials will be found that can actually be described as topological insulators with magnetic impurities.

%\begin{comment}
%\begin{figure}
%\includegraphics[scale=0.55]{./Fig4_MagnetizationOperator_Appendix}
%\caption{Plot of Eq. (\ref{4bandWSM_Appendix_Eq1:Fxx}) for $v_{\rm F} = c / 300$ in red as a function of $\k_{0}$ (units of $\si{\angstrom}^{-1}$) and $E_{\scriptscriptstyle 1/2}$ (units of ${\rm eV}$), which are the range of parameters in reference\cite{Wang2016a,Kubler2016a}. The green and blue planes at $\pm 1$ corresponds to $\mathcal{M}/\mu_{\rm B} = g/2 = \pm 1$ for the electron and positron, respectively.\label{Coupling}}
%\end{figure}
%\begin{figure}
%\begin{minipage}{0.45\textwidth} 
%\begin{tikzpicture}
%\node[anchor=south west, inner sep=0] at (0,0) {\includegraphics[width=0.75\textwidth]{./4BandModelInversionWSM_Coupling}};
%\end{tikzpicture}
%\end{minipage}\hfill
%\begin{minipage}{1\columnwidth} 
%\caption{Plot of Eq. (\ref{4bandWSM_Appendix_Eq1:Fxx}) for $v_{\rm F} \sim c / 300$ in red as a function of $\k_{0}$ (units of $\si{\angstrom}^{-1}$) and $E_{\scriptscriptstyle 1/2}$ (units of ${\rm eV}$), which are the range of parameters in %reference\cite{Wang2016a,Kubler2016a}. The green and blue planes at $\pm 1$ corresponds to $\mathcal{M}/\mu_{\rm B} = g/2 = \pm 1$ for the electron and positron, respectively.\label{Coupling}}
%\end{minipage}
%\end{figure}
%\end{comment}

%-------- Notice: Declare Manuscript to be Masterfile from which compilation is done  ----------------------------------
%%% Local Variables:
%%% mode: latex
%%% TeX-master: "Manuscript"
%%% End:

%section: Statement of Results
%\newpage
\section{Inelastic cross-Section and formalism\label{CrossSectionMagneticFormula_Appendix}}
We will now present the formulae for scattering
cross-sections. These results apply if
the scattering is between two nodes that
are related either by inversion or time reversal symmetry and
that are aligned at (or near) the chemical potential. 
%From now on we will call the positions of the first and second Weyl nodes $-\kb_0$ and $\kb_0$
%respectively, so that $\Delta\kb_0=2\kb_0$.  
Furthermore, we need to assume that
the three parameters $\vb_{0}$ of Eq. (\ref{Kinematics_Appendix_Eq1:Hamiltonian}) are negligible.  These conditions allow the results to be obtained
and interpreted in a relativistic way, as discussed above. 

We will give the cross-section in detail, for arbitrary initial and final
neutron polarization and arbitrary momentum and energy transfer.
To be more precise,
consider incident neutrons of a given momentum $\qb_i$ and spin state represented by a spinor $|\tau_i\rangle$.
Suppose a detector filters the neutrons
according to their final momentum and spin eigenvalue $\pm 1/2$ along
a specific direction and counts only the neutrons with eigenvalue $+1/2$,
described by the state $|\tau_f\rangle$, say.
Then the counting rate %for a specific momentum and eigenvalue $+1/2$
is proportional to the rate of transitions from
the initial neutron state $\lvert i_{\rm n} \rangle = \lvert \qb_{i};\tau_{i} \rangle$ via interactions
with the WSM, defined by the Hamiltonian $H_{0,1} + H_{0,2}$, to the final state  $\lvert f_{\rm n}\rangle=\lvert \qb_f;\tau_f\rangle$.  
The WSM begins in the ground state, $\lvert i_{\rm w}\rangle$,
and ends in $\lvert f_{\rm w}\rangle$  upon absorbing neutron momentum  $\qb = \qb_{i} - \qb_{f}$ and energy $\hbar \omega$. The total differential cross-section is then %given by
\begin{equation}
\frac{\text{d}^{2} \s (\qb,\omega)}{\text{d} \Omega \text{d} E_{f}} \bigg|_{\tau_{i}}^{\tau_{f}} \hspace*{-0.25cm}\approx  \frac{\q_{f}}{\q_{i}} \left(\frac{m_{\rm n}}{2\pi \hbar^2}\right)^{2} \frac{\mu_{0}^{2}}{2\pi\hbar}\hspace*{-0.15cm} \sum_{l,m =1}^{3} \,\m_{\perp,l}^{if}\m_{\perp,m}^{fi} S_{l\,m}(\qb,\omega)  ,\label{DifferentialCrossSection_Appendix_Eq2:DifferentialCrossSectionMagnetic}
\end{equation}
where %$\mu_0$ is the permeability of free space and 
the matrix element of the perpendicular component (with respect to the internode direction) of neutron magnetic moment \footnote{The component of $\mb$ that enters the cross-section should really be the component perpendicular to the momentum transfer $\qb$, but since we focus on low energy scattering, $|\qb-\Delta\kb_0|\ll|\Delta\kb_0|$ the error is negligible, about $\hbar\omega/v_{\rm F} \k_0$.} is $\mb^{fi}_{\perp} = \langle \tau_{f} \rvert \mb_{\perp} \lvert \tau_{i} \rangle$.

The dynamic structure factor $S_{l\,m}(\qb,\omega)$ is the frequency and momentum Fourier transform of the scattering function $S_{l\,m}(\rb,t)$, which can be decomposed into $S_{l\,m}(\rb,t) =  S^{(-)}_{l\,m}(\rb,t) + S^{(+)}_{l\,m}(\rb,t)$ (the contributions of the two processes $\M^{\pm}$ defined in Fig. \ref{Kinematics_Appendix_fig1:Cross-section_Transitions}), since we can ignore intranode scattering. For the $\M^+$ process 
 %The structure factor of a $\tau^{-}$ transition follows trivially from that of a $\tau^{+}$ transition simply by interchange Weyl node labels
%\footnote{In the following, Weyl node indices $1$ and $2$ will not appear explicitly, but only implicitly.  However, interchange $1 \leftrightarrow 2$ is equivalent to interchanging signs $(+) \leftrightarrow (-)$, and conjugate the coupling constants $\F^{\m}_{l} \F^{\nu,*}_{m} \leftrightarrow \F^{\m,*}_{l} \F^{\nu}_{m}$ whenever they appear.} $1 \leftrightarrow 2$, where
\begin{equation}
S^{(+)}_{i\,j}(\rb,t) \equiv V\left< \M^{(-)}_{i}(\rb,t)\,\M^{(+)}_{j}(\mathbf{0},0) \right>_{0}, \nonumber 
%v&=& \m_{\rm B}^{2}\F^{\m,*}_{i} \F^{\nu}_{j} \s^{(+)}_{\m\, \nu}(\rb,t) , \label{DifferentialCrossSection_Appendix_Eq1:structure factor}
\end{equation}
%
%
%\begin{eqnarray}
%S^{(+)}_{i\,j}(\rb,t) &\equiv& V\left< \M^{(-)}_{i}(\rb,t)\,\M^{(+)}_{j}(\mathbf{0},0) \right>_{0}, \nonumber \\
%&=& \m_{\rm B}^{2}\F^{\m,*}_{i} \F^{\nu}_{j} \s^{(+)}_{\m\, \nu}(\rb,t) , \label{DifferentialCrossSection_Appendix_Eq1:structure factor}
%\end{eqnarray}
which expresses the fact that it is a van Hove type correlation function of magnetization operators Eq. (\ref{section:CurrentOperator_Appendix_Eq.1:MagnetizationOperator2}).  The structure factor of an $\M^{-}$ transition follows trivially from that of an $\M^{+}$ transition simply by interchanging Weyl node labels
\footnote{In the following, Weyl node indices $1$ and $2$ will not appear explicitly, but only implicitly.  However, interchange $1 \leftrightarrow 2$ is equivalent to interchanging signs $(+) \leftrightarrow (-)$, and conjugate the coupling constants $\F^{\m}_{l} \F^{\nu,*}_{m} \leftrightarrow \F^{\m,*}_{l} \F^{\nu}_{m}$ whenever they appear.} $1 \leftrightarrow 2$.

The structure factor $S_{l\,m}(\qb,\omega)$ considered as a function of neutron momentum transfer $\qb = 2\kb_{0} - \Db$, will be concentrated in small spheres centered at $2\kb_{0}$ as illustrated in Fig. \ref{Kinematics_Appendix_fig1:Cross-section_Transitions}.  To focus on this region, it is convenient to describe the cross-section in a coordinate system of $\Dtb$. 

The previous expression can be written as
\begin{equation}
S^{(+)}_{ij}(\rb,t)= \m_{\rm B}^{2}\F^{\m,*}_{i} \F^{\nu}_{j} \s^{(+)}_{\m\, \nu}(\rb,t) , \label{DifferentialCrossSection_Appendix_Eq1:structure factor}
\end{equation}
where the
intermediate scattering function 
\begin{equation}
\s^{(+)}_{\mu\, \nu}(\rb,t)=\langle \Psi_1^\dagger(\rb,t)\sigma_\mu\Psi_2(\rb,t)\Psi_2^\dagger(\mathbf{0},0)\sigma_\nu\Psi_1(\mathbf{0},0)\rangle_{0} V, \label{DifferentialCrossSection_Appendix_Eq1:IntermediateStructureFactor}
\end{equation}
is a particle-hole correlator of the relativistic Weyl fermions. It can be related to the absorptive part of the generalized susceptibility $\chi^{(+)}_{\m\, \nu}$ by the fluctuation-dissipation theorem. 
For conventional
neutron scattering, the neutrons interact mainly with the spin degrees of
freedom and hence $\sigma_{\mu\nu}^{(+)}(\qb,\omega)/2\hbar$ describes
the spin susceptibility.  In this case, the states of the Weyl fermions
are pseudospin states, so $\sigma$ does not correspond to the spin.  Instead, $\sigma_{\mu\nu}^{(+)}(\qb,\omega)/2\hbar$ describes the full magnetic susceptibility including both orbital and spin contributions to the magnetic moments,
since we determined the magnetization operator in a way that includes all these contributions. 

%{\color{red} For conventional purely magnetic scattering the inelastic neutron cross-section Eq. (\ref{DifferentialCrossSection_Appendix_Eq6:DifferentialCrossSectionMagnetic}) is proportional to the dynamic magnetic/spin susceptibility $\s^{\m\, \nu}_{(+)}(\qb,\omega)/2\hbar$, which determines the magnetization induced by the probing magnetic field with momentum and frequency from the neutron. Although $\sb$ describes the pseudospin in the Weyl Eq.  (\ref{Kinematics_Appendix_Eq2:HamiltonianAisotropic}), the \enquote{magnetization} operator for the neutron-WSM interaction involves pseudospin matrices $\sb$ which acts on the subspaces of two Weyl nodes, the states of which can be different with respect to both spin and orbital properties. Therefore $\Mb$ is actually \emph{not} describing the magnetization of the WSM, and the susceptibility consequently only describes correlations between the particle-hole Weyl states' spinors generated in the scattering process.} 

The susceptibility can be calculated by integrating over all possible Weyl particle-hole pairs.  At zero temperature we exploit Lorentz invariance to evaluate this analytically (see Appendix \ref{Susceptibility_Appendix}). 
%For scattering between time-reversal (inversion) symmetric Weyl nodes denoted by their chirality $\chi_{i} = \chi$ and $\chi_{f} = \chi$ ($\chi_{f} = - \chi \equiv 
%\bar{\chi}$), 
When the nodes are related by time-reversal symmetry,
they have the same chirality, say $\chi_i=\chi_f=\chi$. The susceptibility for the scattering process is
\begin{eqnarray}
\chi''^{\m\, \nu}_{(+)}(\qb,\omega)  = \s^{\m\, \nu}_{(+)}(\qb,\omega)/2\hbar .  \label{DifferentialCrossSection_Appendix_Eq1:Susceptibility}
\end{eqnarray} 
For time-reversal symmetric nodes it is a Lorentz invariant rank-$2$ tensor with components:
\begin{subequations} 
\label{DifferentialCrossSection_Appendix_Eq1:TR-Susceptibility}
\begin{eqnarray}
a^{-1}\chi''^{0\,0}_{(+)}(\qb,\omega) &=& \lvert\Dtb\rvert^{2}  \label{DifferentialCrossSection_Appendix_Eq1:TR-Susceptibility1} \\
a^{-1}\chi''^{0\,i}_{(+)}(\qb,\omega) &=& a^{-1} \chi''^{i\,0}_{(+)}(\qb,\omega) = \chi (\hbar\omega/v_{\rm F})\Dt_{i} \label{DifferentialCrossSection_Appendix_Eq1:TR-Susceptibility2} \\ 
a^{-1}\chi''^{i\,j}_{(+)}(\qb,\omega) &=& \Dt_{i}\Dt_{j}  + \d_{i\,j} [\left(\hbar\omega/v_{\rm F}\right)^{2} - \lvert\Dtb\rvert^{2} ]  \hspace*{0.25cm} \label{DifferentialCrossSection_Appendix_Eq1:TR-Susceptibility3}
\end{eqnarray} 
\end{subequations} 
with 
\begin{equation}
a=\frac{\pi^2}{3}\frac{V}{v_{\rm F}(2\pi\hbar)^3}.
\end{equation}
When the symmetry between the nodes is inversion, they have opposite
chiralities, which we take to be $\chi_i=-\chi_f=\chi$. In this
case Eq. (\ref{DifferentialCrossSection_Appendix_Eq1:Susceptibility}) breaks up into different tensors:
\begin{subequations} 
\label{DifferentialCrossSection_Appendix_Eq1:I-Susceptibility}
\begin{eqnarray} 
a^{-1} \chi''^{0\,0}_{(+)}(\qb,\omega)  &=& (3/2) [\left(\hbar\omega/v_{\rm F}\right)^2 - |\Dtb|^2] \label{DifferentialCrossSection_Appendix_Eq1:I-Susceptibility1} \\
a^{-1} \chi''^{0\,i}_{(+)}(\qb,\omega) &=& a^{-1} \chi''^{i\,0}_{(+)}(\qb,\omega) = 0 \label{DifferentialCrossSection_Appendix_Eq1:I-Susceptibility2} \\
a^{-1} \chi''^{i\,j}_{(+)}(\qb,\omega) 
&=& \d_{i\,j}[\left(\hbar\omega/v_{\rm F}\right)^{2} + |\Dtb|^{2} ]/2 - \Dt_{i} \Dt_{j} \nonumber \\
&+& \chi \,i\,\epsilon_{i\,j\,k} \left(\hbar\omega/v_{\rm F}\right) \Dt_{k} \label{DifferentialCrossSection_Appendix_Eq1:I-Susceptibility3}
\end{eqnarray} 
\end{subequations}
Clearly, $\chi''^{0\,0}_{(+)}$ is a Lorentz scalar.  The other tensor
does not look Lorentz covariant since it has only spatial indices, but
it actually is a usual type of tensor, see Appendix \ref{Susceptibility_Appendix}.

Now, by combining Eq. (\ref{DifferentialCrossSection_Appendix_Eq2:DifferentialCrossSectionMagnetic}) and (\ref{DifferentialCrossSection_Appendix_Eq1:structure factor}) with 
either the time-reversal or inversion-symmetric susceptibility,
Eq. (\ref{DifferentialCrossSection_Appendix_Eq1:TR-Susceptibility}) or (\ref{DifferentialCrossSection_Appendix_Eq1:I-Susceptibility}), we get
the general expressions for scattering with both a polarized beam and a polarized detector.
All these results are in the isotropic coordinate system obtained from the physical
one by applying the transformation $\Dtb=T\mathbf{\Delta}$.
Section \ref{Interpretation_Appendix:Dispersion} explains how to find the appropriate transformation $T$ \emph{experimentally}.\\
%These results have to be transferred to the physical
%coordinate system by applying the transformation $\Dtb=T\mathbf{\Delta}$.
%The transformation $T$ can be found experimentally in a simple way.

In realistic neutron scattering experiments, the initial neutron beam of $N$ neutrons has an average polarization vector $\mathbf{P}$, which can be described by a density matrix $\bm{\rho}  = \left(\tau_{0} + \mathbf{P}\cdot \bm{\tau} \right)/2 $, where $\bm{\tau}$ is a vector of Pauli matrices and $\tau_{0}$ the identity matrix in neutron spin basis. The inelastic cross-section Eq. \eqref{DifferentialCrossSection_Appendix_Eq2:DifferentialCrossSectionMagnetic} of the scattered beam measured by an \emph{unpolarized} detector is given by\cite{Lovesey1984b,Hirst1997a}
\begin{subequations}
\label{DifferentialCrossSection_Appendix_Eq6:DifferentialCrossSectionMagnetic}
\begin{eqnarray}
\frac{\text{d}^{2} \s^{(+)} (\qb,\omega;\Pb)}{\text{d} \Omega \text{d} E_{f}} &=&  \frac{\q_{f}}{\q_{i}} \left(\frac{g r_{0}}{4}\right)^{2} 
\left[\Sigma^{(+)}(\qb,\omega) + \Pb \cdot \bm{\Sigma}'^{(+)}(\qb,\omega)\right] \nonumber  
\end{eqnarray}
where $\Sigma^{(+)}$ and $\bm{\Sigma'}{(+)}$ can be found using Eq. \eqref{DifferentialCrossSection_Appendix_Eq1:structure factor},
\begin{eqnarray}
\Sigma^{(+)}(\qb,\omega) &\equiv& \hspace*{0.205cm} \left< \Mb^{(-)}_{\perp}(-\qb,-\omega) \hspace*{0.075cm} \cdot \hspace*{0.075cm} \Mb^{(+)}_{\perp}(\qb,\omega) \right>/2\pi \hbar \mu_{\rm B}^{2}, \nonumber \label{DifferentialCrossSection_Appendix_Eq6b:DifferentialCrossSectionMagnetic_Unpolarized} \\
 &=& \hspace*{0.125cm} \Fb^{\m,*}_{\perp} \cdot \Fb^{\nu}_{\perp} \chi^{(+)}_{\m\, \nu}(\qb,\omega)/\pi , \label{DifferentialCrossSection_Appendix_Eq6:DifferentialCrossSectionMagnetic_Unpolarized}\\ 
\bm{\Sigma}'^{(+)}(\qb,\omega) &\equiv& i \left< \Mb^{(-)}_{\perp}(-\qb,-\omega) \times \Mb^{(+)}_{\perp}(\qb,\omega) \right>/2\pi \hbar \mu_{\rm B}^{2}, \nonumber %\hspace*{0.05cm}
\label{DifferentialCrossSection_Appendix_Eq6b:DifferentialCrossSectionMagnetic_Polarized} \\ 
&=&  i\Fb^{\m,*}_{\perp} \times \Fb^{\nu}_{\perp} \chi^{(+)}_{\m \,\nu}(\qb,\omega)/\pi .\label{DifferentialCrossSection_Appendix_Eq6:DifferentialCrossSectionMagnetic_Polarized}
%\Sigma^{(+)}(\qb,\omega) &\equiv& \left< \Mb^{(-)}_{\perp}({\color{red}-\qb},{\color{red}-\omega}) \cdot \Mb^{(+)}_{\perp}{\color{red}(\qb,\omega)} \right>_{0}/2\pi \hbar \mu_{\rm B}^{2}, \label{DifferentialCrossSection_Appendix_Eq6b:DifferentialCrossSectionMagnetic_Unpolarized} \\
% &=& \Fb^{\m,*}_{\perp} \cdot \Fb^{\nu}_{\perp} \chi^{(+)}_{\m\, \nu}(\qb,\omega)/\pi , \label{DifferentialCrossSection_Appendix_Eq6:DifferentialCrossSectionMagnetic_Unpolarized}\\ 
%\bm{\Sigma}'^{(+)}(\qb,\omega) &\equiv& i \left< \Mb^{(-)}_{\perp}({\color{red}-\qb},{\color{red}-\omega}) \times \Mb^{(+)}_{\perp}{\color{red}(\qb,\omega)} \right>_{0}/2\pi \hbar \mu_{\rm B}^{2}, \hspace{1.5cm}  
%\label{DifferentialCrossSection_Appendix_Eq6b:DifferentialCrossSectionMagnetic_Polarized} \\ 
%&=&  i\Fb^{\m,*}_{\perp} \times \Fb^{\nu}_{\perp} \chi^{(+)}_{\m \,\nu}(\qb,\omega)/\pi .\label{DifferentialCrossSection_Appendix_Eq6:DifferentialCrossSectionMagnetic_Polarized}
\end{eqnarray}
\end{subequations}
%One can think of $\Fb^{\m,*}_{\perp} \cdot \Fb^{\nu}_{\perp}$ and $\Fb^{\m,*}_{\perp} \times \Fb^{\nu}_{\perp}$ as acting as polarizers of the bands.
The coefficients $\Fb^{\m,*}_{\perp} \cdot \Fb^{\nu}_{\perp}$ and $\Fb^{\m,*}_{\perp} \times \Fb^{\nu}_{\perp}$ select which components of $\chi_{\mu\nu}$ are measured
by neutron scattering. 
%{\color{red}They are somewhat analogous to a polarizer for light--just as light with a certain strength electric field is modulated when
%t is rotated relative to a polarizer, the signal of the Weyl fermions in
%neutron scattering oscillates when the direction of $\qb$ is rotated relative to $\F$, even though the tensor $\chi_{\mu\nu}$ just rotates.}
The $(\mu,\nu) = (0,0)$ component give rise to no angular $\Dtb$ dependence. However, the remaining 
hermitian ($i,j=1,2,3$) parts do and can be written in their spectral decompositions 
\begin{subequations}\label{DifferentialCrossSection_Appendix_Eq1:EffectiveCoupling}
\begin{eqnarray}
\Fb^{i,*}_{\perp} \cdot \Fb^{j}_{\perp} &=& \sum_{l=1}^{2}\, \alpha_{l}\, \widehat{\ab}^{l}_{j}\, \widehat{\ab}^{l*}_{i},  \label{DifferentialCrossSection_Appendix_Eq1:EffectiveCoupling_Unpolarized}  \\
\Fb^{i,*}_{\perp} \times \Fb^{j}_{\perp} &=&  -i \widehat{\kb}_{0} \sum_{l=1}^{2}\, \beta_{l}\, \widehat{\bb}^{l}_{i}\, \widehat{\bb}^{l*}_{j},  \label{DifferentialCrossSection_Appendix_Eq1:EffectiveCoupling_Polarized}
\end{eqnarray}
\end{subequations}
where $\alpha_{l} (\beta_{l})$ and $\widehat{\ab}^{l} (\widehat{\bb}^{l})$ are the $l^{th}$ eigenvalue and normalized eigenvector of matrix $\Fb^{i,*}_{\perp} \cdot \Fb^{j}_{\perp}$ ($i \widehat{\kb}_{0} \cdot \Fb^{i,*}_{\top} \times \Fb^{j}_{\top}$). To prove these, we used the fact that $\Fb^{i}_{\perp} \cdot \widehat{\kb}_{0} = 0$ for each $i$, hence $\mathrm{det}[\mathbf{F}^i\cdot \hat{j}] =0$ and therefore Eq. (\ref{DifferentialCrossSection_Appendix_Eq1:EffectiveCoupling}) will have a zero eigenvalue.

\section{Experimental predictions and interpretation\label{Interpretation_Appendix}}
The results of the last section have several conceptually and experimentally
interesting special cases.  Although there are many
parameters describing the coupling of neutrons to Weyl fermions,
there are some universal predictions contained in these formulae.  In addition, one can observe spin-momentum locking even without using polarized neutron beams or measuring the polarization of the scattered neutrons. 
Furthermore, \emph{with} a polarized measurement, it is possible to determine the chiralities of the Weyl fermions in the inversion-symmetric case, without knowing the coupling
parameters.
	
The scattering process is distinguished by whether the symmetry relation between the two nodes involved is inversion or time-reversal. %which are assumed to be aligned with the chemical potential and the three parameters $\vb_{0}$ of Eq. (\ref{Kinematics_Appendix_Eq1:Hamiltonian}) are negligible.
While the density of states is the same for either
type of symmetry, the cross-sections differ, for two reasons.
First, the chiralities are different in the two cases and hence the relativistic susceptibilities have different forms, see
Eq. (\ref{DifferentialCrossSection_Appendix_Eq1:TR-Susceptibility}) and (\ref{DifferentialCrossSection_Appendix_Eq1:I-Susceptibility}). Second, the symmetry constraints on the
coupling between neutrons and Weyl nodes are different for
time-reversal and inversion symmetry.
Appendix \ref{Interaction_Appendix} shows that
\begin{equation}
\Fb^{0} = \bm{0} \quad , \quad \Fb^{j}\in \mathds{C}^{3} \text{ with } j=1,2,3 \label{TR_Coupling:Interpretation_Appendix}
\end{equation}
for time-reversal symmetric nodes, whereas 
\begin{equation}
\Fb^{\mu} \in \mathds{R}^{3} \text{ with } \mu=0,1,2,3 \label{I_Coupling:Interpretation_Appendix}
\end{equation} 
for inversion symmetric nodes. As the predictions will be different for time-reversal and inversion symmetric nodes, they will  be discussed separately.

\subsection{Measurement of dispersion, principal axes and velocities\label{Interpretation_Appendix:Dispersion}} 
The rate of neutron scattering depends on what final electron-hole states
can be produced in the material.  This is determined by the number of final
states and the matrix element for creating
the particle-hole pair.
We will begin by describing the possible final states and estimating the
density of states (DOS). Understanding the density of states will help
to understand a few features of the scattering cross-section, and in particular
will show how one can measure the linearity of
the Weyl fermion dispersion and determine its principal axes and the velocities along them. 

\begin{figure}
\includegraphics[width=0.75\columnwidth]{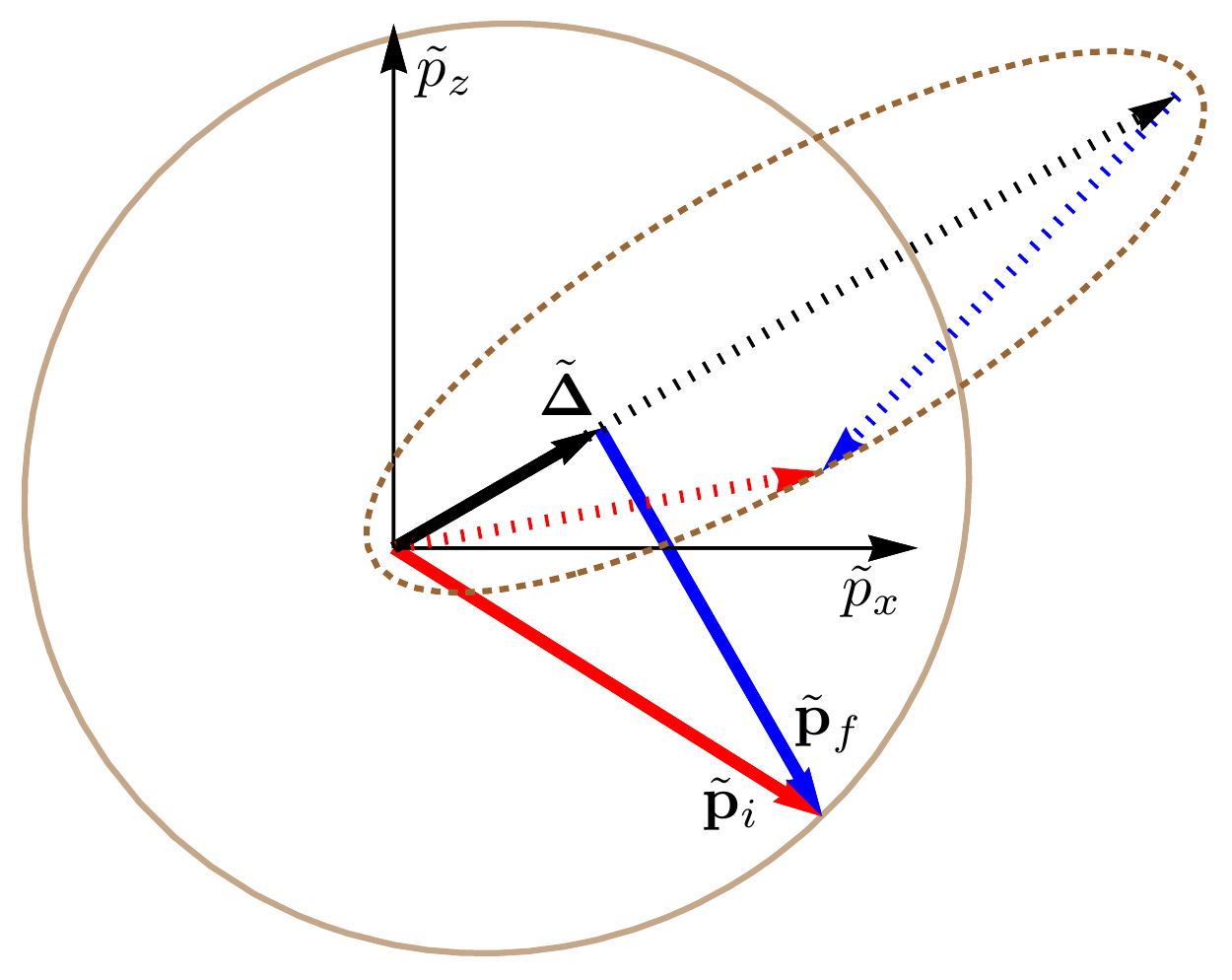} 
\caption{Contour of constant energy transfer $\hbar \omega$ for Weyl excitations produced in an scattering event: A prolate spherioid (brown line) in the $\pt_{x},\pt_{z}$-plane with symmetry axis (black arrow) along $\Dtb$ and foci at origin of initial $\pbt_{i}$ (red arrow) and final $\pbt_{f}$ (blue arrow) excitations for a given $|\Dtb|$. Full (dottted) lines are for $|\Dtb| = 0.25 \hbar \omega/v_{\rm F}$ ($|\Dtb| = 0.95 \hbar \omega/v_{\rm F}$). \label{EnergyMomentumConservation_d005_d075_d095}}
%\caption{Contour of constant energy: prolate spherioid (brown line) in the $\pt_{x},\pt_{z}$-plane with symmetry axis (black arrow) along $\Dtb$ and foci at origin of $\pbt_{i}$ (red arrow) and $\pbt_{f}$ (blue arrow) for given $|\Dtb|$. Full lines are for $|\Dtb| = 0.25 \hbar \omega/v_{\rm F}$ and dotted lines are for $|\Dtb| = 0.95 \hbar \omega/v_{\rm F}$. %Both plots are for $\Dtb = |\Dtb|(\sqrt{3},0,1)/2$.\label{EnergyMomentumConservation_d005_d075_d095}}
\end{figure}

%The DOS is defined as an integral over all initial and
%final states of the electron that conserve energy and momentum:
The DOS is defined as an integral over all internal states that conserve energy and momentum:
\begin{equation}
D(\Dtb,\omega)=\iint \frac{d^3\pbt_i d^3\pbt_f}{(2\pi\hbar)^3} \delta^3(\pbt+\Dtb)\delta(\hbar\omega-\Delta\xi^{\rm w}). \label{Eq1:Interpretation_Appendix:Dispersion}
\end{equation}
The set of allowed momenta have a simple geometric description,
see Fig. \ref{EnergyMomentumConservation_d005_d075_d095}.
Plot a point at the origin and a point displaced from this by $\Dtb$.
If the initial electron momentum is represented
by a point $P$ displaced from the origin by $\pbt_i$, then the 
\emph{final} momentum is the vector from $\Dtb$ to $P$, according
to conservation of momentum.
The change in energy is $v_{\rm F}(|\tilde{\pb}_f|+|\tilde{\pb}_i|)$, so
conservation of energy forces $P$ to lie on a prolate ellipsoid
with foci at $\mathbf{0}$ and $\Dtb$. % Thus, the DOS is an integral over this ellipsoid.  
When $|\Dtb|=0$, the ellipsoid turns into a sphere; when $|\Dtb|=\hbar\omega/v_{\rm F}$, the ellipsoid degenerates into a line
segment connecting the two foci; and for any smaller ratio of $\hbar\omega/v_{\rm F}$ to $|\Dtb|$
there are no final states compatible with conservation laws.  %This agreeswith the discussion in Section \ref{Kinematics_Appendix} for the two limiting cases, and shows that
Hence, the region of nonzero DOS is defined by $|\Dtb| \leq \hbar\omega/v_{\rm F}$ and within this region the density of states is found to be %{\color{red}In Appendix \ref{DOS_Appendix} we calculate the DOS explicitly, finding} 
\begin{equation}
D(\Dtb,\omega) = \frac{\pi}{2v_{\rm F}(2\pi\hbar)^3}[\left(\hbar\omega/v_{\rm F}\right)^2 - (1/3)|\Dtb|^2] \label{Interpretation_Appendix:Eq1_DOScalculation} % \Theta(\frac{\hbar\omega}{v_{\rm F}} - |\Dtb|)
\end{equation}
We remark, first of all, that this shows that the scattering cross-section
scales as the square of the transferred energy like the DOS for a single node. 
This makes the scattering cross-section small at low energies.  This can be problematic, since experiments must be restricted to energies small enough that the Weyl Hamiltonian is correct.  In particular, the momentum transfer can be at most of order $|\kb_0|$ since beyond that distance from one Weyl point, the other Weyl Hamiltonian becomes a better approximation.  Luckily, the small size of the cross-section at small energies can be compensated by the possibility that the coupling to the neutrons is larger than the usual $g$-factor of the electrons. To illustrate this, we employed a WSM toy model in Section IIIB.  The factors $\F$ are enhanced and even diverge as the spacing between the Weyl nodes approaches zero, which can compensate for the small DOS. This is actually more general than this specific model. In Eq. (\ref{section:MagnetizationOperator_Appendix_AnomalousCoupling}),
the current matrix element  $\bm{\mathcal{J}}(2\kb_0)$ depends on two contributions to the current\cite{Balcar1989a}, 
orbital and spin current.  The orbital Schr{\"o}dinger current is proportional to the velocity, represented by the operator
$\frac{\hbar}{m_ei}\bm{\nabla}_\mathbf{R}$ where $\mathbf{R}$ is the position 
of the electron, and hence the current at a specific point is $\mathbf{J}_{orbital}(\mathbf{r})=\frac{e\hbar}{2m_e i}\{\bm{\nabla}_\mathbf{R},\delta(\rb-\mathbf{R})\}$. (Here $\{a,b\}$ represents the anticommutator of the two operators.) The spin current is described
by an infinitesimal spinning sphere, which can be represented by the gradient of a delta-function,
$\mathbf{J}_{spin}(\rb)=\frac{g\mu_\B}{2}\bm{\sigma}\times\bm{\nabla}_\rb\delta(\rb-\mathbf{R})$.  The matrix element of the spin current comes out to be the structure factor that usually determines neutron cross-sections: taking the Fourier transform causes the delta function to be replaced by $e^{-2i\kb_0\cdot\mathbf{R}}$ and the gradient gives a factor of $2\kb_0$ that cancels the factor in the denominator of Eq.  (\ref{section:MagnetizationOperator_Appendix_AnomalousCoupling}). However, in the orbital current, the gradient acts on the electron position $\mathbf{R}$ rather than $\rb$, hence
this produces a factor of $1/d$ where $d$ is the length scale for variation of the phase of the electronic
wave functions, which, if the imaginary parts of the wave-functions, due to spin-orbit coupling for example, are large, can be the same as the size of an atom.  Thus, $\mathrm{F}_{x,x;\perp}$ is of order $1/k_0d$, so if accidentally the two Weyl points happen to be close to one another, the coupling is large. Even if the Weyl points are separated by an amount on the order of the Brillouin zone, $1/k_0d$ will be large if a unit cell contains many atoms. To get a real estimate one needs to know in detail the form of the wave functions; in particular, the wave-functions might have small imaginary parts, or the orbitals at the two Weyl points might be separated in space, and then $\F$ would be small because of the small overlap integral of the orbitals.

To give a concrete estimate of the unpolarized cross-section, Eq. (\ref{DifferentialCrossSection_Appendix_Eq6:DifferentialCrossSectionMagnetic_Unpolarized}), we return to the 4-band model. For internode scattering it has magnitude
\begin{eqnarray}
\frac{1}{V}\frac{\q_i}{\q_f}\frac{\text{d}^{2} \s^{(+)} (\qb,\omega)}{\text{d} \Omega \text{d} E_{f}} &=&  \left(\frac{g r_{0}}{4}\right)^{2}\frac{\pi}{3}\frac{(\hbar\omega)^2}{(2\pi\hbar v_{\rm F})^3} \F^{x}_{x,\perp} \F^{x}_{x,\perp}, \hspace*{0.5cm}\label{CS_Estimate:Eq1} \\
&\lesssim& 5 \cdot 10^{-2} \frac{v_{\rm F}}{c}  \; \frac{\rm mb}{\rm meV\, \r{A}^3 \,Sr} . \label{CS_Estimate:Eq2}
\end{eqnarray}
The expression Eq. (\ref{CS_Estimate:Eq1}) is a generally applicable expression with coupling given by Eq. (\ref{4bandWSM_Appendix_Eq1:Fxx}), whereas Eq. (\ref{CS_Estimate:Eq2})
is an estimate for the 4-band model.  We made the following substitutions.
Since $\chi_{\mu\nu}$ was derived in the isotropic coordinate system, the factor of $v_{\rm F}$ is not the physical velocity.  The physical Weyl nodes have three eigen-velocities; the two perpendicular to the internode direction are equal to $v_{\rm D}$ whereas that parallel is smaller, and $v_{\rm F}$ should be the geometric mean of all three.
In the above, we conservatively took
all three velocities to be identical, i.e., $v_{\rm D} = v_{\rm F}$. The intensity would be higher than Eq. (\ref{CS_Estimate:Eq2}) if one took account of the anisotropy.  Further, the energy transfer $\hbar\omega$ has been expressed in terms of the displacement of the momenta of the excitations from the Weyl point.  
We have taken the value $\mathrm{k}_0$, which is the largest possible
as explained above.  Since the result scales as $\omega^2$, the cross-section
decreases quickly for momenta below this optimistic value.
Finally $\mathrm{F}_{x,x;\perp}^2$
is taken as $(m_e v_{\rm D}/\mathrm{k}_0)^2$.  Despite the fact that
$\chi_{\mu\nu}$ is suppressed by a factor $(\hbar\omega)^2/v_{\rm F}^3\propto \mathrm{p}^2/v_{\rm F}$
from the DOS, the coupling squared, $\mathrm{F}_{x,x;\perp}^2$, partly cancels this
suppression leaving the product to have an order $\lesssim v_{\rm F}$ resulting in Eq. (\ref{CS_Estimate:Eq2}).
This implies that a higher node velocity leads to a higher intensity of the cross-section. For a typical Fermi velocity $v_{\rm F} = c/300$ Eq. (\ref{CS_Estimate:Eq2}) is $\lesssim 1.7 \times 10^{-4} \; {\rm mb} / {\rm meV\, \r{A}^3 \,sr}$. Now assuming a typical unit cell has volume $V = (5\rm \r{A})^3$, the intensity  $\q_{i}/\q_{f} \times \text{d}^{2} \s^{(+)}(\qb,\omega)/\text{d} \Omega \text{d} E_{f} \lesssim 2 \times 10^{-2} \;  {\rm mb} / {\rm meV\,sr\,f.u.}$. As anticipated for a semimetal the intensity is low, but much higher than the early estimates\cite{Silver1984aa} of the neutron cross-section for one-electron metallic band structures, which were of order $10^{-4} - 10^{-3} {\rm mb/meV\,sr}$. Our estimate for the $4$-band model is only of order $10^{-2} - 1$ smaller than what has been observed in scattering off spin-$\frac12$ particle-hole pairs~\cite{Goremychkin2018aa,Vignolle:2007aa,Walters:2009aa,Fujita2012aa,Janoschek2015aa}.

One other property of the Weyl scattering cross-section
may also help it to be visible--namely at the maximum
momentum transfer the DOS is still nonzero, and then there is a sharp jump down to zero.
A sharp jump can be separated out when there is a smooth background, even
if the background is large, by differentiating.

Let us understand why the DOS has a sharp jump. Imagine fixing the transferred momentum and lowering the energy. 
%(which is easier to think about than fixing the energy and raising the momentum). 
The set of final states is always a prolate ellipsoid with
the same foci $\mathbf{0}$ and $\tilde{\Db}$, that
eventually degenerates to a line segment at the minimum possible energy transfer.
Because there is a whole line segment rather than a single final state, the
DOS is larger than usual in this limit.
To be more precise, let $\Delta\xi^{\rm w}(\pbt_i)$ be the change in energy
of the electron as a function of the initial momentum ($\pbt_f = \pbt_i-\Dtb$ since $\Dtb$ is fixed), $\Delta\xi^{\rm w}(\pbt_i) = v_{\rm F}(|\pbt_i|+|\pbt_i-\Dtb|)$.
%\begin{equation}
%\Delta\xi^{\rm w}(\pbt_i) = v_{\rm F}(|\pbt_i|+|\pbt_i-\Dtb|).
%\end{equation}
The DOS of the particle-hole pair is given by $(2\pi\hbar)^{-3} \int d^3\pbt_i \, \delta(\Delta\xi^{\rm w}(\pbt_i)-\hbar\omega)$, which is the same formula used to calculate the DOS of a single particle whose dispersion happens to be given
by $\Delta\xi^{\rm w}(\tilde{\pb_i})$. We will use this analogy to understand the 
behaviour of the particle-hole pair DOS at the surface of the spherical scattering region. Here its behaviour corresponds to a van Hove singularity.
To see this, notice that the function $\Delta\xi^{\rm w}$ has
a minimum value $\Delta\xi^{\rm w}_{min}=v_F|\Dtb|$. Increasing $|\Dtb|$
with a fixed $\omega$, beyond the surface
of the scattering region, is equivalent to letting $\hbar\omega$ fall
below this minimum value. Generically, in three-dimensions
the DOS close to a minimum should have
the van Hove dependence of $\sqrt{\hbar\omega-\Delta\xi^{\rm w}_{min}}$.
This assumes that the minimum is at an isolated point. However,
for the pair of Weyl excitations, there is a line which $\Delta\xi^{\rm w}$ is minimum on,
the line connecting the foci of the ellipsoid.
The DOS may be found by integrating over layers
perpendicular to the line connecting foci.  For example, if $\Dtb$ is parallel to the $z$-axis, $D(\omega)=\int \frac{\dd \pt_z}{2\pi\hbar} D_\perp(\pt_z)$
%\begin{equation}
%D(\omega)=\int \frac{\dd \pt_z}{2\pi\hbar} D_\perp(\pt_z)
%\end{equation}
where $D_\perp$ is the DOS in one of these planes. For each fixed $\pt_z$, $D_\perp$ has the van Hove singularity one expects in \emph{two} dimensions (this function is quadratic near its minimum except in the planes passing
through the foci), that is, it should jump from 0 to a nonzero value. Since the minimal values of $\omega$ are equal for all planes between the foci, there is still a discontinuous jump after integrating over $\pt_z$ and thus also in $D(\omega)$. %, which is plotted in Fig. \ref{Kinematics_Appendix_fig1:Cross-section_TransitionPlus}.  

If the transferred energy is fixed, the region of nonzero scattering is a sphere\footnote{Note that for scattering between two Weyl points that are not related
by symmetry,  the shape of this region is more complicated because
it arises from a combination of two different dispersions.  Also, one
does not expect a sharp jump in the scattering cross-section, because
$\Delta\xi^{\rm w}(\tilde{\pb}_i)$ will have a unique minimum then.} of radius $\hbar\omega/v_{\rm F}$. Thus, by measuring the radius
of this sphere as a function of the transferred energy one may deduce
the dispersion velocity of the Weyl fermions.  Furthermore, the linear relationship between the radius of the sphere and the energy reflects the linear dispersion
of the Weyl fermions.  Now this region is spherical only because we began by rescaling all momenta to make the dispersion isotropic.  In general,
the dispersion of Weyl fermions is likely to be anisotropic; it has
the form $\sqrt{\sum_{i\,j}(v^2)_{i\,j}\pb_i \pb_j}$ where $v^2$
is a certain matrix.  Indeed, when one diagonalizes Eq. (\ref{Kinematics_Appendix_Eq2:HamiltonianAisotropic}), one finds that the energy of the excitation
has this form, with 
$v^2=v_{\rm F}^2\lambda^T\lambda$.  By the inversion or time-reversal symmetry, both Weyl particles have the same dispersion.  One can then show that the region of allowed momentum and
energy transfers is 
$\hbar\omega\geq v_{\rm F}\sqrt{\sum_{i\,j} (\lambda^T\lambda)_{i\,j}\pb_i\pb_j}$,
which is an ellipsoid for each fixed $\omega$ rather than a sphere. It has the same shape as the equal-energy contours of a single particle.  The directions and lengths of the principle axes give the eigenvectors and eigenvalues of $v^2$.  Let $T$
be any linear transformation that distorts this ellipsoid to a sphere; then
the dispersion becomes isotropic upon redefining $\tilde{\pb}=T\pb$. The
Weyl equation then takes the form\footnote{More precisely, the Weyl Eq. takes the form $v_{\rm F}\tilde{\lambda}_{lm}\sigma_l\tilde{p}_m$ where $\tilde{\lambda}^T\tilde{\lambda}$ is a multiple of the identity; i.e., $\lambda$ is a multiple of a rotation matrix. Now a change of basis of the two states spanning the pseudospin space
corresponds to a rotation of the Pauli matrices.  Thus one can choose
a transformation that converts $\lambda$ into a scalar matrix.} in Eq. (\ref{Kinematics_Appendix_Eq1:HamiltoniannoLin}).
In this way, one can measure from the cross-section the principal axes,
velocities of the dispersion as well as the transformation $T$ that will
be important to be able to see the ``universal" predictions of this theory below.

\begin{figure}
\includegraphics[width=0.75\columnwidth]{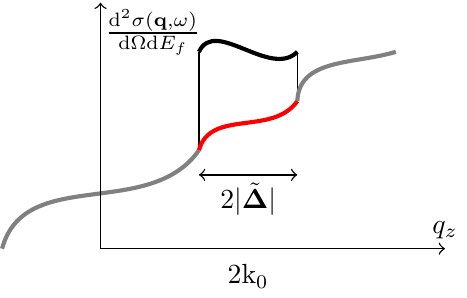}\label{Kinematics_Appendix_fig1:Cross-section_TransitionPlus_a}
%\subfigure[]{\includegraphics[width=0.75\columnwidth]{./Fig2c_Interpretation_Appendix}\label{Kinematics_Appendix_fig1:Cross-section_TransitionPlus_b}}
\caption{Sketch of cross-section including background scattering along $q_{z}$ in Fig. \ref{Kinematics_Appendix_fig1:Cross-section_Transitions} for the $\M^{+}$ process. The intensity jumps discontinuously at the boundary between the region describing internode scattering (black curve) and that which does not (gray curve), while there might be background scattering (red curve) in the region of interest.
%Density of states [subfig. \subref{Kinematics_Appendix_fig1:Cross-section_TransitionPlus_b}] Eq. (\ref{Interpretation_Appendix:Eq1_DOScalculation}) centered at $2\kb_{0}$ as a function of $|\Dtb|$ for fixed $\omega$.  Black (dashed) line is for $\alpha = 0$ ($\alpha \ll 1$).
\label{Kinematics_Appendix_fig1:Cross-section_TransitionPlus}}
\end{figure}

The discontinuous jump is unique to the case where $\vb^{(i)}_{0} = \bm{0}$ in Eq. (\ref{Kinematics_Appendix_Eq2:HamiltonianAisotropic}).
When the vector $\vb^{(i)}_{0}$ is nonvanishing then its values at the
two nodes are negative of one another by symmetry (either inversion or
time reversal), i.e., $\mathbf{v}_{0}^{(1)} = - \mathbf{v}_{0}^{(2)}=\mathbf{v}_0$. 
By transforming the coordinates, one 
can still make $\lambda_{i\,j}^{(1)} = \pm \lambda_{i\,j}^{(2)} = \delta_{i\,j}$ and additionally make $\mathbf{v}_{0}$ parallel to any direction one prefers. 
One then sees that there is only one parameter in the Hamiltonian that is important: the ratio $|\mathbf{v}_{0}|/v_{\rm F} \equiv \alpha$, which for a type-I WSM
\footnote{In fact, it is possible to calculate the cross-section analytically, for arbitrary type-I WSM nodes with $\alpha < 1$. The Lorentz symmetry method we used, when $\alpha=0$, does not work because $H = v_{\rm F}\sb\cdot \pbt + \alpha v_{\rm F} \pt_z$ is not Lorentz invariant.  However, one can evaluate the contribution to the cross-section from fermions with a fixed  $\pt_{z}$ using \emph{two}-dimensional Lorentz symmetry, and then, the resulting expressions can be integrated over $\pt_{z}$. There are many terms to evaluate (since now there is no symmetry between $\sigma_z$ and the other $\sigma^\mu$'s) \label{alpha}}
takes values\cite{Soluyanov2015a} $0 \leq \alpha<1$.
%\footnote{A type-II WSM\cite{Soluyanov2015a} has $\alpha > 1$.
%%A type-II WSM\cite{Soluyanov2015a} has $\alpha > 1$. Then there are states with zero energy and nonzero momentum. Neutrons can then be scattered by the Weyl fermions 
%%through a finite momentum with an infinitesimal transfer of energy, so the region of neutron scattering is extended over the whole Brillouin zone, even for small energy transfer.
%} 
The parameter $\alpha$ upsets Lorentz invariance more seriously than the coupling parameters $\mathbf{F}^\mu$. It changes the kinematics, such that the constant energy contour is \emph{not} an ellipsoid any longer\cite{TurnerUnpublished}.  It also appears in a nontrivial way in the structure factor Eq. (\ref{DifferentialCrossSection_Appendix_Eq1:IntermediateStructureFactor}). A specific effect is that the cross-section will not jump suddenly to zero at the edge (see Fig. \ref{Kinematics_Appendix_fig1:Cross-section_TransitionPlus} and \ref{ConventionalMag_zaxis_xyplane}); it vanishes continuously.  If $\alpha$ is small, this jump happens in a layer of a thickness proportional to $\alpha$, so when $\alpha$ is very small, it seems to be a sharp jump. On the other hand, the spin-momentum locking could still be observed; it would still cause the cross-section to vary strongly as a function of the angle around the center of the  region.  The formula for the variation would not be so simple as that given here.%in Eq. (\ref{DifferentialCrossSection_Appendix_Eq1:TR-Susceptibility}) and (\ref{DifferentialCrossSection_Appendix_Eq1:I-Susceptibility}).

\subsection{Probing spin-momentum locking in a fully unpolarized experiment\label{DiffCrossSectionMagnetic_Appendix:ConventionalMagneticScattering}}
We have previously just quoted the susceptibility. Now we turn to an intuitive explanation of it in terms of simple concepts of spin matrix elements and spin-momentum locking, 
thereby enabling us to understand how a fully unpolarized measurement can probe the spin-momentum locking of Weyl spinors, which at first seems like a contradiction. Appendix \ref{Susceptibility_Appendix} gives a different interpretation of the results in terms of Lorentz transformations of spinors.%, as well as the detailed derivation.

To guide our intuition, we will explain it here for the case of coupling strengths that most closely resemble conventional purely magnetic scattering\footnote{This expression includes components of $\F$ that are not transverse to $\Delta \kb_0$.  This does not
affect the results however, since only the transverse components enter into Eq. (\ref{DifferentialCrossSection_Appendix_Eq2:DifferentialCrossSectionMagnetic}).}, i.e. $\Fb^{0} = \bm{0}$ and $\F_{i\,j} = \d_{i\,j}$.
 We will assume that the nodes are on the $z$-axis, $\kb_{0,2}= -\kb_{0,1} = \k_0\hat{z}$.  
Then the cross-section Eq. (\ref{DifferentialCrossSection_Appendix_Eq6:DifferentialCrossSectionMagnetic})  becomes $\pi \Sigma^{(+)}(\qb,\omega) = \sum_{i=1}^{2} \chi''^{(+)}_{i\, i}(\qb,\omega)$. This clearly highlights the fact, see Eq. (\ref{section:MagnetizationOperator_Appendix_AnomalousCoupling}), that neutrons couple only to components of the coupling vectors that are 
perpendicular to the internode direction. This has the desirable consequence that the cross-section will have angular $\Dtb$ dependence, which is a signature of probing spin-momentum locking of Weyl spinors.  

A consequence of momentum conservation is that initial $\lvert i_{\rm w} \rangle =  \lvert \widehat{\pbt}_{i}; -\chi_{i} \rangle$ 
and final $\lvert f_{\rm w} \rangle =  \lvert \widehat{\pbt}_{f}; +\chi_{f} \rangle$ Weyl states are related by $\pbt_{f} = \pbt_{i} - \Dtb$, and energy conservation dictates that any pair $\pbt_{i}$ and $\pbt_{f}$ are restricted to the ellipsoid 
constant energy contour in Fig. \ref{EnergyMomentumConservation_d005_d075_d095}. In the limit $\Dtb = \bm{0}$, the allowed initial and final states are pairs $\pbt_{f} = \pbt_{i}$ 
on a sphere of radius $\hbar \omega/ 2 v_{\rm F}$,  and the polarization
vectors of the Weyl spinors are thus related by
\begin{eqnarray}
\langle f_{\rm w} \rvert \sb \lvert f_{\rm w} \rangle &=& \mp \langle i_{\rm w} \rvert \sb  \lvert i_{\rm w} \rangle  \text{ for } \chi_{f} =  \pm \chi_{i},  \label{DifferentialCrossSection_Appendix_Eq1:Spinors_DMinimum} 
\end{eqnarray}
i.e. initial and final spinors are antiparallel (parallel) for same (opposite) chirality. (To understand this, remember that the initial state has a negative energy and the final state has a positive energy.) 
%{\color{blue}(The fact that the initial state is a negative-energy eigenfunction
%of the Weyl equation causes the spin to be backwards.)} 
All these different spinors just contribute to the cross-section at a single point,
so there is no signature that distinguishes between $\chi_f=\pm\chi_i$ apart from a constant factor of 2 (which cannot be measured anyway unless one
knows the values of the $\mathrm{F}$'s).

For increasing $\lvert \Dtb \rvert$ the energy conserving contour takes a more extreme prolate spheroid form and the cross-section will have angular $\Dtb$ dependence because the Weyl state contributions depend on the direction 
of $\Dtb$. 

In the extreme limit  $|\Dtb| \approx \hbar \omega/ v_{\rm F}$ the energy conserving contour becomes an extremely slim, elongated prolate spheroid, 
which degenerates to a line at maximum $|\Dtb| = \hbar \omega/v_{\rm F}$. The initial and final unit vectors along the momenta are therefore approximately $\widehat{\pbt}_{i} \approx \widehat{\Dtb} \approx - \widehat{\pbt}_{f}$, so states are $\lvert i_{\rm w} \rangle \approx 
\lvert \widehat{\Dtb}; -\chi_{i} \rangle$ and $\lvert f_{\rm w} \rangle \approx  \lvert \widehat{\Dtb}; -\chi_{f} \rangle$, which means that spinors are related as
\begin{eqnarray}
\langle f_{\rm w} \rvert \sb  \lvert f_{\rm w} \rangle & = & \pm  \langle i_{\rm w} \rvert \sb \lvert i_{\rm w} \rangle  \text{ for } \chi_{f} = \pm \chi_{i}, \label{DifferentialCrossSection_Appendix_Eq1:Spinors_DMaximum}
\end{eqnarray}
which is the reverse of Eq. (\ref{DifferentialCrossSection_Appendix_Eq1:Spinors_DMinimum}). Because of this, the momenta of the particle and hole are opposite to each other while the spins Eq.  (\ref{DifferentialCrossSection_Appendix_Eq1:Spinors_DMaximum}) are parallel or antiparallel to one another depending on the type of symmetry. 

We saw above that only the transverse components of the neutron and of the Weyl
fermion are coupled. To understand how this causes the
cross-section to become anisotropic, we note that the interaction
of electrons and neutrons is proportional to $\sigma_x\tau_x+\sigma_y\tau_y$
where $\bm{\tau}$ is the Pauli spin matrices of the neutron.  The cross-section is proportional
to the integral of the interaction matrix element $|\langle \tau_f;\chi_f;\widehat{\tilde{\pb}}_f|\sigma_x\tau_x+\sigma_y\tau_y|\widehat{\tilde{\pb}}_i;-\chi_i;\tau_i\rangle|^2$ over
all possible final states of the electron.  Even for the unpolarized neutrons
averaging this over all initial and final neutron states gives  $|\langle \chi_f;\widehat{\pbt}_f|\sigma_x|\widehat{\pbt}_i;-\chi_i\rangle|^2+|\langle \chi_f;\widehat{\pbt}_f|\sigma_y|\widehat{\pbt}_i;-\chi_i\rangle|^2$ which is still asymmetric.
The effect of this interaction, in which $\sigma_x$ or $\sigma_y$
are applied to
the Weyl fermion's pseudospin, is different depending on the initial
direction of the pseudospin: for some directions it is more likely
to flip it and for others more likely not to.
%If the initial spinor is along $+\hat{z}$, then both $\sigma_x$ and $\sigma_y$ flip it, so the
%spinor must flip for the matrix element to be nonzero. {\color{blue} Hence there is
%no scattering with a momentum transfer
%of $\Dtb=\pm\hbar\omega/v_{\rm F}\hat{z}$ if $\chi_f=\chi_i$.   The situation is different if the spin starts out in the $xy$-plane or the chiralities are the same.}% If the initial spinor is along the $x$-axis, the sum is the same whether the spinor flips
%or not, because $\sigma_y$ flips the spin and $\sigma_x$ does not.
This causes the cross-section to oscillate over the surface of the sphere.  This oscillation has a
different form for the time-reversal and inversion-symmetric cases.
For example, in the time-reversal symmetric case, the spin directions
before and after scattering must be parallel, so the
cross-section is zero when $\Dtb\rightarrow \pm\hbar\omega/v_{\rm F}\hat{\mathbf{z}}$ (in which case both $\sigma_x$ and $\sigma_y$ flip the spin), while the cross-section is maximum on this
axis in the inversion-symmetric case.

\begin{figure}
\includegraphics[width=0.85\columnwidth]{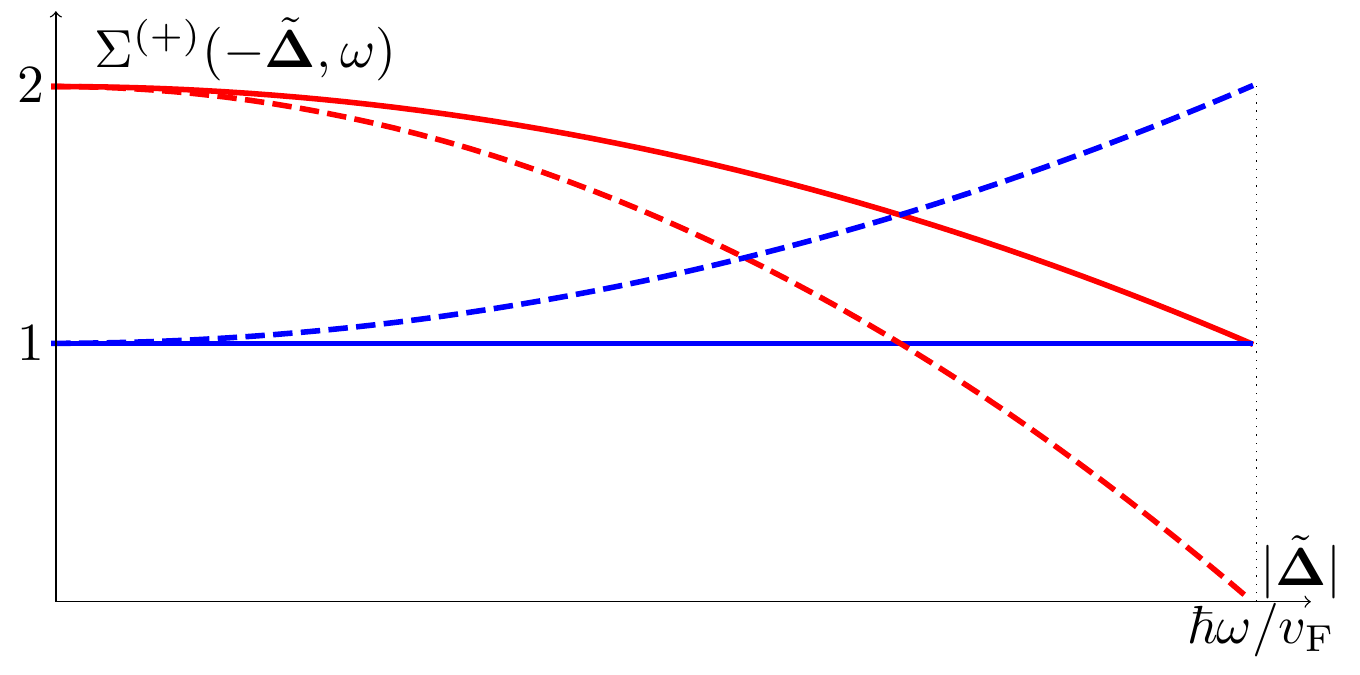}
%\caption{Plot of cross-section as a function of $\lvert \Dtb\rvert$ where blue (red) curve is for scattering between nodes of chirality $\chi_{f} = -(+) \chi_{i}$. Full (dashed) lines are for $\theta_{\Dtb} = \pi/2$ ($\theta_{\Dtb} = 
%0$). \label{ConventionalMag_zaxis_xyplane}}
\caption{Cuts of the cross-section for a coupling $\Fb^{0} = \bm{0}$ and $\F_{i\,j} = \d_{i\,j}$. 
The cross-section is plotted as a function of $\lvert \Dtb\rvert$  for scattering between nodes of chirality $\chi_{f} = -(+) \chi_{i}$ in blue (red) along  $\theta_{\Dtb} = \pi/2$ ($\theta_{\Dtb} =  0$) in full (dashed) lines.
\label{ConventionalMag_zaxis_xyplane}}
\end{figure}

\begin{figure*}
\hspace*{-1.0cm}
\centering
%-- upper row
% d001 pp
\begin{minipage}[c]{0.3\textwidth}
\centering
\subfigure[]{\includegraphics[width=0.55\textwidth]{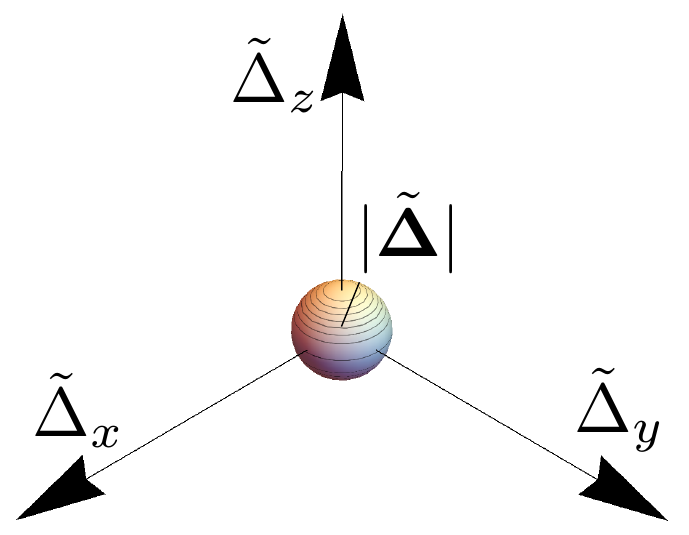}\label{Unpol_E_F0_a1_a2(02_0_x_y)pm}} \\
\subfigure[]{\includegraphics[width=0.55\textwidth]{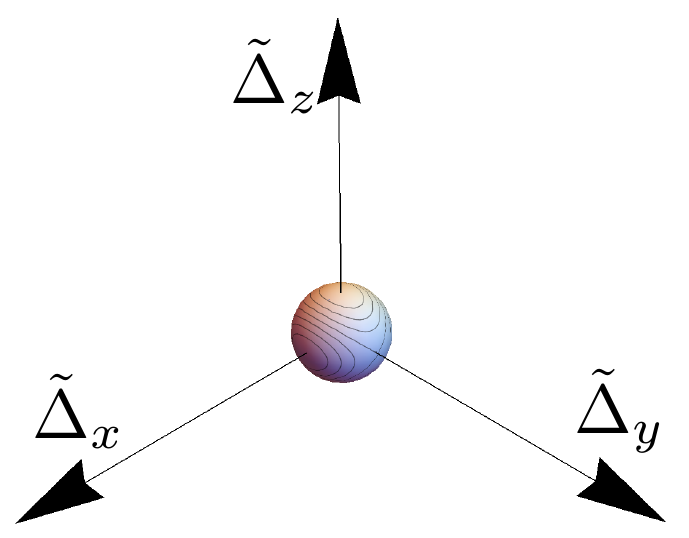}\label{Unpol_E_F0_a1_a2(02_0_x_05y)pm}} \\
\subfigure[]{\includegraphics[width=0.55\textwidth]{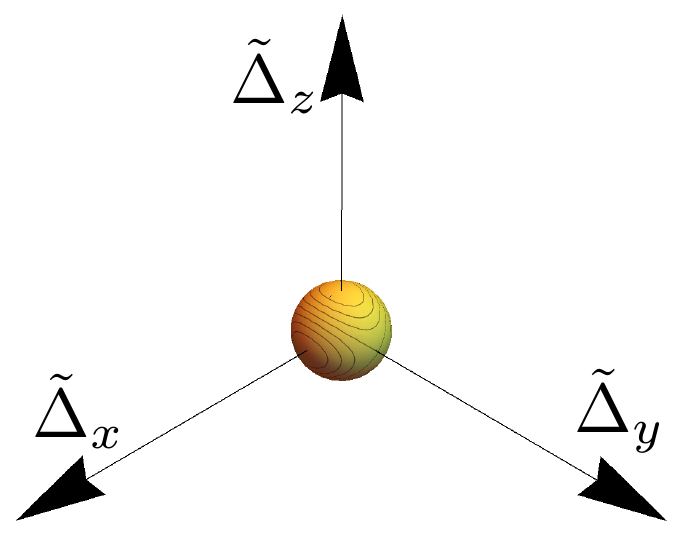}\label{Unpol_E_F0_a1_a2(02_05_x_05y)pm}}
\end{minipage}  \hspace*{-0.55cm} %\hfill
\begin{minipage}[c]{0.3\textwidth}
\centering
\subfigure[]{\includegraphics[width=0.55\textwidth]{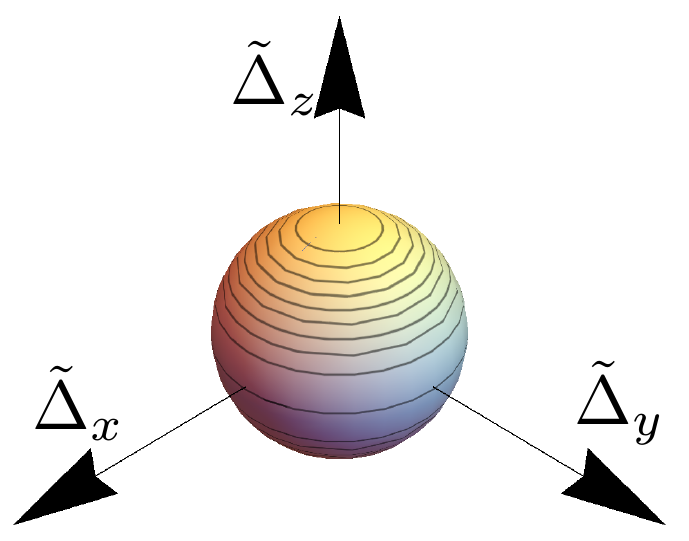}\label{Unpol_E_F0_a1_a2(05_0_x_y)pm}} \\
\subfigure[]{\includegraphics[width=0.55\textwidth]{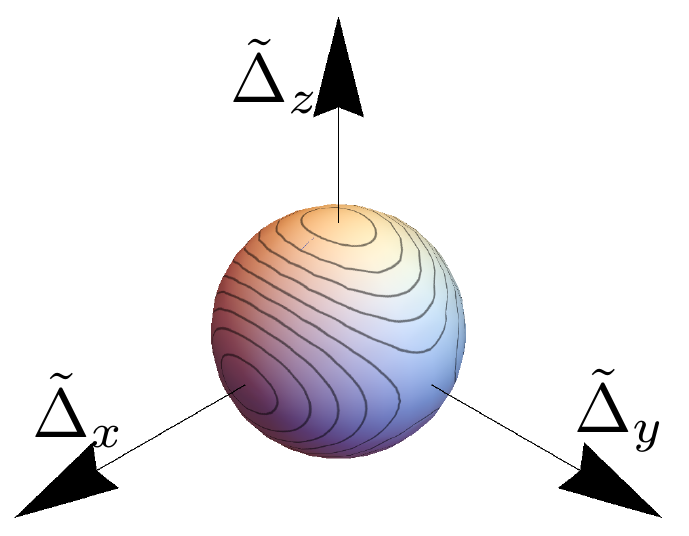}\label{Unpol_E_F0_a1_a2(05_0_x_05y)pm}} \\
\subfigure[]{\includegraphics[width=0.55\textwidth]{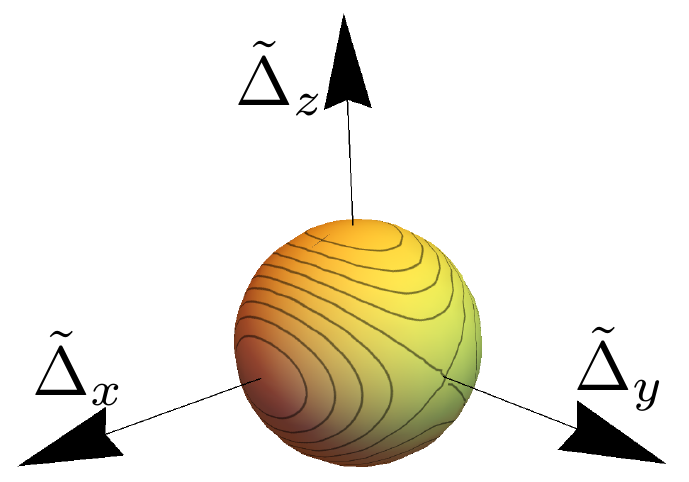}\label{Unpol_E_F0_a1_a2(05_05_x_05y)pm}}
\end{minipage}  \hspace*{-0.55cm} %\hfill
\begin{minipage}[c]{0.3\textwidth}
\centering
\subfigure[]{\includegraphics[width=0.55\textwidth]{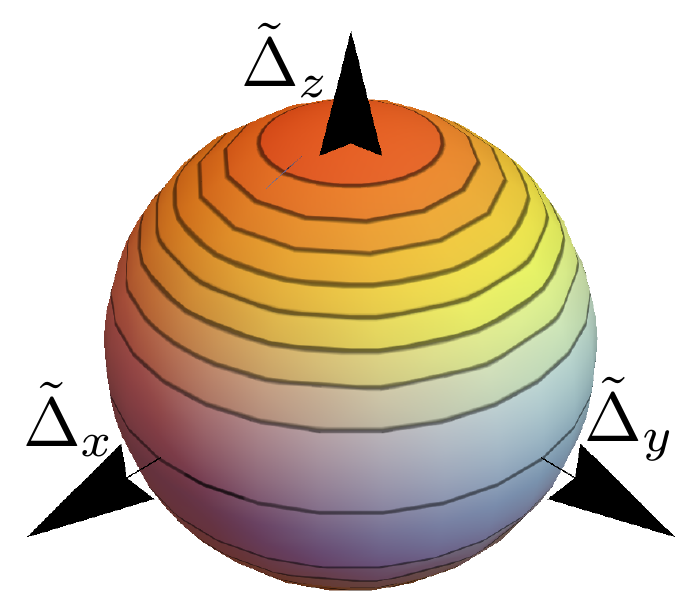}\label{Unpol_E_F0_a1_a2(095_0_x_y)pm}} \\
\subfigure[]{\includegraphics[width=0.55\textwidth]{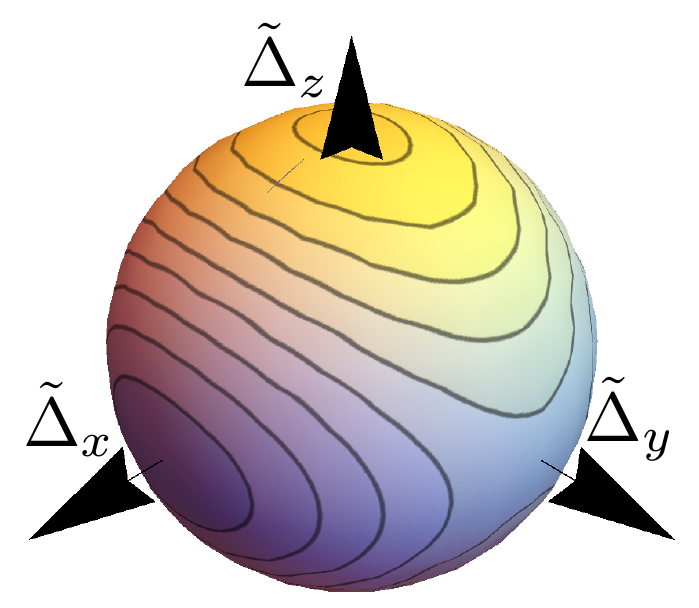}\label{Unpol_E_F0_a1_a2(095_0_x_05y)pm}} \\
\subfigure[]{\includegraphics[width=0.55\textwidth]{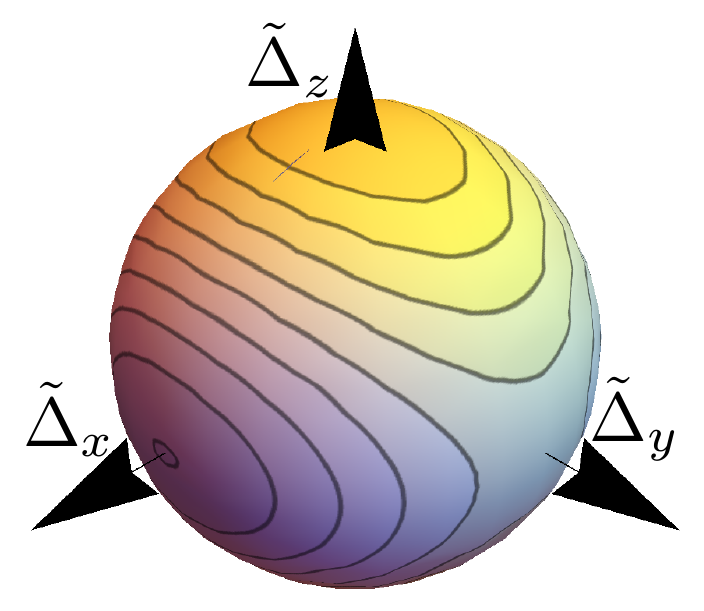}\label{Unpol_E_F0_a1_a2(095_05_x_05y)pm}}
\end{minipage} \hspace*{-0.55cm}
\begin{minipage}[c]{0.075\textwidth}
\centering
\includegraphics[width=0.05\columnwidth]{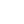} \\
\vspace{-0.25cm}
\subfigure[]{\includegraphics[width=0.35\columnwidth]{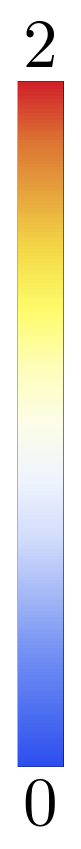}\label{colorscale}} \\
\includegraphics[width=0.05\columnwidth]{./Fig_Blank} 
\end{minipage} \\ %\hspace*{-0.55cm}
\begin{minipage}[c]{2.1\columnwidth}
%\caption{Comparison of cross-sections for various couplings. Plot of cross-section $\Sigma^{(+)}(-\Dtb,\omega)$ centered at $2\kb_{0}$ for inversion symmetric nodes in the case where coupling eigendirections Eq. %(\ref{DifferentialCrossSection_Appendix_Eq1:EffectiveCoupling_Unpolarized}) are $\widehat{\ab}_{1} = \hat{\mathbf{x}}$ and $\widehat{\ab}_{2} = \hat{\mathbf{y}}$. The columns are cuts of $|\Dtb|$ for a fixed $\hbar\omega = 1$ with the left, middle, and right %columns at $|\Dtb| v_{\rm F}/\hbar \omega = 0.2 , 0.5$, and $0.95$, respectively. The upper row [\subref{Unpol_E_F0_a1_a2(02_0_x_y)pm},\subref{Unpol_E_F0_a1_a2(05_0_x_y)pm},\subref{Unpol_E_F0_a1_a2(095_0_x_y)pm}] is for $\F^{0}_{\perp}=0$ and %$\alpha_{1} = \alpha_{2} = 1$,
% The middle row [\subref{Unpol_E_F0_a1_a2(02_0_x_05y)pm},\subref{Unpol_E_F0_a1_a2(05_0_x_05y)pm},\subref{Unpol_E_F0_a1_a2(095_0_x_05y)pm}] is for $\F^{0}_{\perp}=0$ and $\alpha_{1} = 2 \alpha_{2} = 1$. The lower row %[\subref{Unpol_E_F0_a1_a2(02_05_x_05y)pm},\subref{Unpol_E_F0_a1_a2(05_05_x_05y)pm},\subref{Unpol_E_F0_a1_a2(095_05_x_05y)pm}] is for $\F^{0}_{\perp} = 1/2$ and $\alpha_{1} = 2 \alpha_{2} = 1$. Intensity is given by the temperature scale in %subfigure \subref{colorscale}. \label{CrossSectionFigNEW}}
\caption{Comparison of cross-sections for different couplings.
The cross-section $\Sigma^{(+)}(-\Dtb,\omega)$ for scattering between inversion symmetric nodes is plotted in isotropic coordinates $\Dtb$ for a given energy transfer $\hbar\omega$. Columns are cuts of $|\Dtb|$ with the left, middle, and right columns at $|\Dtb| v_{\rm F}/\hbar \omega = 0.2 , 0.5$, and $0.95$, respectively. All rows are the same case of coupling eigendirections, Eq. (\ref{DifferentialCrossSection_Appendix_Eq1:EffectiveCoupling_Unpolarized}),  $\widehat{\ab}_{1} = \hat{\mathbf{x}}$ and $\widehat{\ab}_{2} = \hat{\mathbf{y}}$ but different eigenvalues $\alpha_{1}$ and $\alpha_{2}$ and constant $\F^{0}_{\perp}$.
The upper row [\subref{Unpol_E_F0_a1_a2(02_0_x_y)pm},\subref{Unpol_E_F0_a1_a2(05_0_x_y)pm},\subref{Unpol_E_F0_a1_a2(095_0_x_y)pm}] is for $\F^{0}_{\perp}=0$ and $\alpha_{1} = \alpha_{2} = 1$,
The middle row [\subref{Unpol_E_F0_a1_a2(02_0_x_05y)pm},\subref{Unpol_E_F0_a1_a2(05_0_x_05y)pm},\subref{Unpol_E_F0_a1_a2(095_0_x_05y)pm}] is for $\F^{0}_{\perp}=0$ and $\alpha_{1} = 2 \alpha_{2} = 1$. The lower row [\subref{Unpol_E_F0_a1_a2(02_05_x_05y)pm},\subref{Unpol_E_F0_a1_a2(05_05_x_05y)pm},\subref{Unpol_E_F0_a1_a2(095_05_x_05y)pm}] is for $\F^{0}_{\perp} = 1/2$ and $\alpha_{1} = 2 \alpha_{2} = 1$. Intensity is given by the temperature scale in subfigure \subref{colorscale}. \label{CrossSectionFigNEW}} 
\end{minipage}
\end{figure*}

Figure \ref{ConventionalMag_zaxis_xyplane} illustrates the variation
of the cross-section as a function of $|\Dtb|$ on the $z$-axis $\theta_\Dtb=0$
and the $xy$-plane $\theta_\Dtb=\pi/2$ for the two types of symmetry.  Figure 
\ref{Unpol_E_F0_a1_a2(02_0_x_y)pm},\subref{Unpol_E_F0_a1_a2(05_0_x_y)pm} and \subref{Unpol_E_F0_a1_a2(095_0_x_y)pm} plots the full $\Dtb$ dependence of the cross-section centered around $2\kb_{0}$ for the case of inversion symmetric nodes. %For $|\Dtb| 
%> \hbar \omega/v_{\rm F}$ the cross-section drops sharply to zero because no states can satisfy the the energy-momentum conservation. 

In summary, due to energy and momentum constraints of the excitations, the scattering channels are effectively those of a polarized measurement for any $|\Dtb|>0$ with the degree of polarization being maximum for maximal momentum transfer $|\Dtb|=\hbar 
\omega/ v_{\rm F}$. Hence by sweeping $\Dtb$, i.e., by sweeping external neutron momentum transfer $\qb$, one indirectly performs a polarized experiment despite not using polarized neutrons.\\

\begin{comment}
\begin{figure}[h]
\begin{minipage}{0.45\textwidth} %right.  
\begin{tikzpicture}
\node [anchor=south west, inner sep=0] at (0,4) {\includegraphics[width=.7\textwidth]{./ConventionalMag_zaxis_xyplane}};
%-- delta
\node [above right] at (6,3.75) {$\lvert\Dtb\rvert $};
\draw [->] (0.1,4.0) -- (5.95,4.0);%horizontal
\draw [dotted] (5.575,4.1) -- (5.575,6.37); % omega line
\node [below right] at (0,4) {$0$};
\node [below left] at (6.25,4) {$\hbar \omega/v_{\rm F}$};
%-- intensity
\draw [->] (0.1,4.0) -- (0.1,6.6);%vertical
\node [above right] at (0,6.6) {$\tilde{\Sigma}^{(+)}(-\Dtb,\omega)$}; 
\node [above left] at (0,4) {$0$};
\node [above left] at (0,6.1) {$2$};
\end{tikzpicture}
\end{minipage}
\begin{minipage}{1\columnwidth}
\caption{Plot of cross-section as a function of $\lvert \Dtb\rvert$ where blue (red) curve is for scattering between nodes of chirality $\chi_{f} = -(+) \chi_{i}$. Full lines (dashed) are for $\theta_{\Dtb} = \pi/2$ ($\theta_{\Dtb} = 
0$). \label{ConventionalMag_zaxis_xyplane}}
\end{minipage}
\end{figure}
\end{comment}

The angular dependence of the cross-section of unpolarized neutrons results
from a combination of two facts: first, the electron polarization is
dependent on the transferred momentum, and second, the $\F_{ij}^\perp$ is
anisotropic so it is possible to see the variation of the electron-polarization
even with unpolarized neutrons. If, hypothetically it had been the case that $\F^{\perp}_{i\,j} = \d_{i\,j}$, then the cross-section would have no angular $\Dtb$ dependence, but would be spherical symmetric as a function of $|\Dtb|$ for a given $\omega$. 
However, $\F_{\perp}$ can \emph{never} be diagonal because in a coordinate system where $\widehat{\kb}_{0} = \hat{\mathbf{z}}$, $\F_{\perp}$ would have two columns orthogonal to $\widehat{\kb}_{0}$ because $\Fb^{i}_{\perp}\cdot \widehat{\kb}_{0} = 0$ holds 
\emph{always}. This generally implies angular dependence. However, although
this condition rules out $\F^{\perp}_{i\,j} = \d_{i\,j}$, there \emph{is}
a way that the spin-momentum locking could be hidden in the
time-reversal symmetric case.  The coupling
\begin{equation}
\F_{\perp} = \begin{pmatrix} 1 & 0 & 0 \\ 0 & 1 & 0 \\ i & 0 & 0\end{pmatrix},
\end{equation}
gives a cross-section $\pi \Sigma^{(+)}(\qb,\omega) = \sum_{j=1}^{3} \chi''^{(+)}_{j\, j}(\qb,\omega)$,
which has the same effect as if $\F^{\perp}_{i\,j} = \d_{i\,j} $. Such a coupling is allowed, though probably not very likely to occur since it is very specific.  
%Consequently, any angular dependence of the cross-section 
%\emph{implies} probing spin-momentum locking, while the reverse is not necessarily true. 
Consequently, any angular dependence of the cross-section 
\emph{implies} probing spin-momentum locking. The reverse statement \emph{is} necessarily true for inversion-symmetric nodes, whereas it is \emph{not} necessarily true for time-reversal symmetric nodes. 
%Consequently, any angular dependence of the cross-section 
%\emph{guarantee} probing spin-momentum locking for inversion-symmetric nodes. For time-reversal symmetric nodes any angular dependence of the cross-section 
%\emph{implies} probing spin-momentum locking, while the reverse is not necessarily true.  
The strong angular $\Dtb$ dependence of the cross-section reflects that the spherical harmonics Eq. (\ref{DifferentialCrossSection_Appendix_Eq1:TR-Susceptibility}) 
and (\ref{DifferentialCrossSection_Appendix_Eq1:I-Susceptibility}) change rapidly as a function of $\Dtb$.  When $\omega$ is small, the cross-section
varies just as strongly with the angle on the surface
of the sphere $|\Dtb|=\hbar\omega/v_{\rm F}$.  This is a large variation for a small change in momentum. That is because the Weyl particle and hole have their momentum locked to spin, or equivalently, it reflects the singularity of the wavefunctions $|\pbt;\eta\chi\rangle$ at $\pbt=0$.  This differs from scattering
between two pockets of a narrow gap semiconductor, where there would be no angular $\Dtb$ dependence because the wavefunctions are continuous.
% no 1begin: canged
%%\onecolumngrid
\subsection{Universal features of the cross-section with an unpolarized detector}

The form of the cross-section can change a great deal depending on the values of
the coupling parameters, suggesting in particular that it might
not be possible to observe the chiralities at the two Weyl nodes, or even
whether they are the same or different.  With an unpolarized detector one loses information about how the neutron's spin is affected by coupling to the electron
so the situation is worse.

To understand the situation better, both theoretically and experimentally, a result that is independent of sample-parameters is desirable (e.g., a sum-rule).
However the usual sum-rules involve sums over all bands, obscuring the
relevant low energy physics of a WSM.
However, by taking the ratio of spherically averaged %\footnote{The average of some function $f(\Dtb)$ over the solid angle with respect to $\Dtb$ is
%$<f(\Dtb)>_{4\pi} = (1/4\pi)\int_{4\pi} \text{d}\Omega_{\Dtb}\, f(\Dtb) $
%where $\text{d}\Omega_{\Dtb} = \text{d} \theta_{\Dtb}\,  \text{d} \phi_{\Dtb} \, \sin \theta_{\Dtb}$. In the original coordinates this would be an average over an ellipsoid.} 
cross-sections we will get a prediction, which for time-reversal symmetric nodes (see Sec. \ref{DiffCrossSectionMagnetic_Appendix:Section:TR_nodesUnpolarized} and \ref{DiffCrossSectionMagnetic_Appendix:Section:TR_nodesPolarized}) is a universal expression capturing only the relevant relativistic Weyl fermion physics measured in 
internode scattering. Hence this expresses exactly the information we seek from a sum-rule. For inversion symmetric nodes (see Sec. \ref{DiffCrossSectionMagnetic_Appendix:Section:I_nodesUnpolarized}), the averaging method does not lead
to a completely universal expression, because of the coupling $\mathbf{F}^0$ may be present in this case. However, there is another universal property of the cross-section.

\subsubsection{Time-reversal symmetric Weyl nodes: Unpolarized incident neutrons\label{DiffCrossSectionMagnetic_Appendix:Section:TR_nodesUnpolarized}}
%%\label{DiffCrossSectionMagnetic_Appendix:Section:TR_nodes\label{DiffCrossSectionMagnetic_Appendix:Section:I_nodes}
Time-reversal symmetry has two consequences: the chiralities of the Weyl nodes is
the same, and the couplings are restricted by Eq. (\ref{TR_Coupling:Interpretation_Appendix}).
The inelastic cross-section Eq. (\ref{DifferentialCrossSection_Appendix_Eq6:DifferentialCrossSectionMagnetic}) is determined by Eq. (\ref{DifferentialCrossSection_Appendix_Eq1:TR-Susceptibility}). In spite
of the large number of coupling parameters, the averaging method mentioned above gives some universal predictions, and these reflect the two nodes's handedness being identical.  On the other hand,
the chirality cannot be measured.  The chirality appears only in the $\chi''^{i\,0}_{(+)}$ ($i=1,2,3$) components of the susceptibility, but since $\Fb^{0}_{\perp} = \bm{0} $, such terms do not 
appear in the cross-section.

For unpolarized incident neutrons Eq. (\ref{DifferentialCrossSection_Appendix_Eq6:DifferentialCrossSectionMagnetic_Unpolarized}) is
\begin{equation}
\Sigma^{(+)}(\qb,\omega) = \frac{a}{\pi} \Fb^{i,*}_{\perp} \cdot \Fb^{j}_{\perp} [\{\left(\hbar \omega/v_{\rm F}\right)^{2} - \vert \Dtb \vert^2 \} \d_{ij} + \Dt_{i}\Dt_{j} ]. \label{DifferentialCrossSection_Appendix_Eq1:Unpolarized_TR} 
\end{equation}
The tensor $\chi''^{ij}_{(+)}$ has no antisymmetric part,
so it consists only of terms that transform as a spherical tensor with angular momentum $l=0$ and $2$.
% (see Appendix \ref{DecompositionCrossSection_Appendix}).

\begin{figure}
\includegraphics[scale=0.75]{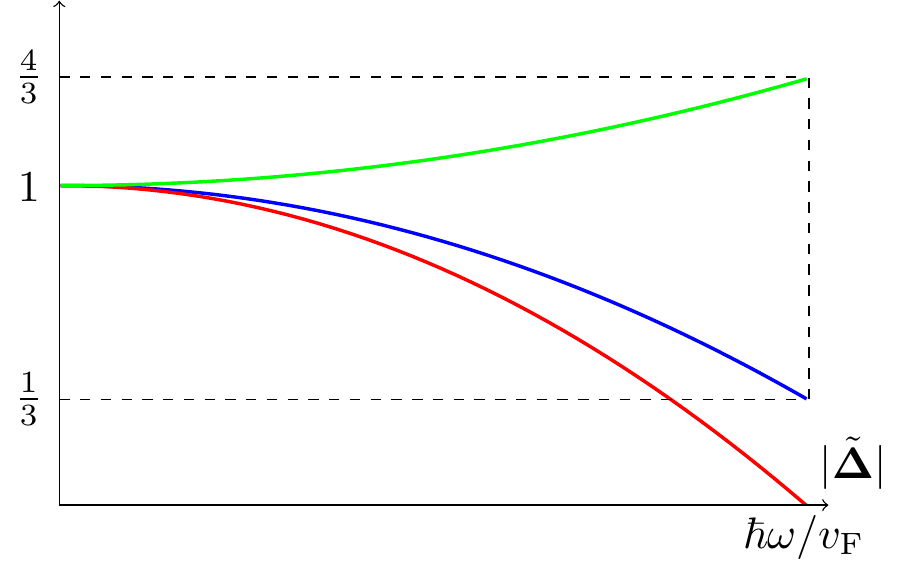}
%\caption{The solid angle averaged cross-section ratio Eq. \eqref{DifferentialCrossSection_Appendix_Eq2:AveragedRatioUnpolarized_TR}, \eqref{DifferentialCrossSection_Appendix_Eq1:AveragedRatioUnpolarized_I}, and \eqref{DifferentialCrossSection_Appendix_Eq2:AveragedRatioUnpolarized_I} in blue, red, and green, respectively.\label{DifferentialCrossSection_Appendix_fig:AveragedCrossSectionRatios}}
\caption{Universal predictions. The solid angle averaged cross-section ratio Eq. \eqref{DifferentialCrossSection_Appendix_Eq2:AveragedRatioUnpolarized_TR}, \eqref{DifferentialCrossSection_Appendix_Eq1:AveragedRatioUnpolarized_I}, and \eqref{DifferentialCrossSection_Appendix_Eq2:AveragedRatioUnpolarized_I} in blue, red, and green, respectively, as a functions of isotropic coordinate $\Dtb$.\label{DifferentialCrossSection_Appendix_fig:AveragedCrossSectionRatios}}
\end{figure}

 As previously stated, we can extract information by averaging the cross-section Eq. (\ref{DifferentialCrossSection_Appendix_Eq1:Unpolarized_TR}) over the solid angle\footnote{The average of some function $f(\Dtb)$ over the solid angle with respect to $\Dtb$ is
$<f(\Dtb)>_{4\pi} = (1/4\pi)\int_{4\pi} \text{d}\Omega_{\Dtb}\, f(\Dtb) $
where $\text{d}\Omega_{\Dtb} = \text{d} \theta_{\Dtb}\,  \text{d} \phi_{\Dtb} \, \sin \theta_{\Dtb}$.} One must do this
average with respect to $\Dtb=T\Db$, in which
the dispersion is isotropic. In terms of the original coordinates this is an average over an ellipsoid.
Averaging gives
$\left< \pi \Sigma^{(+)}(\qb,\omega) \right>
= a [\left(\hbar \omega/v_{\rm F}\right)^{2} - \frac{2}{3}\lvert \Dtb \rvert^{2}] \left(\alpha_{1} + \alpha_{2}\right)$ where all the sample-specific information $\alpha_{1,2}$ factors out. Hence, we can divide by $\left< \Sigma^{(+)}(\qb',\omega') \right>_{4\pi}$ for any arbitrary reference (within the low energy window) $\qb' = 2\kb_{0} - \Dtb'$ and $\omega'$ to get a result \emph{independent} of $\alpha_{1,2}$, i.e.,
\begin{equation}
\frac{\left< \Sigma^{(+)}(\qb,\omega) \right>_{4\pi}}{\left< \Sigma^{(+)}(\qb',\omega') \right>_{4\pi}}
=  \frac{\left(\hbar \omega/v_{\rm F}\right)^{2} - (2/3)\lvert \Dtb \rvert^{2}}{\left(\hbar \omega'/v_{\rm F}\right)^{2} - (2/3)\lvert \Dtb' \rvert^{2} } ,  \label{DifferentialCrossSection_Appendix_Eq1:AveragedRatioUnpolarized_TR} 
\end{equation}
which is a \emph{universal} function of $\Dtb,\Dtb',\omega$ and $\omega'$  that are all controlled in experiment.
For example, choosing the reference cross-section to be of same energy but with $\tilde{\Delta}=0$, the ratio of averaged cross-sections in $\Dtb$-coordinates centered on $2\kb_{0}$ is 
\begin{equation}
\frac{\left<\Sigma^{(+)}(-\Dtb,\omega) \right>_{4\pi}}{\left<\Sigma^{(+)}(\bm{0},\omega) \right>_{4\pi}}
= 1 - \frac{2}{3} \left(\frac{\lvert \Dtb \rvert}{\hbar \omega/v_{\rm F}}\right)^{2}, \label{DifferentialCrossSection_Appendix_Eq2:AveragedRatioUnpolarized_TR} 
\end{equation}
which is a \emph{universal} function of $\Dtb$ and $\omega$ plotted in Fig. \ref{DifferentialCrossSection_Appendix_fig:AveragedCrossSectionRatios}.
In particular, the averaged cross-section before the jump is $\frac13$ the cross-section at $\Dtb=0$.  This is a combined result of the density of states
decreasing by a factor of $\frac23$ with increasing $\Dtb$ and the interaction matrix elements decreasing
when the spins go from being antiparallel to parallel. This has to do with the fact that the interaction is more likely to flip
than not to flip the electron spin, as explained in Sec. \ref{DiffCrossSectionMagnetic_Appendix:ConventionalMagneticScattering} for the case $\F^i_j\propto\delta_{ij}$.  Note that the result surprisingly applies to \emph{any} $\F$ after averaging.

\begin{comment}
\begin{figure}
\begin{minipage}{0.95\columnwidth} 
\centering
\begin{tikzpicture}
\node at (0,1) {\includegraphics[width=0.65\textwidth]{./AveragedCrossSectionRatios}};
\node at (-3,1.55) {$1$};
\node at (-3,0.10) {$1/3$};
\draw [dashed] (-2.6,0.110) -- (2.55,0.110);
\node at (-3,2.3) {$4/3$};
\draw [dashed] (-2.6,2.30) -- (2.55,2.30);
\node at (2.5,-0.9) {$\frac{\hbar \omega}{v_{\rm F}}$};
\node at (3.4,-0.65) {$|\Dtb|$};
\draw [dashed] (2.55,-0.5) -- (2.55,2.3);
\draw [->] (-2.6,-0.65) -- (-2.6,2.7);
\draw [->] (-2.6,-0.65) -- (2.8,-0.65);
\end{tikzpicture}
\end{minipage}
\begin{minipage}{1\columnwidth}
\caption{The solid angle averaged cross-section ratio Eq. \eqref{DifferentialCrossSection_Appendix_Eq2:AveragedRatioUnpolarized_TR} for inversion symmetric
 nodes, in blue, and for time reversal symmetric nodes, Eqs. \eqref{DifferentialCrossSection_Appendix_Eq1:AveragedRatioUnpolarized_I}, and \eqref{DifferentialCrossSection_Appendix_Eq2:AveragedRatioUnpolarized_I} in red, and green, respectively.\label{DifferentialCrossSection_Appendix_fig:AveragedCrossSectionRatios}}
\end{minipage}
\end{figure}
\end{comment}

\subsubsection{Time-reversal symmetric Weyl nodes: Polarized incident neutrons\label{DiffCrossSectionMagnetic_Appendix:Section:TR_nodesPolarized}}
For polarized incident neutrons Eq. (\ref{DifferentialCrossSection_Appendix_Eq6:DifferentialCrossSectionMagnetic_Polarized}) is
\begin{eqnarray}
\pi \Pb \cdot \bm{\Sigma}'^{(+)}(\qb,\omega) \label{DifferentialCrossSection_Appendix_Eq1:Polarized_TR}
&=&  ia \, (\Pb\cdot \widehat{\kb}_{0}) \widehat{\kb}_{0} \cdot (\Fb^{i,*}_{\perp} \times \Fb^{j}_{\perp}) \\ &\times& [\{ (\hbar \omega/v_{\rm F})^{2} - \vert \Dtb \vert^2 \} \d_{ij} + \Dt_{i}\Dt_{j}] . \nonumber 
\end{eqnarray}
The cross-section for any material depends only on the component of $\mathbf{P}$
along $\kb_0$; in fact, in \emph{any} neutron scattering experiment the cross-section at low energies depends
only on the component of $\mathbf{P}$ in the direction of the momentum
transfer, because of the condition $\bm{\mathcal{J}}\cdot \bm{\Delta}\kb=0$.
We can take a ratio between any two solid angle averages of Eq. (\ref{DifferentialCrossSection_Appendix_Eq1:Polarized_TR}) to get a result \emph{independent} of $\beta_{1,2}$, i.e.
\begin{equation}
\frac{\left< \Pb\cdot \bm{\Sigma}'^{(+)}(\qb,\omega) \right>_{4\pi}}{\left< \Pb'\cdot \bm{\Sigma}'^{(+)}(\qb',\omega') \right>_{4\pi}} 
=  \frac{\Pb\cdot \widehat{\kb}_{0}}{\Pb' \cdot \widehat{\kb}_{0}}  \frac{\left< \Sigma^{(+)}(\qb,\omega) \right>_{4\pi}}{\left< \Sigma^{(+)}(\qb',\omega') \right>_{4\pi}},
\end{equation}
which is a \emph{universal} function of $\Pb,\Pb',\Dtb,\Dtb',\omega$ and $\omega'$ that are controlled in experiment. This function is that of Eq. (\ref{DifferentialCrossSection_Appendix_Eq1:AveragedRatioUnpolarized_TR}) weighted by the ratio of polarization 
vectors' projection onto the internode direction.

%-------------------- Experimental signatures: Inversion symmetric Weyl nodes ---------------
\subsubsection{Inversion symmetric Weyl nodes: Unpolarized incident neutrons\label{DiffCrossSectionMagnetic_Appendix:Section:I_nodesUnpolarized}}%\label{DiffCrossSectionMagnetic_Appendix:Section:I_nodes}
%{\color{blue}For inversion symmetric Weyl nodes, one can derive analogous results.}
For the inversion symmetric case, the inelastic cross-section Eq. (\ref{DifferentialCrossSection_Appendix_Eq6:DifferentialCrossSectionMagnetic}) is determined by Eq. (\ref{DifferentialCrossSection_Appendix_Eq1:I-Susceptibility}) and the coupling is restricted by Eq. 
(\ref{I_Coupling:Interpretation_Appendix}). 
Two differences from the time-reversal symmetric case are
that $\mathbf{F}^0$ can be nonzero which makes it more complicated
to obtain a ``universal prediction".
%for unpolarized scattering,and that the chirality appears in the cross-section.  This suggests
%that it will be possible to measure the chirality of the nodes. 
Furthermore, the chirality of the node where a hole is created \emph{can} enter the cross-section  through the antisymmetric part of the susceptibility $\chi''^{i\,j}_{(+)}$ ($i,j=1,2,3$), which
allows the chirality to be measured, although polarized neutrons and detectors are required for this.
In this section, we will illustrate the use of the spectral 
decomposition of the effective coupling Eq. (\ref{DifferentialCrossSection_Appendix_Eq1:EffectiveCoupling}).

For unpolarized incident neutrons Eq. (\ref{DifferentialCrossSection_Appendix_Eq6:DifferentialCrossSectionMagnetic_Unpolarized}) is
\begin{eqnarray}
\pi &\Sigma^{(+)}&(\qb,\omega)
= \alpha_{0}\frac{3 a}{2}[\left(\hbar\omega/v_{\rm F}\right)^{2} - \lvert \Dtb \rvert^{2}] \label{DifferentialCrossSection_Appendix_Eq1:DifferentialCrossSectionMagnetic_Unpolarized_I} \\ 
&+&  \frac{a}{2}\sum_{m=1}^{2}\,\alpha_{m}\,[\left(\hbar\omega/v_{\rm F}\right)^{2} + \lvert \Dtb \rvert^{2} - 2|\Dtb \cdot \abh_{m}|^{2}],\nonumber 
\end{eqnarray}
where $\hat{\mathbf{a}}_{1,2}$ are the two orthogonal, real unit vectors from the spectral decomposition\footnote{The vectors $\hat{\mathbf{a}}_{1,2}$ are real since ${\rm F}$ is real for inversion symmetric nodes.} of $\F$, and $\alpha_{1,2}$ are the corresponding parameters as in Eq. (\ref{DifferentialCrossSection_Appendix_Eq1:EffectiveCoupling_Unpolarized}). In this expression appears
a term
$\alpha_{0} = \Fb^{0}_{\perp}\cdot \Fb^{0}_{\perp}$ which
is generically nonzero.
This term gives a $\chi''^{0\,0}_{(+)}$ contribution to the cross-section with no angular $\Dtb$ dependence.   As $\chi''^{i\,0}_{(+)} = 0$ and  $\Fb^{i}_{\perp} \cdot \Fb^{j}_{\perp}$ is symmetric in spin-indices, only the symmetric part of $\chi''^{i\,j}_{(+)}$ contributes, which we have seen does
not depend on $\chi$.  It is therefore not possible
to measure the chirality of the nodes with unpolarized neutrons\footnote{This can be proved without
calculation, by a general symmetry argument--the symmetric part consists only of terms, that transform as spherical tensors with angular momentum $l=0$ and $2$, and any
such tensor function of $\Dtb$ is an ordinary tensor, not an axial tensor.
Therefore this term is independent of the chirality of the nodes.}.

\begin{figure}
\includegraphics[scale=0.65]{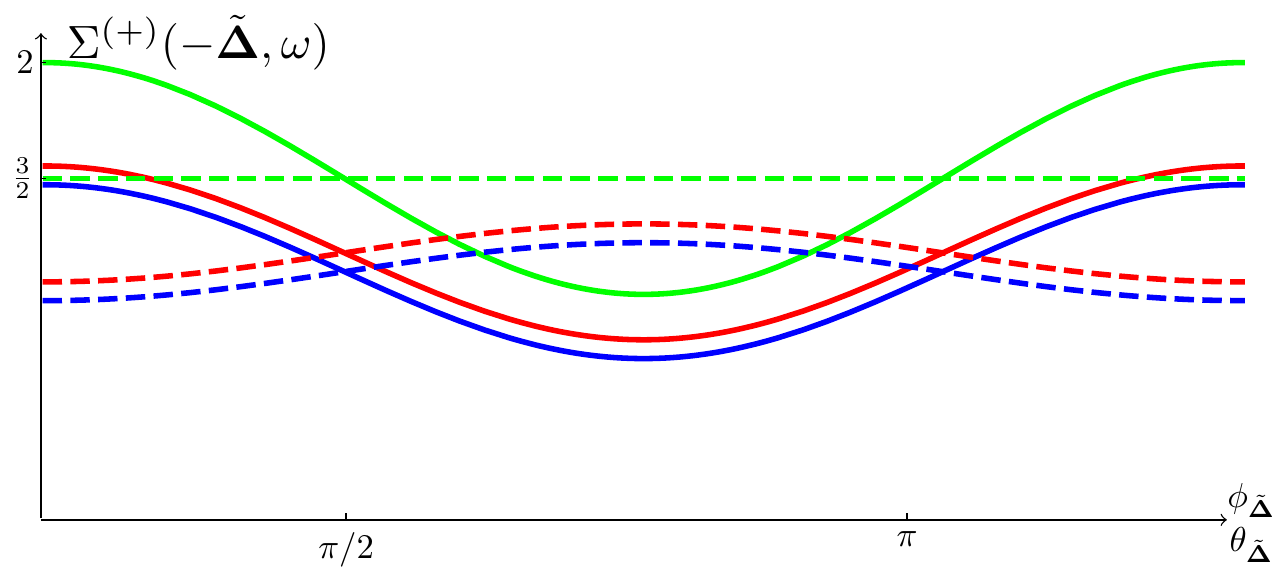} %[scale=0.55]
%\caption{Plot of cross-section $\Sigma^{(+)}(-\Dtb,\omega)$ centered on $2\kb_{0}$ for inversion symmetric nodes for coupling eigendirections Eq. (\ref{DifferentialCrossSection_Appendix_Eq1:EffectiveCoupling_Unpolarized}) $\widehat{\ab}_{1} = \hat{\mathbf{x}}$ and $\widehat{\ab}_{2} = \hat{\mathbf{y}}$ for $|\Dtb| = 0.95 \hbar \omega / v_{\rm F}$. The green, blue, and red curves corresponds to that of Fig. \ref{Unpol_E_F0_a1_a2(095_0_x_y)pm}, \subref{Unpol_E_F0_a1_a2(095_0_x_05y)pm},%\subref{Unpol_E_F0_a1_a2(095_05_x_05y)pm}, respectively, as a function of $\theta_{\Dtb}$ at $\phi_{\Dtb} = \pi/4$, whereas the dashed curves are functions of $\phi_{\Dtb}$ at $\theta_{\Dtb} = \pi/4$.\label{Unpol_E_F0_a1_a2(d09_thetaPi4_psiPi4)pm}}
\caption{Angular variation of cross-section for different couplings. The cross-section for scattering between inversion symmetric nodes at a given energy transfer $\hbar\omega$ is plotted as a function angles $\phi_{\Dtb}$ and $\theta_{\Dtb}$ on a sphere with radius $|\Dtb| = 0.95 \hbar \omega / v_{\rm F}$. The green, blue, and red curves corresponds to that of Fig. \ref{Unpol_E_F0_a1_a2(095_0_x_y)pm}, \subref{Unpol_E_F0_a1_a2(095_0_x_05y)pm},\subref{Unpol_E_F0_a1_a2(095_05_x_05y)pm}, respectively, as a function of $\theta_{\Dtb}$ at $\phi_{\Dtb} = \pi/4$, whereas the dashed curves are functions of $\phi_{\Dtb}$ at $\theta_{\Dtb} = \pi/4$.\label{Unpol_E_F0_a1_a2(d09_thetaPi4_psiPi4)pm}}
\end{figure}

The cross-section Eq. (\ref{DifferentialCrossSection_Appendix_Eq1:DifferentialCrossSectionMagnetic_Unpolarized_I}) is plotted in Fig. \ref{CrossSectionFigNEW} as a function of $\Dtb$ for the case where coupling eigendirections of Eq. 
(\ref{DifferentialCrossSection_Appendix_Eq1:EffectiveCoupling_Unpolarized}) are $\widehat{\ab}_{1} = \hat{\mathbf{x}}$ and $\widehat{\ab}_{2} = \hat{\mathbf{y}}$ for various values of $\alpha_{0,1,2}$. Figure \ref{Unpol_E_F0_a1_a2(d09_thetaPi4_psiPi4)pm} plots cuts of the cross-section plotted in Fig. \ref{Unpol_E_F0_a1_a2(095_0_x_y)pm},\subref{Unpol_E_F0_a1_a2(095_0_x_05y)pm} and \subref{Unpol_E_F0_a1_a2(095_05_x_05y)pm}, from which one sees that the intensity variation is substantial. The $4$-band toy model (see Sec. \ref{4bandWSM_Appendix}) corresponds to couplings with $\widehat{\ab}_{1} = \hat{\mathbf{x}}$, $\widehat{\ab}_{2} = \hat{\mathbf{y}}$, $\alpha_{1} = \alpha_{2} = \F_{\perp,xx}^2$ and $\alpha_{0} = 0$, the cross-section of which therefore has the same angular dependence as the top row of Fig. \ref{CrossSectionFigNEW} %\ref{Unpol_E_F0_a1_a2(02_0_x_y)pm},%\subref{Unpol_E_F0_a1_a2(05_0_x_y)pm},\subref{Unpol_E_F0_a1_a2(095_0_x_y)pm}, 
but the intensity is a factor $2 \F_{\perp,xx}^2$ amplified by the value in Fig. \ref{Coupling}.

The angular average of Eq. (\ref{DifferentialCrossSection_Appendix_Eq1:DifferentialCrossSectionMagnetic_Unpolarized_I}) is
\begin{eqnarray}
\left< \pi \Sigma^{(+)}(\qb,\omega) \right>_{4\pi} &=&\alpha_{0}\frac{3 a}{2}[\left(\hbar\omega/v_{\rm F}\right)^{2} - \lvert \Dtb \rvert^{2}] \label{DifferentialCrossSection_Appendix_Eq2:DifferentialCrossSectionMagnetic_Unpolarized_I} \\ 
&+ &\frac{a}{2} [\left(\hbar\omega/v_{\rm F}\right)^{2} + \lvert \Dtb \rvert^{2}/3]\left(\alpha_{1} + \alpha_{2}\right), \nonumber 
\end{eqnarray}
where the sample specific information $\alpha_{0,1,2}$ does not factor out. Hence we \emph{cannot} divide by $\left< \Sigma^{(+)}(\qb',\omega') \right>_{4\pi}$ for any arbitrary reference $\qb'$ and $\omega'$ to get a universal result independent of $\alpha_{0,1,2}$.  However, if the coupling $\Fb_{\perp}^{i} = \mathbf{0}$ vanish for all $i =1,2,3$ then $\alpha_{1,2} = 0$ and we can get a result \emph{independent} of $\alpha_{0}$.  For example, choosing the reference cross-section to be of same energy but direct internode  scattering, the ratio of averaged cross-sections in $\Dtb$-coordinates centered on $2\kb_{0}$ is 
\begin{equation}
\frac{\left< \Sigma^{(+)}(-\Dtb,\omega) \right>_{4\pi}}{\left< \Sigma^{(+)}(\bm{0},\omega) \right>_{4\pi}} = 1 - \left(\frac{\lvert \Dtb \rvert}{\hbar\omega/v_{\rm F}}\right)^{2} ,\label{DifferentialCrossSection_Appendix_Eq1:AveragedRatioUnpolarized_I} 
\end{equation}
which is a  monotonically attenuating function plotted in Fig. \ref{DifferentialCrossSection_Appendix_fig:AveragedCrossSectionRatios}. If, on the other hand, the coupling $\Fb_{\perp}^{0} = \bm{0}$ vanishes then $\alpha_{0} = 0$ and
we can get a result \emph{independent} of $\alpha_{1,2}$.  For example, choosing the reference dataset to be the same as above, the ratio of averaged cross-sections in $\Dtb$-coordinates centered on $2\kb_{0}$ is 
\begin{equation}
\frac{\left< \Sigma^{(+)}(-\Dtb,\omega) \right>_{4\pi}}{\left< \Sigma^{(+)}(\bm{0},\omega) \right>_{4\pi}} = 1 + \frac{1}{3}\left(\frac{\lvert \Dtb \rvert}{\hbar\omega/v_{\rm F}}\right)^{2}  ,\label{DifferentialCrossSection_Appendix_Eq2:AveragedRatioUnpolarized_I}
\end{equation}
which is a monotonically increasing function plotted in Fig. \ref{DifferentialCrossSection_Appendix_fig:AveragedCrossSectionRatios}.
Hence the ratio with $\alpha_{1,2} = 0$ can be distinguished from the ratio with $\alpha_{0} = 0$. In the general case with $\alpha_{0,1,2} \neq 0$ one does not obtain a universal ratio of solid angle averaged cross-sections. Although Eq. (\ref{DifferentialCrossSection_Appendix_Eq2:DifferentialCrossSectionMagnetic_Unpolarized_I}) is non-universal, the functional dependence, $c_1(\hbar\omega/v_{\rm F})^2+c_2|\Dtb|^2$ with constants $c_{1,2}$ is very specific.

In fact, there \emph{is} a more quantitative universal prediction as well.
 Eq. \eqref{DifferentialCrossSection_Appendix_Eq1:DifferentialCrossSectionMagnetic_Unpolarized_I} can be written as
\begin{equation}
\pi\Sigma^{(+)}(\qb,\omega)=\frac{\bar{\alpha}a}{2}\left[ (\hbar\omega/v_F)^2+\sum_{m=1}^3 c_m(\Dtb\cdot \hat{\mathbf{a}}_m)^2\right],
\label{InversionSymmetricCrossSectionCombined}
\end{equation}
where $\bar{\alpha}$ and $c_i$'s are certain parameters
and  $\hat{\bm{a}}_3$ is a unit vector completing a basis with $\hat{\mathbf{a}}_1$
and $\hat{\mathbf{a}}_2$; i.e., it is the direction in pseudospin space
that is not coupled to the neutron spin. That such a direction exists follows
from the fact that there is a direction in \emph{neutron spin space} that is not
coupled to the pseudospin, as seen more formally in the derivation of the spectral decomposition, see Sec. \ref{CrossSectionMagneticFormula_Appendix}.
 Eq. \eqref{InversionSymmetricCrossSectionCombined} follows from $|\Dtb|^2=\sum_{m=1}^3 (\Dtb\cdot \hat{\mathbf{a}}_m)^2$.
The $\Dtb$-dependence of this expression is a quadratic function
of $\Dtb$; although with respect to the $\hat{\mathbf{a}}$ basis it is diagonal, in the coordinate system of the experiment, it could be an arbitary quadratic function of $\Dtb$.
Consider the cross-section at the maximum possible transfer momentum,
$|\Dtb|=\hbar\omega/v_{\rm F}$.  A quadratic form on
the surface of a sphere has two maxima, two minima and two saddle-points (at
diametrically opposite pairs of points).  The prediction is that, the cross-section
always has the property that the value at the maximum is the sum of
the value at the saddle point and the minimum.
In fact, the extrema always correspond to the eigendirections of the quadratic
form, namely $\hat{\mathbf{a}}_1,\hat{\mathbf{a}}_2$ and $\hat{\mathbf{a}}_3$.
The values of the cross-sections at these points are $2\alpha_2(\hbar\omega/v_{\rm F})^2$, $2\alpha_1(\hbar\omega/v_{\rm F})^2$, and $2(\alpha_1+\alpha_2)(\hbar\omega/v_{\rm F})^2$ respectively.  The
last is the largest since $\alpha_1,\alpha_2\geq 0$.  This prediction
can be understood qualitatively by noting that the initial and final spins of the
electron are antiparallel to one another.  Hence there is no contribution
to the cross-section at maximal $|\Dtb|$
due to the $\mathbf{F}^0$ coupling, which does not
cause spin flips, while the cross-section due
to the other interactions is greatest when the momentum
is along $\mathbf{\hat{a}}_3$ because
the neutron couples to both components of the spin perpendicular to this, namely $\hat{\mathbf{a}}_1$ and $\hat{\mathbf{a}}_2$ and so each term in the
interaction\footnote{\label{clamshell}The spectral decomposition, $\unexpanded{\Fb_\perp^i\cdot \Fb_\perp^j=\sum_{m=1}^2 \alpha_m \hat{\mathbf{a}}_m^i \hat{\mathbf{a}}_m^j}$, implies $\unexpanded{\F_{\perp,j}^{i}}=\sum_{m=1}^2 \sqrt{\alpha_m}\hat{\mathbf{a}}_{m}^i \hat{\mathbf{a}}_{m}^{\prime j}$ for a certain pair of orthonormal vectors 
$\unexpanded{\hat{\mathbf{a}}^{\prime}_1,\hat{\mathbf{a}}^{\prime}_2}$, according to
the theory of singular value decompositions.
Thus the interaction is
$\unexpanded{\F_{\perp,j}^i\tau_j\sigma_i=\sum_{m=1}^2\sqrt{\alpha_{m}}
(\hat{\mathbf{a}}_{m}\cdot\bm{\sigma}) (\hat{\mathbf{a}}^{\prime}_{m}\cdot\bm{\tau})}$.}
 induces the spin and also the momentum to flip.

\subsubsection{Inversion symmetric Weyl points: Polarized incident neutrons}

For polarized incident neutrons Eq. (\ref{DifferentialCrossSection_Appendix_Eq6:DifferentialCrossSectionMagnetic_Polarized}) is
%\begin{subequations} \label{DifferentialCrossSection_Appendix_Eq6:DifferentialCrossSectionMagnetic_Polarized_I} 
\begin{eqnarray}
\pi&\Pb&\cdot \bm{\Sigma}'^{(+)}(\qb,\omega)  \nonumber \\
&=& -\chi a \left(\hbar \omega/v_{\rm F}\right) \Pb \cdot (\Fb^{i}_{\perp} \times \Fb^{j}_{\perp}) \epsilon_{ijk} \Dt_{k}. \label{DifferentialCrossSection_Appendix_Eq6a:DifferentialCrossSectionMagnetic_Polarized_I} 
%&=& \chi a \left(\hbar \omega/v_{\rm F}\right)  \Pb\cdot \widehat{\kb}_{0}  \sum_{m=1}^{2}\,\beta_{m}\, i \Dtb \cdot \widehat{\bb}_{m}\times \widehat{\bb}_{m}^{*}.  
%\label{DifferentialCrossSection_Appendix_Eq6b:DifferentialCrossSectionMagnetic_Polarized_I}
\end{eqnarray}
%\end{subequations}
%where we used the fact that $\Fb^i_\perp\times\Fb^j_\perp$ is along $\kb_0$ to
%show that this depends only on the component of $\mathbf{P}$ along $\kb_0$.
%(Compare to
%Eq. (\ref{DifferentialCrossSection_Appendix_Eq1:Polarized_TR}).)
Despite the possibility that $\Fb_{\perp}^{0} \neq \bm{0}$ there is no $\chi''^{0\,0}_{(+)}$ contribution because $\widehat{\kb}_{0} \cdot \Fb^{0}_{\perp} \times \Fb^{0}_{\perp} = 0$. As $\chi''^{i\,0}_{(+)} = 0$ and $\Fb^{i}_{\perp} \times \Fb^{j}
_{\perp}$ is antisymmetric in spin-indices, only the antisymmetric part of $\chi''^{i\,j}_{(+)}$ contributes, which is  a term that transforms as a spherical tensor with angular momentum $l=1$. From the antisymmetric part of $\chi''^{i\,j}_{(+)}$
one sees that this measures \enquote{chiral} fluctuations $\langle \sb(\qb,\omega)\times \sb(-\qb,-\omega) \rangle \cdot \Dtb$ originating in the axial-vector of the interaction.% for a $\chi \to \bar{\chi}$ scattering process.

Now $\mathbf{F}^i_\perp\times \mathbf{F}^j_\perp$ is always parallel to $\kb_0$. % because $\mathbf{F}^i_\perp$ is perpendicular to $\kb_0$.
This implies $(\mathbf{F}^i_\perp\times\mathbf{F}^j_\perp)\cdot \mathbf{P}$
depends only on the component of $\mathbf{P}$ in the internode direction $\kb_0$;
further it is antisymmetric between $i$ and $j$,
hence it can be written $\epsilon_{ijk}\gamma_k (\mathbf{P}\cdot\hat{\kb}_0)$, for
some numbers $\gamma_k$. (Explicitly, $\gamma_3=\hat{\kb}_0\cdot (\Fb_1\times\Fb_2$, etc.)
Hence
\begin{eqnarray}
\pi\Pb \cdot \bm{\Sigma}'^{(+)}(\qb,\omega)
&=& -\chi a \left(\hbar \omega/v_{\rm F}\right) (\bm{\gamma}\cdot\Dtb) (\Pb\cdot\hat{\kb}_0).%\left(\gamma_{1} \Dt_{x} + \gamma_{2} \Dt_{y} + \gamma_{3} \Dt_{z} \right), \nonumber
\label{DifferentialCrossSection_Appendix_Eq6b:DifferentialCrossSectionMagnetic_Polarized_I} 
\end{eqnarray}
This part is linear in $\Dtb$, so the angular average is
\begin{equation}
\left< \Pb\cdot \bm{\Sigma}'^{(+)}(\qb,\omega) \right>_{4\pi}= 0.
\end{equation}
Now although this result depends on the chirality, the coefficients $\gamma_i$
are not known because they depend on $\F$, so it is not possible to measure
the chirality even with polarized neutrons when the detector is unpolarized. 
The next section explains that the polarization-independent
and dependent cross-sections $\Sigma^{(+)},\bm{\Sigma}^{\prime(+)}$ are not enough to determine $\chi$; there is always at least one choice of $\F$ that matches
the data for each of $\chi=\pm 1$.

\subsection{Polarized measurement}\label{PolarizedMeasurement}
%We will now consider polarized neutrons and polarized detectors;
%one of the results is that it \emph{is} possible to measure the chirality
%for inversion-symmetric WSMss in such an experiment.
We will now consider polarized neutrons and detector;
the main result is that it \emph{is} possible to measure the chirality
for inversion-symmetric WSMs.

\subsubsection{Pure States of Scattered Neutrons}

Consider directing an incident fully polarized beam of neutrons with polarization vector $\Pb^{i}$ on a WSM and measuring the polarization vector $\Pb^{f}$ of the scattered beam. The Blume-Maleyev polarization matrix describes the relationship between them. Instead of calculating this, we simplify the discussion and consider, for the moment, a single incident neutron in spin state $\lvert \tau_{i} \rangle$ and  measuring whether the scattered neutron is in the state $\lvert \tau_{f} \rangle$ or in the orthogonal one. %The  scattering amplitude is then $\langle f_{\rm n};f_{\rm w} \rvert H_{\Bb} \lvert i_{\rm w}; i_{\rm n} \rangle \propto  \langle \tau_{f} \rvert  \taub \lvert  \tau_{i} \rangle \cdot  \langle f_{\rm w} \rvert \bm{\mathcal{M}}_{\perp}(\qb) \lvert i_{\rm w} \rangle$.

The Weyl states are not eigenvectors of $\s_{z}$, but are dependent on the direction and magnitude of $\Dtb$. For a given scattering process, i.e. a fixed initial and final neutron state,  the cross-section Eq. (\ref{DifferentialCrossSection_Appendix_Eq2:DifferentialCrossSectionMagnetic}) sums up all internal particle-hole pair Weyl states which fulfill the energy and momentum constraints of the system (see Fig. \ref{EnergyMomentumConservation_d005_d075_d095}). Intuitively one expects that each of these pairs affects the scattered neutron in a different way.%, which is also apparent from the transition current. 
%Consequently, the spin-flip rule is not expected to apply, as the final neutron state is not expected to be a pure state but rather in a mixed state.%, because one probes particle-hole pairs which can have many internal states for a given energy and momentum, and not conventional magnetic excitations.

For small amounts of transferred momentum it \emph{is} correct (as this
reasoning suggests) that the scattered neutron will be in a mixed state.  
However,
consider the case where the momentum transfer is the maximum that is possible for the given energy transfer, $|\Dtb| = \hbar \omega/ v_{\rm F}$. %; in this case, the constant energy contour degenerates to a line. 
For a given pure initial spin state of the neutron and a fixed momentum transfer, the cross-section can be shown (see below)
to take the form $\frac{\text{d}^{2} \s (\qb,\omega)}{\text{d} \Omega \text{d} E_{f}} \big|_{\tau_{i}}^{\tau_{f}} \propto  \left|\langle\tau_{f} \vert \phi \rangle\right|^2$ where the auxiliary state $\vert \phi \rangle$ depends on the initial neutron state $\lvert \tau_{i} \rangle$ and direction $\Dtb$. In other
words, the scattered neutron is in a pure state $|\phi\rangle$. This can be demonstrated experimentally by measuring that there is a certain final state for which the scattering rate into that state is zero. This final neutron state is the time-reversed ket\footnote{Time-reversal operator $T$ on a single-particle state is $T = \theta K$, where $K$ is conjugation operation and $\theta = \s_{y}$, as explained in Appendix \ref{Interaction_Appendix}.} of $\rvert \phi\rangle$, i.e. $\lvert\tau_{f}^{\rm TR} \rangle = T \lvert \phi \rangle$, since $\langle\tau_{f}^{\rm TR} \vert \phi \rangle =  0$. %Because the Hilbert space is two-dimensional, the \emph{only} scattering channel available is the final state $\lvert \tau_{f} \rangle = \lvert \phi \rangle / |\phi| = c_{1} \lvert \uparrow \rangle + c_{2} \lvert \downarrow \rangle $. %Consequently, the spin-flip $\D S$ does not have to be quantized in integers despite that the final neutron is in a pure state. %The inelastic cross-section (see Appendix \ref{FullyPolarized_Appendix}) is therefore 
%\begin{equation}
%\frac{\text{d}^{2} \s (\qb,\omega)}{\text{d} \Omega \text{d} E_{f}} \bigg|_{\tau_{i}}^{\tau_{f}} 
%= \frac{\q_{f}}{\q_{i}} \left(\frac{\gamma r_{0}}{2}\right)^{2} \frac{a}{\pi} \left(\frac{\hbar \omega}{v_{\rm F}}\right)^2
%|\phi|^2. \label{PolarizationFull_Appendix:Eq1.CrossSection}
%\end{equation}

The fact that there is only one transition available for a given momentum transfer is \emph{direct} evidence of spin-momentum locking. 
The reason for the perfect polarization, in more detail, is that in the extreme limit $|\tilde{\bm{\Delta}}|\approx\hbar\omega/v_{\rm F}$, the set of possible
internal momenta degenerates from an ellipsoid to a line. All the
possible values are parallel and thus %the particle-hole pair momenta are forced to be opposite to each other (see Eq. (\ref{DifferentialCrossSection_Appendix_Eq1:Spinors_DMaximum})), so the states of the electron and hole are eigenstates along the same direction $\tilde{\bm{\Delta}}$.  
 the electron and hole spin states are the same throughout the particle-hole continuum. The current matrix element is the same for all pairs, so the integral
over the state of the electrons and holes just gives a multiplicative factor and
%the scattering rate is proportional to the integral of the interaction operator,
%$\bm{\mathcal{M}}\cdot\bm{\mu}$ (where $\bm{\mu}$ is the spin of the neutron
%and $\bm{\mathcal{M}}$ is the magnetization)
%between all kinematically allowed initial and final states of electron and neutron.  At the maximal momentum transfer, the initial and final momenta of
%the electron always point in a fixed direction, which the spin of the electron
%is locked to. Hence the matrix element does not
%vary over the range of the integral, and the cross-section is proportional 
the cross section is proportional to  %|\langle \tau_f|\bm{\mu}|\tau_i\rangle\cdot\langle \chi_f;\hat{\Dtb}|\bm{\mathcal{M}}|-\hat{\Dtb}; -\chi_i\rangle|^2$ 
\begin{equation}
\frac{d\sigma}{d\Omega}\propto|\langle \tau_f;\chi_f;-\widehat{\Dtb}|\bm{\tau}\cdot\bm{\mathcal{M}}|\widehat{\Dtb}; -\chi_i;\tau_i\rangle|^2
\label{PolarizationFull_Appendix:Eq1Transition}
%\label{PolarizationFull_Appendix:EqTransition}
\end{equation}
where the magnetization $\bm{\mathcal{M}}$ is given by
Eq. (\ref{section:CurrentOperator_Appendix_Eq.1:MagnetizationOperator1})
and $\bm{\tau}$ is, as above, the neutron spin operator.  The dynamics of the neutron spin may be understood as a precession of the neutron in a magnetic field that depends on how the electron transitions.
To see this, we factor this expression as $|\langle \tau_f|\bm{\tau}|\tau_i\rangle\cdot\langle \chi_f;-\widehat{\Dtb}|\bm{\mathcal{M}}|\widehat{\Dtb}; -\chi_i\rangle|^2$, then define the c-number 
$\bm{\mathcal{M}}_{fi}=\langle\chi_f;-\widehat{\Dtb}|\bm{\mathcal{M}}|\widehat{\Dtb};-\chi_i\rangle$.  The transition probability can now be written as
$\frac{d\sigma}{d\Omega}\propto |\langle \tau_f|\bm{\tau}\cdot\bm{\mathcal{M}}_{fi}|\tau_i\rangle|^2$
 Thus, we may
define $|\phi\rangle=\bm{\mathcal{M}}_{fi}\cdot\bm{\tau}|\tau_i\rangle$, and
the cross-section is given by $|\langle\tau_f|\phi\rangle|^2$ as claimed above.  Intuitively, when  
the electron's spin flips in a particular way, the scattered beam ends up in a fully polarized state if the beam was initially fully polarized; the final state is obtained by applying the operator $\bm{\mathcal{M}}_{fi}\cdot\bm{\tau}$ to
the initial state.  This mechanism is due to the constant energy contour degenerating into a line and to perfect spin-momentum locking; if there were curvature, the electron spinors would not be all aligned and the final neutron beam would not be fully polarized.

%{\color{green} MAYBE HAVE TO ADD A LITTLE AND CHANGE A LITTLE TO APPENDIX E TO MAKE IT FIT
%WITH DISCUSSION HERE}\\

In a neutron experiment, in which a beam of $N \gg1$ neutrons having a polarization $\Pb^{i}$ is incident to the target, all neutrons scattered to a certain momentum have the same  available scattering channel $\lvert \phi \rangle=\bm{\mathcal{M}}_{fi}\cdot\bm{\tau}|\tau_i\rangle$ if the initial neutron beam is fully polarized. This state has an expansion
\begin{equation}
 c_{1} = \langle \uparrow \rvert \phi \rangle/\lvert \phi \rvert \quad , \quad 
 c_{2} = \langle \downarrow \rvert \phi \rangle/\lvert \phi \rvert. \label{PolarizationFull_Appendix:Eq2.Polarization}
\end{equation}
 The emitted neutrons in this direction are  \emph{fully polarized} and specified by $\Pb^{f} \cdot \taub \lvert \tau_{f} \rangle = \lvert \tau_{f} \rangle$
where the polarization vector has the components
\begin{equation}
\P^{f}_{x} = 2 \Re\left[c_{1}^{*}c_{2}\right] \> , \>  \P^{f}_{y} = 2 \Im \left[c_{1}^{*}c_{2}\right] \> , \> \P^{f}_{z} = |c_{1}|^2 - |c_{2}|^2. \label{PolarizationFull_Appendix:Eq1.Polarization}
\end{equation}
The polarization vector Eq. (\ref{PolarizationFull_Appendix:Eq1.Polarization}) is to be understood as a field $\Pb^{f}(\Dtb/|\Dtb|)$ on the surface of the sphere of transferred maximum momentum.
The matrix element of the magnetization
can be evaluated explicitly %{\color{red}(see Appendix \ref{FullyPolarized_Appendix})}
for time-reversal and inversion symmetric nodes, $\bm{\mathcal{M}}_{fi} = \widehat{\Dt}_{i} \Fb^{i}_{\perp}$ and  $\bm{\mathcal{M}}_{fi} = (\hat{u}_{j}+i\chi \hat{v}_j)\Fb_{\perp}^{j}$, respectively. Here $\hat{\mathbf{u}},\hat{\mathbf{v}}$ are some pair of vectors making a right-handed coordinate system\footnote{The latter expression follows from  \unexpanded{$\langle-\hat{\mathbf{n}}|\bm{\sigma}|\hat{\mathbf{n}}\rangle=\hat{\mathbf{u}}+i\hat{\mathbf{v}}$},  where  \unexpanded{$|\hat{\mathbf{n}}\rangle$} is the spin-1/2 state aligned with \unexpanded{$\hat{\mathbf{n}}$}.  Changing to a different pair of vectors \unexpanded{$\hat{\mathbf{u}},\hat{\mathbf{v}}$} just turns out to multiply the right-hand side by a phase, which matches the ambiguity in the phase of the left-hand side; the phases of \unexpanded{$|\pm\hat{\mathbf{n}}\rangle$} can be chosen independently of one another.} together with $\Dtb$.
The \emph{total} cross-section is proportional to $|\phi|^2$, which gives
\begin{subequations}\label{PolarizationFull_Appendix:Eq1.auxiliarystateNorm}
\begin{eqnarray}
|\phi|^2 &=& \widehat{\Dt}_{i}\widehat{\Dt}_{j} (\Fb^{i,*}_{\perp} \cdot \Fb^{j}_{\perp} +  \Pb^{i} \cdot \widehat{\kb}_{0} i\widehat{\kb}_{0}\cdot\Fb^{i,*}_{\perp} \times \Fb^{j}_{\perp}), \hspace*{0.45cm} \label{PolarizationFull_Appendix_TR:Eq1.auxiliarystateNorm} \\
|\phi|^2 &=& (\d_{ij} - \widehat{\Dt}_{i}\widehat{\Dt}_{j}) \Fb_{\perp}^{i}\cdot \Fb_{\perp}^{j} \nonumber \\  
  &&\hspace*{0.05cm} - \chi \Pb^{i} \cdot \widehat{\kb}_{0} \widehat{\Dt}_{k} \epsilon_{k\,i\,j} \widehat{\kb}_{0} \cdot \Fb_{\perp}^{i} \times \Fb_{\perp}^{j},\label{PolarizationFull_Appendix_I:Eq1.auxiliarystateNorm}
\end{eqnarray} 
\end{subequations}
for time-reversal and inversion symmetric nodes, respectively, in agreement
with our general expressions [see Eqs. (\ref{DifferentialCrossSection_Appendix_Eq1:Unpolarized_TR}), (\ref{DifferentialCrossSection_Appendix_Eq1:Polarized_TR}),
(\ref{DifferentialCrossSection_Appendix_Eq1:DifferentialCrossSectionMagnetic_Unpolarized_I}),(\ref{DifferentialCrossSection_Appendix_Eq6a:DifferentialCrossSectionMagnetic_Polarized_I})] for the cross-section in the case where $|\Dtb|=\hbar\omega$.   Notice that $\vert \phi \rangle$ is not of unit norm.

Notice that in this result, the chirality $\chi$ appears
only for inversion-symmetric nodes, suggesting that it is possible
to measure the chirality for inversion-symmetric but not time-reversal symmetric materials.  This is true as shown in the next section, but it is not possible
to determine the chirality from a measurement of the total cross-section,
although this formula seems to suggest it.  The problem is that the $\mathbf{F}$ parameters are unknown. It is possible to compensate for a change in sign of $\chi$
by changing the $\mathbf{F}$'s. If two materials have scattering
cross-sections as a function of $\tilde{\mathbf{\Delta}}$ that look the same except
that the cross-section pattern is reflected through the $z$-axis (whenever
neutrons polarized in the \emph{same} way are passed through the material),
then it looks as if the materials have the opposite sign of $\chi$.  However,
there is an alternative explanation:  suppose $\mathbf{k}_0$ is
parallel to the $z$-axis, $\hat{\mathbf{z}}\cdot \Fb^{i}=0$.  If
one material has $\mathbf{F}^i=\F^i_x\hat{\mathbf{x}}+\F^i_y\hat{\mathbf{y}}$ while
the other has $\mathbf{F}^i=\F^i_x\hat{\mathbf{x}}-\F^i_y\hat{\mathbf{y}}$ for each $i$,
this would also explain the reflection of the cross-section 
in the $xy$-plane.

%--in particular, suppose that $\hat{\mathbf{k}}_0$ is along the $z$-axis, and take $\hat{\mathbf{y}}\cdot \mathbf{F}^i_\perp\rightarrow -\hat{\mathbf{y}}\cdot\mathbf{F}^i_\perp$; this reverses the sign
%of the second term. (In other words, $\mathbf{F}$ is reflected in $xz$-plane
%of the degree of freedom that couples to the neutron spin, but not in the degree of freedom coupled to the electron spin.) Nevertheless, with a polarized measurement, it is possible to determine the chirality of the Weyl nodes.

%Thus for time-reversal symmetric nodes, it is not possible to measure the chirality. 
%However, for inversion symmetric nodes, the chirality $\chi$ of the scattering process $\chi \to \bar{\chi}$ can be measured by the second term in Eq. (\ref{PolarizationFull_Appendix_I:Eq1.auxiliarystateNorm}) and separated from the first term by being linear in 
%momentum or by subtracting a cross-section with $\Pb^{i}$ directed along $\widehat{\kb}_{0}$ with the reverse polarization.

To summarize, the scattering Eq. (\ref{PolarizationFull_Appendix:Eq1.auxiliarystateNorm}) is dependent on the initial neutron 
beam polarization vector, the scattering direction, and the \emph{a priori} unknown coupling constants. Measuring the polarization vector of the final neutron beam at $|\Dtb| = \hbar \omega/ v_{\rm F}$, one finds that $|\Pb^{f}| = 1$ for all scattering directions 
and  any incident fully polarized neutron beam. This is quite remarkable and counter-intuitive as one is probing particle-hole Weyl pairs and not conventional magnetic excitations. %An example of a possible measured polarization vector Eq. 

\subsubsection{Measuring Chiralities}

%ANSWER Q: WHAT DOES IT MEAN TO DET. WHETHER MOMENTUM IS PAR. OR ANTIPAR.
%TO PSEUDOSPIN, SINCE PSEUDOSPIN IS IN A DIFFERENT COORDINATE SYSTEM (MAYBE
%SAY IT AT END OF SEC. ABOUT SX SY=iSZ, but probably not, probably at beginning)
%GIVE EXAMPLE OF FLIPPING SIGN OF INTERACTION AND NOT GETTING THE SAME RESULTS AS FLIPPING CHIRALITY)

It is possible to measure the chirality of the nodes in an inversion symmetric WSM, although it is not straightforward because of the unknown $\mathbf{F}$ parameters.

First, it is clear that it is not possible to measure the chirality for scattering between two nodes related by time-reversal symmetry,
since the chirality does not appear in the cross-section, Eqs. (\ref{DifferentialCrossSection_Appendix_Eq1:TR-Susceptibility}) and (\ref{PolarizationFull_Appendix_TR:Eq1.auxiliarystateNorm}). %(It actually does appear in $\chi_{0i}$, but as mentioned above, this is multiplied by $\mathbf{F}_0$ in the cross-section, which is zero on account of time-reversal symmetry.)
This seems at first surprising since the two Weyl points are either both left-handed or right-handed, which should be distinguishable. One can understand
why, nevertheless, it is impossible to distinguish them with neutron scattering
from the following point of view: The scattering produces a particle-hole pair.  The hole and particle at a Weyl point have the opposite handedness. So the two cases are essentially the same, with one excitation of each
handedness in both cases.  The only difference is how the charges of the excitations is correlated to their handedness.  This does not affect the cross-section since the sign of the charge does not appear in the cross-section, which depends on the square of the matrix
elements.  On the other hand, in the inversion symmetric case, either two left-handed excitations are produced (if the Weyl point at $-\mathbf{k}_0$ is right-handed and the excitation at $+\mathbf{k}_0$ is left-handed) or two right-handed excitations are produced, explaining why $\chi$ enters into the cross-section. We will now explain
how to measure the chirality in this case. %, or more precisely how to determine
%for a portion of the scattering data near $\Delta\mathbf{k}_0$ corresponds
%to scattering from a right-handed node to a left-handed node or vice versa.

We will focus on the case discussed in the last section, where $|\Dtb|=\hbar\omega$. Because the spin and momentum of the electron are locked, we may ignore
the momentum of the electron.  We can simply consider the electron
as fixed in space with a neutron scattering off of it.  The expression for the cross-section, Eq. (\ref{PolarizationFull_Appendix:Eq1Transition}),
is then interpreted as the cross-section for scattering in which
the electron's spin changes from $-\chi_i\Dtb$ to $-\chi_f\Dtb$, which
is always a spin-flip scattering since $\chi_i=-\chi_f$. 
%\begin{equation}
%f(\tau_f,\tau_i,\hat{\Dtb})=|\langle\tau_f,-\Dtb,\chi_f|\bm{\mu}\cdot\bm{\mathcal{M}}|-\chi,\Dtb,\tau_i\rangle|^2.
%\end{equation}
The interaction operator can be written $\bm{\tau}\cdot\bm{\mathcal{M}}=\tau_x(\mathbf{a}\cdot\bm{\sigma})+\tau_y(\mathbf{b}\cdot\bm{\sigma})$ where $a_i=\hat{\mathbf{x}}\cdot\Fb_i,b_i=\hat{\mathbf{y}}\cdot\Fb_i$.  No $z$-component appears because
$\kb_0\cdot \Fb_i=0$.  The $\Fb_0$ term of $\mathcal{M}$ is omitted because it does not contribute to the matrix element for an event in which the electron's
spin flips.
%The cross-section for inversion-symmetric scattering is given by Eq. (\ref{PolarizationFull_Appendix_I:Eq1.auxiliarystateNorm})
%\label{PolarizationFull_Appendix:EqTransition}. We will
%focus on the situation described in the last section, where $|\Dtb|= \hbar\omega / v_{\rm F}$. This can also be visualized
%as the probability of a given transition in the state of one spin scattering
%off another spin and interacting
%according to the interaction $\bm{\tau}\cdot\bm{M}=\tau_x\mathbf{a}\cdot\bm{\sigma}+\tau_y\mathbf{b}\cdot\bm{\sigma}$ where the vectors $\mathbf{a}$ and $\mathbf{b}$
%collect together some components of the $\F$ parameters, $a^i=\F^i_{\perp x},b^i=\F^i_{\perp y}$.  The energy does not depend on $\tau_z$,
%because 
%we are assuming that $\mathbf{k}_0$ is in the $z$-direction, hence only
%components perpendicular to this appear in the current. 
%Also, there is no term proportional to $\sigma_0 \mathbf{F}_0$ in the cross-section for scattering
%at $|\Dtb|=\hbar\omega/v_{\rm F}$ due to Lorentz contraction. 

%There is also no term proportional to $\sigma_0 \mathbf{F}_0$ because this
%interaction does not appear in the cross-section for scattering
%at $|\Dtb|=\hbar\omega/v_{\rm F}$.
One can do an experiment where one focuses on events
where the neutron's spin changes from a given $\lvert \tau_{i} \rangle$ to another $\lvert \tau_{f} \rangle$.
%In an experiment one can impose a certain change on the spin of the neutron.
If one could measure how the spin of the electron changes, one would expect
(because of the form of the interaction above) a certain correlation of this measurement to the
way the spin of the neutron has changed.   Now an experiment actually measures how the \emph{momentum} of the electron
changes (by measuring momentum transfer), which is locked to the spin of the electron up to the sign
that we wish to find.  If the way the electron's \emph{momentum} changes is reversed from the behavior
one expects from the spin, it must be because $\chi_f=-\chi_i=-1$ so that
the spin is antiparallel to the momentum.
   If the interaction were $\sigma_x\tau_x+\sigma_y\tau_y$, this is clear;
it is easy to work out how the electron's spin is affected by scattering.
With the arbitrary $\mathbf{a}$ and $\mathbf{b}$ the expectation
of how the electron spin should flip would be distorted and it is
not clear that it is possible to determine the sign of $\chi_f,\chi_i$
if they are unknown.
%would have a simple relationship, but for the actual interaction one has to rotate one's expectations of how the electron spin should flip (besides possibly inverting it).  This suggests that it might not be possible to determine
%the sign of $\chi_f,\chi_i$.

In principle, we could prepare the neutron in any initial state and measure its final state along any axis.
However, to trim the problem down, we will focus on just the
rate of spin-flip scattering of the neutron.
So consider an experiment where the neutron is prepared with a certain polarization direction $\hat{\mathbf{N}}$ and one measures the cross-section $f_{\hat{\mathbf{N}}}(\widehat{\Dtb})$ that it flips to $-\hat{\mathbf{N}}$ as a function
of the direction of $\Dtb$.
First consider the case where $|\mathbf{a}|=|\mathbf{b}|$ and they
are orthogonal to one another for simplicity.  We will calculate
the probability as a function of the direction of the initial $\emph{spins}$ of
the neutron and electron, $\hat{\mathbf{N}}$ and $\hat{\mathbf{E}}$ respectively (rather than \emph{momentum}); the relation is $\hat{\mathbf{E}}=-\chi_i\widehat{\Dtb}$. One can evaluate
$|\langle -\hat{\mathbf{N}},-\hat{\mathbf{E}}|\mathbf{a}\cdot\bm{\sigma}\tau_x+\mathbf{b}\cdot\bm{\sigma}\tau_y|\hat{\mathbf{E}},\hat{\mathbf{N}}\rangle|^2$ by means of 
the formula given in the previous section, $\langle -\hat{\mathbf{n}}|\bm{\sigma}|\hat{\mathbf{n}}\rangle=\mathbf{u}+i\mathbf{v}$
where $|\hat{\mathbf{n}}\rangle$ is a spinor 
oriented along the $\hat{\mathbf{n}}$ direction
of the Bloch sphere, and $\hat{\mathbf{u}}$ and $\hat{\mathbf{v}}$ are any unit vectors
that make a right-handed coordinate system together with $\hat{\mathbf{n}}$. 
%{\color{green}(See Appendix \ref{FullyPolarized_Appendix}: I WILL ADD A LITTLE MORE TO THIS APPENDIX)} 
We apply this formula to both the electron and neutron by introducing
vectors $\hat{\mathbf{u}}_e,\hat{\mathbf{v}}_e,\hat{\mathbf{u}}_n,\hat{\mathbf{v}}_n$.
 The probability of the electron flipping
from $\hat{\mathbf{E}}$ to $-\hat{\mathbf{E}}$ and the neutron flipping from $\hat{\mathbf{N}}$ to $-\hat{\mathbf{N}}$ comes out\footnote{For this
calculation, use the completeness identity $\hat{\mathbf{u}}_e\hat{\mathbf{u}}_e^T+\hat{\mathbf{v}}_e\hat{\mathbf{v}}_e^T+\hat{\mathbf{E}}\hat{\mathbf{E}}^{T}=\mathds{1}$ and $\hat{\mathbf{u}}_e\times\hat{\mathbf{v}}_e=\hat{\mathbf{E}}$ and analogous identities for the neutron.} to be 
\begin{equation}
f_{\hat{\mathbf{N}}}(\widehat{\Dtb})\propto ({\rm N}_x{\rm E}_a+{\rm N}_y{\rm E}_b)^2+({\rm N}_z-{\rm E}_c)^2
\label{eq:umbrella}
\end{equation}
where ${\rm E}_a,{\rm E}_b,{\rm E}_c$ are the components of $\hat{\mathbf{E}}$ along the directions
of $\mathbf{a},\mathbf{b}$ and a third direction making a coordinate
system with them, $\hat{\mathbf{c}}=\hat{\mathbf{a}}\times\hat{\mathbf{b}}$.

A striking effect is that for any direction of the initial neutron spin, there
are two initial spin directions of the electron for which spin-flips of the neutrons have zero probability. 
Thus in an experiment, one can map out the cross-section for a neutron spin-flip with a fixed $\hat{\mathbf{N}}$ as a function
of momentum transfer and then search for these nodes.  
If the initial spin of the neutron is 
in the $xy$-plane, the two nodes of $f_{\widehat{\hat{\mathbf{N}}}}$ are in the $ab$ plane
at opposite points of the equatorial circle.
As the neutron spin rotates in the $xy$-plane the nodes of the electron
spin rotate in the $ab$ plane.
The corresponding nodes of the electron spin are related to $\hat{\mathbf{N}}$ as follows.
Rotate the $xy$-plane onto the $ab$-plane so that $\hat{\mathbf{x}}$ maps to $\hat{\mathbf{a}}$
and $\hat{\mathbf{y}}$ maps to $\hat{\mathbf{b}}$.  The initial
neutron spin maps to a certain point on the great circle in the $ab$-plane and
the two points $90^\circ$ away from this point on the great circle are the nodes. This follows from Eq. \eqref{eq:umbrella} by noting that $f$ is zero if ${\rm E}_a=\pm {\rm N}_y$; ${\rm E}_b=\mp {\rm N}_x$, ${\rm E}_c=0$.
Comparing this prediction to experimental data would allow one to determine the directions of the $\hat{\mathbf{a}}$ and $\hat{\mathbf{b}}$ vectors up
to a common sign.

Now if the neutron spin is moved out of the $xy$-plane, the nodes for the electron move out of the $ab$-plane, as $f=0$ when ${\rm E}_c={\rm N}_z$ and ${\rm E}_a/{\rm E}_b=-{\rm N}_y/{\rm N}_x$. Note that these nodes are not antipodal to one another: they both move into the same hemisphere, toward the $\hat{\mathbf{c}}$-axis.

\begin{figure}
%\begin{minipage}[1\columnwidth]
%\includegraphics[width=.45\textwidth]{./mirrorenwithdots}
\includegraphics[width=.45\textwidth]{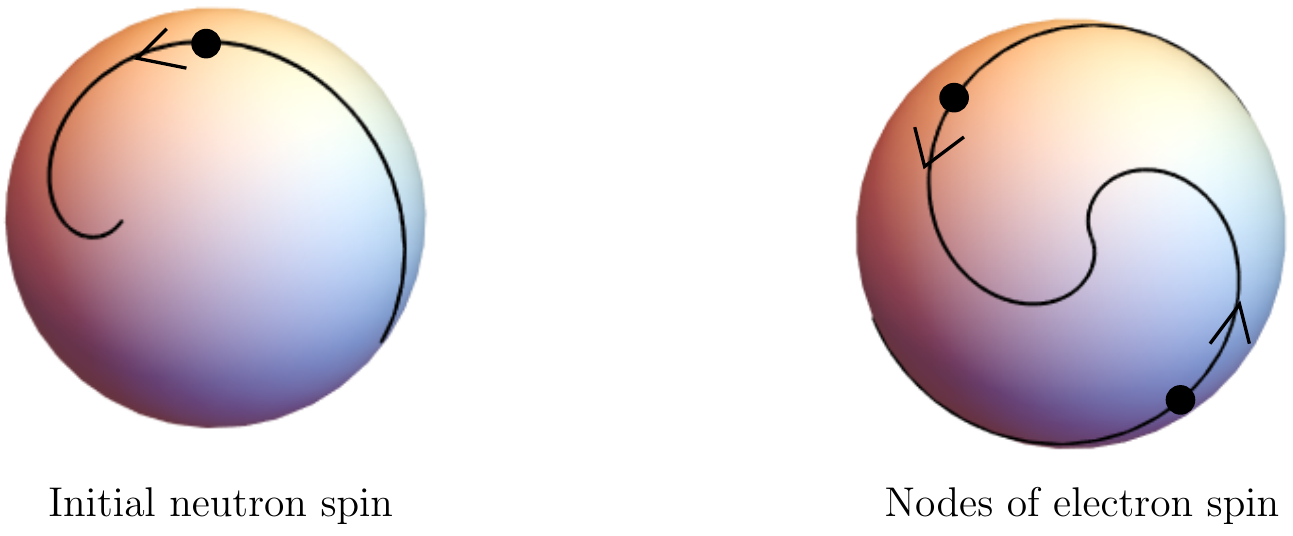}
\caption{How to measure chirality. Consider scattering
with a maximal $|\Dtb|$. For each initial spin of the neutron find the momentum
transfer direction where spin-flip scattering has zero cross-section. There are always two such nodes. 
At the left is a sphere representing the possible
initial spin directions of the neutron; as the spin
moves along a spiral from the equator perpendicular to $\mathbf{k}_0$ up to
one of the axes parallel to $\mathbf{k}_0$, the corresponding nodes (shown
on the right) also spiral in some direction, with an arbitrary
rotation and possibly a distortion due to the interaction parameters.  However, no matter what the parameters are, if the nodes form a double-spiral with the same
handedness as the neutron-spin's spiral, then the scattering is from a right-handed to a left-handed Weyl point, and if with the opposite handedness, then
the scattering is from a left-handed to a right-handed Weyl point.\label{Chirality_figure}}
%\end{minipage}
\end{figure}

To describe this in a geometrical way that is independent of knowing
$\hat{\mathbf{a}},\hat{\mathbf{b}},\hat{\mathbf{c}}$
 %{\color{green}Question: should be $\hat{\mathbf{a}},\hat{\mathbf{b}},\hat{\mathbf{c}}$?} 
 and their signs correctly, return to the first case where the neutron spin is rotating around the $xy$-plane and
the electron nodes are rotating around the $ab$-plane.  Find from which
hemisphere of the electron's Bloch sphere the nodes can be seen rotating
with the same handedness as the neutron's initial spin rotates when
seen from the positive $z$-axis.  That is the hemisphere the nodes
will move into. Figure \ref{Chirality_figure} illustrates this:
if the neutron's initial spin follows a helix starting on the great circle
in the $xy$-plane and spirals in toward the $z$-axis, the two nodes
for the electron's initial spin both form a helix of the \emph{same} handedness
contracting towards one of the poles of the Bloch sphere.

Now in the neutron scattering experiment one measures $\widehat{\Dtb}$ rather
than $\bm{\sigma}$.  $\widehat{\Dtb}$ is parallel to the initial spin if $\chi_f=1$
and to the final spin if $\chi_f=-1$. Hence if $\chi_f=1$, the helix formed by the
nodes has the same handedness as the helix which the neutron's spin
is made to rotate along, and it has the opposite handedness if $\chi_f=-1$.

The fact that the nodes move into the same hemisphere when
the neutron's spin moves out of the $xy$-plane can be understood
by noting that $\tau_x\sigma_a+\tau_y\sigma_b$ is an ordinary 
``easy-axis" coupling of two spins, except that each is measured relative to its own coordinate system.
 So the sum of the component of each spin along the axis perpendicular to the plane where the coupling is, $\sigma_c+\tau_z$, should
be conserved; if the neutron flips from up to down it is not possible
for the electron spin to also flip from up to down, since then the net
spin would change by $2\hbar$.  It is surprising that there is a correlation
between flips of the neutron's spin along the $z$-axis and flips of the
electron's spin along the $c$-axis, since these components of spin
do not appear in the interaction Hamiltonian.  The direction of the $c$-axis is
determined however by the relation $\sigma_c=i\sigma_a\sigma_b$.  %(More intuitively, a state with $\sigma_c=+1$ is a superposition of states with $\sigma_a=\pm 1$.  Knowing that a spin-flip scattering of the neutron from $\hat{\mathbf{x}}\cos\phi+\hat{\mathbf{y}}\sin\phi$ to its negative is never accompanied by a spin-flip
%of the electron from $\hat{\mathbf{a}}\sin\phi-\hat{\mathbf{b}}\cos\phi$ breaks the symmetry
%between scattering from $\hat{\mathbf{c}}\rightarrow -\hat{\mathbf{c}}$ and the reverse
%because of how these states are related.)

As a brief remark on how to carry out such an experiment,
it might seem as if measuring the nodes in the cross-section
as a function of momentum-transfer for three polarizations
of the neutrons, along the $x$, $y$, and $z$ axis, should
be sufficient.  This seemingly allows one to determine the momentum
directions that are locked to the $a$,$b$ and $c$ axes of spin and whether
they are right- or left-handed.  However, the nodes for initial
neutron spin $\hat{\mathbf{x}},\hat{\mathbf{y}}$ are at $\pm\hat{\mathbf{b}}$
and $\pm\hat{\mathbf{a}}$, so it is not possible to determine the signs
of these axes, leaving the handedness indeterminate.  This is solved by identifying the nodes for a few additional spin directions intermediate between
$\hat{\mathbf{x}}$ and $\hat{\mathbf{y}}$.

If $\mathbf{a}$ and $\mathbf{b}$ are more general (not
orthogonal and with different magnitudes), the same routine
would allow one to measure the chirality; the only difference
is that the nodes of the electron's spin do not slide all
the way to the $c$-axis
when
the spin of the neutron moves to the $z$-axis--they still
spiral with the same handedness though\footnote{To show this, first note that it is always possible to find coordinates for the $\tau_x, \tau_y$ plane such that 
$\hat{\mathbf{a}},\hat{\mathbf{b}}$ are orthogonal, although not normalized.
This follows from footnote [47].  To show that there are two nodes of Eq. \eqref{PolarizationFull_Appendix:Eq1Transition},
we must find for which electron and neutron spin directions the scattering cross-section, or more simply, the matrix element that it is the square of, vanishes.
We use a representation for the electron and neutron spinors leads to simple expressions, e.g. for the electron $\psi_e=A_e(1,\lambda_e)^T$.
The orientation of the spinor in space is determined by the complex parameter $\lambda_e$  via $\lambda_e=\tan\frac{\theta_e}{2}e^{i\phi_e}$, 
and $A_e$ is a normalization constant that cancels out.
The condition that the matrix element vanishes is a quadratic polynomial in $\lambda_e,\lambda_n$.
Thus for each neutron direction there are two nodes $\lambda_e$, which spiral when represented on a sphere as described.}.
\section{Conclusion\label{Discussion_Appendix}}
%--------------- Begin: Interaction_Appendix -----------------------------------------------------------------------------
In this paper, we have shown that INS can probe bulk excitations of type-$1$ Weyl nodes. These where assumed to be aligned at (or near) the chemical potential\cite{Ruan2016aa,Buchhold2018aa,Nandkishore:2014aa,Huang:2013aa,Skinner:2014aa,Syzranov:2015aa} with realistic anisotropy, but a negligible scalar term $\alpha$. Footnote [$40$] outlines how any $|\alpha| < 1$ is analytically tractable.

%This article explained how INS can probe bulk excitations of type-$1$ Weyl nodes, which where assumed to be aligned at (or near) the chemical potential\cite{Buchhold2018aa,Nandkishore:2014aa,Huang:2013aa,Skinner:2014aa,Syzranov:2015aa} with realistic anisotropy, but a negligible scalar term $\alpha$. Footnote [$40$] outline how any $|\alpha| < 1$ is analytically tractable.

The analysis separated the cross-section into a Lorentz invariant susceptibility and a symmetry breaking coupling of neutrons to Weyl fermions determined by material specific $g$-factors.  
This had advantages: first, Lorentz invariant properties of the susceptibility, describing the excitations' dynamics, are reflected in the cross-section.  This leads to several universal quantitative predictions, and furthermore, the possibility of measuring chirality for inversion symmetric nodes despite arbitrary material parameters. Noticeably, the chirality of a Weyl point
can be seen through the distortions produced by the unknown form of the neutron-electron interaction, which reflects its topological character. Secondly, anisotropy of these $g$-factors is actually helpful, as they render spin-momentum locking observable even in a fully unpolarized experiment. Furthermore, the $g$-factors can enhance the cross-section intensity as they, in principle, can take any value from zero to diverging, which differs from the bare coupling value $g/2 = 1$.
As a proof of concept, we estimated the intensity under optimistic conditions $\q_{i}/\q_{f} \times \text{d}^{2} \s^{(+)}(\qb,\omega)/\text{d} \Omega \text{d} E_{f} \lesssim 2 \times 10^{-2} \;  {\rm mb} / {\rm meV\,f.u.\,sr}$ for a toy model. This is low but remarkably only of order $10^{-2} - 1$ smaller than what has been observed in scattering off spin-$\frac12$ particle-hole pairs\cite{Goremychkin2018aa,Vignolle:2007aa,Walters:2009aa,Fujita2012aa,Janoschek2015aa}.

INS can thus provide a platform to understand the intrinsic behavior of WSMs, for example, the spin and orbital effects discussed here. 
It can test the form of the Weyl equation in materials, including monitoring changes in it such as relocation in energy and momentum space, distortion of dispersion, redistribution of occupation numbers, due to applied fields, currents or elastic and magnetic deformations as predicted Ref. \onlinecite{Chan2016aa,Akbar2018aa,Cortijo2015aa,Ghimire2019aa,Araki2019aa,Sekine2017aa}.

Some of the details that have appeared in this study could give new information about Weyl materials. For example, the many $g$-factors describing the emergent magnetic moment of the Weyl fermions induced by an external magnetic field (which does not have to be from a neutron). In particular, it would be interesting to know how these parameters evolve as a magnetic Weyl semimetal approaches the transition point\cite{Crippa2020aa,Chang2018aa,Wan2011a,Ivanov2019aa}. The $4$-band model above shows that they depend on the strength of the spontaneous magnetic ordering and hybridization between bands. Such an endeavor can be done theoretically by use of numerically realistic band structure calculations and experimentally be measured by neutron scattering. 

Besides the specific problem of neutron scattering, particle-hole correlators (as calculated here) are relevant to WSMs' intrinsic properties.  For example,
particle-hole bound states (like plasma waves) might form, and their self-energy
is closely related to the spin-susceptibility. If a particle-hole bound state from excitations at distinct Weyl points can form, we would expect angle-dependent properties, for example, it should have an effective mass that is proportional to the matrix elements
between the two Weyl points and hence would be strongly momentum dependent. Also, just as there is an emergent magnetic moment of Weyl excitations, there could be an emergent electric dipole moment, which is not ruled out by symmetry unlike the case of an electron in free space\cite{Khriplovich1997aa}. This could influence the bound state through pseudospin dipole-dipole interactions. It is necessary to understand carefully what properties of Weyl fermions are universal for such analyses, and the relativistic method developed here should be useful. 

Furthermore, it would be interesting to extend our method to derive the cross-section for scattering between emergent BdG Weyl nodes induced in a monopole superconducting WSM\cite{Sun2019aa}.

%{\color{red}Since the probing mechanism is measuring the particle-hole correlation of Weyl spinors, which are guaranteed if nodes are present, INS does not require a full resolution of a node's dispersion.}

%--------------- End: Interaction_Appendix -----------------------------------------------------------------------------

%-------- Notice: Declare Manuscript to be Masterfile from which compilation is done  ----------------------------------
%%% Local Variables:
%%% mode: latex
%%% TeX-master: "Manuscript"
%%% End:

%\newpage
\begin{acknowledgments}
We are grateful for the many discussions with Collin Broholm and his group throughout all stages of this project. In particular, Wesley Fuhrman, Guy Marcus, 
Youzhe Chen and Jonathan Gaudet for numerous explanations of neutron scattering and Yi Li, Shu-Ping Lee  and Yishu Wang for suggestions of further extensions of 
this work.

%BG has been supported by Huygens scholarship. MB and AT have been supported by the U.S. DOE Basic Energy Sciences, Materials Sciences and Engineering Award DE-SC$0019331$ and 
%DE-FG$02$-$08$ER$46544$, respectively.

BG has been supported by Huygens scholarship. MB has been supported by the U.S. DOE Basic Energy Sciences, Materials Sciences and Engineering Award DE-SC$0019331$. AMT has been supported by the U.S. DOE Basic Energy Sciences, Materials Sciences and Engineering Award DE-FG$02$-$08$ER$46544$ and the Israel Science Foundation grant number $1939/18$.

%BG was supported by Huygens scholarship whereas MB and AT was supported by the Institute for Quantum Matter under DOE EFRC grant DE-SC$0019331$. 
\end{acknowledgments}

\begin{appendix}
%%section: Principal axis transformation
\section{Principal axis transformation\label{PrincipalAxisTransformation_Appendix}}
%--------------- Begin: PrincipalAxisTransformation_Appendix -----------------------------------------------------------------------------
When the three parameters $\vb_{0}^{(i)}$ of Eq. (\ref{Kinematics_Appendix_Eq2:HamiltonianAisotropic}) are negligible, the total Hamiltonian is 
\begin{eqnarray}
H_{0} &=& \sum_{i = 1,2} \sum_{\kb}\, c^{\dagger}_{\kb;i} H_{0,i}(\pb) c_{\kb;i} , \nonumber \\
H_{0,i}(\pb) &=&  v_{\rm F} \s_{l} \lambda^{(i)}_{l\,m}\p_{m} , \label{HamiltonEq1:PrincipalAxisTransformation_Appendix}
\end{eqnarray}
with $\pb = \kb-\kb_{0,i}$. The Hamiltonian has to be Hermitian, which means that $\lambda^{(i)}_{l\,m}\in \mathds{R}$ for both nodes $i=1,2 $. Now the symmetry, either time-reversal or inversion symmetry exchange the nodes.  This symmetry
takes a particle at the second Weyl point in the first-quantized state $\psi_2$
to $\psi_1=\theta K\psi_2$ or $\psi_1=\theta\psi_2$ at the first Weyl point,
 in the case of time-reversal and inversion respectively, where $\theta$ is a unitary matrix and $K$ is complex conjugation. The matrix $\theta$ can be chosen \emph{arbitrarily} since the two states at each Weyl point are pseudospin states.  One can choose them in some way
at the second Weyl point and define the two states at the first Weyl point
by the transforms of these states under the appropriate
symmetry combined with a convenient $\theta$\footnote{It \emph{is}
required that $T^2=-1$ and $I^2=1$, but rather than constraining $\theta$,
they determine the way that $T$ and $I$ transform the first Weyl point
back to the second.  For example, for inversion symmetry $\theta$ does not
have to square to $1$; as long as $\psi_1\rightarrow \psi_2=\theta^\dagger\psi_1$, $I^2=1$.}.  We choose $\theta=\sigma_0,\sigma_y$ for inversion and time-reversal respectively. With this choice, the requirement that the Hamiltonian be invariant under inversion or time-reversal symmetry dictates that $\lambda^{(1)}_{l\,m}=\pm\lambda^{(2)}_{l\,m}$ with $+(-)$ for the former (latter) symmetry.
%This and the requirement of either time-reversal\footnote{Time-reversal operator $T = \theta K $ is antiunitary, where $K$ is a complex conjugation operator, and the unitary operator is chosen (see Appendix \ref{Interaction_Appendix:TimeReversalSymmetry}) to be $\theta = \s_{y}$. The requirement of time-reversal symmetric nodes is $H_{0,1}(-\kb) = T H_{0,2}(\kb)  T^{-1}$, where $H_{0,i}(\kb)$ is the single-particle Hamiltonian Eq. (\ref{HamiltonEq1:PrincipalAxisTransformation_Appendix}).} or inversion\footnote{Inversion operator $I = \theta$ is unitary and chosen (see  Appendix \ref{Interaction_Appendix:InversionSymmetry}) to be $\theta = \s_{0}$. The requirement of inversion symmetric nodes is $H_{0,1}(-\kb) = I H_{0,2}(\kb)  I^{-1}$, where $H_{0,i}(\kb)$ is the single-particle Hamiltonian Eq. (\ref{HamiltonEq1:PrincipalAxisTransformation_Appendix}).} symmetry 
%dictate that $\lambda^{(1)}_{l\,m} = \pm \lambda^{(2)}_{l\,m}$ with $+(-)$ for the former (latter) symmetry. 
Hence, in order to transform the nodes into isotropic form, it is sufficient to perform a singular value decomposition on $\lambda^{(2)}_{l\,m}$ only.  A singular value decomposition is a general way to
diagonalize non-Hermitian matrices; it is a representation
in the form $\lambda^{(2)} = O D V^{T}$ with orthogonal matrices $O$ and $V$ and a diagonal matrix $D_{a\,b} = \delta_{a\,b}d_{b}$ the elements of which are the singular values, i.e. the square root of the eigenvalues of the velocity tensor $\lambda^{(2),\rm T}\lambda^{(2)} $. To get the Hamiltonian in an isotropic
form one then transforms both momentum and spin degrees of freedom.  The new coordinate for momentum $\pbt = T\pb$ is obtained from the original $\pb$ by a transformation $T_{ab} = V_{ba} d_{a}$, which is a coordinate transformation $V^{\rm T}$ and scaling $d_{a} >0$. The same transformation must act on the momenta of both nodes simultaneously, since in the constraint that momentum
is conserved in a scattering between modes, momenta from the two nodes are subtracted. If spin is
 transformed by a unitary matrix such that $O_{a\,i} \s_{i} = U^{\dagger} \s_{a} U$, then the transformation $H_{0,2}(\pb)\to UH_{0,2}(T^{-1}\pbt)U^\dagger$
brings the second node and thus both nodes into isotropic form,
\begin{equation}
H_{0,i}(\pb) \to H_{0,i}(\pbt) =  \chi_{i} v_{\rm F} \sb \cdot \pbt . \label{HamiltonEq2:PrincipalAxisTransformation_Appendix}
\end{equation}
Here chirality is $\chi_{1} = \chi_{2}$ for time-reversal symmetric nodes, and $\chi_{1} = -\chi_{2}$ for inversion symmetric nodes, or alternatively $\chi_{i} = sign{|\lambda^{(i)}|} $. For example, the effective low energy Hamiltonian of the $4$-band toy model (see Section \ref{4bandWSM_Appendix}) will be transformed to isotropic nodes by $T_{ab} = d_{a} \d_{ab}$ where $\mathbf{d} = (1,1,\sqrt{1-(|\d|/m)^2})^{\rm T}$.

\begin{comment}
\subsection{Time-reversal symmetric Weyl nodes}\label{PrincipalAxisTransformation_Appendix:TimeReversalSymmetry}
\Time-reversal operator $T = \theta K $ is antiunitary, where $K$ is a complex conjugation operator, and the unitary operator is chosen to be $\theta = \s_{y}$, as explained in Appendix \ref{Interaction_Appendix:TimeReversalSymmetry}. The requirement of time-reversal symmetric nodes is $H_{0,1}(-\kb) = T H_{0,2}(\kb)  T^{-1}$, where $H_{0,i}(\kb)$ is the single-particle Hamiltonian Eq. (\ref{HamiltonEq1:PrincipalAxisTransformation_Appendix}). This gives  $H_{0,1}(-\kb) = \theta H_{0,2}^{*}(\kb) \theta^{-1}$, which together with $\lambda^{1(2)}_{i\,j}\in \mathds{R}$, establish the relation $\lambda^{(1)}_{i\,j} = \lambda^{(2)}_{i\,j}$ between nodes.

\subsection{Inversion symmetric Weyl nodes}\label{PrincipalAxisTransformation_Appendix:InversionSymmetry}
Inversion operator $I = \theta$ is unitary and chosen to be $\theta = \s_{0}$, as explained in Appendix \ref{Interaction_Appendix:InversionSymmetry}. The requirement of inversion symmetric nodes is $H_{0,1}(-\kb) = I H_{0,2}(\kb)  I^{-1}$, where $H_{0,i}(\kb)$ is the single-particle Hamiltonian Eq. (\ref{HamiltonEq1:PrincipalAxisTransformation_Appendix}). This and $\lambda^{1(2)}_{i\,j}\in \mathds{R}$ establish the relation $\lambda^{(1)}_{i\,j} = -\lambda^{(2)}_{ij}$ between nodes.

\end{comment}

%--------------- End: PrincipalAxisTransformation_Appendix -----------------------------------------------------------------------------

%%section: Interaction - Time+reversal and Inversion symmetry
\section{Interaction for symmetry related nodes\label{Interaction_Appendix}}
%--------------- Begin: Interaction_Appendix -----------------------------------------------------------------------------
In $2^{nd}$ quantization, the Hamiltonian of the $i^{th}$ Weyl node is $H_{i} =  \int\text{d}\rb\,  \Psi_{i}^{\dagger}(\rb,t) H_{0,i}(-i\bm{\nabla}_{\rb})  \Psi_{i}(\rb,t)$, where $\Psi_{i}(\rb,t)$ is the $2^{nd}$ quantized Weyl fermion field, and $H_{0,i}$ has the $1^{st}$ quantized isotropic form Eq. (\ref{HamiltonEq2:PrincipalAxisTransformation_Appendix}). The interaction for scattering between nodes is given by $H_{\Bb}  =  \int \text{d}\rb\, \mathcal{H}_{\Bb}(\rb,t)$, where interaction density is $\mathcal{H}_{\Bb}(\rb,t)  = - \Mb(\rb,t) \cdot \Bb(\rb)$ with Eq. (\ref{section:CurrentOperator_Appendix_Eq.1:MagnetizationOperator2}). For scattering between nodes related by either time-reversal or inversion symmetry, the coupling is constrained, as will be explained in Appendix \ref{Interaction_Appendix:TimeReversalSymmetry} and \ref{Interaction_Appendix:InversionSymmetry}, respectively.

%% NOTICE: we changed from action to the Hamiltonian because the action of Weyl fermions also has a time derivative term in it (although action is a little nicer because you can think of psi as being a grassman
%field rather than an operator in it, which makes it easier to understand the effects of time-reversal.  But I think it is clear enough this way.

\subsection{Time-reversal symmetric Weyl nodes}\label{Interaction_Appendix:TimeReversalSymmetry}
Let $\tauh$ denote the antiunitary time-reversal operator acting\cite{Ludwig2016a} on $\Psi_{i}$. Time-reversal symmetry transforms the spinors at the two Weyl points via\footnote{The transformation of $\psi_1$ follows
from the transformation of $\psi_2$ and the fact that $\tauh^2\psi_2\tauh^{-2}=-\psi_2$.} $\tauh\psi_2\tauh^{-1}=\theta^\dagger\psi_1$,$\tauh\psi_1\tauh^{-1}=-\theta^*\psi_2$.  The standard isotropic form of
the Hamiltonian occurs only if $\theta=\sigma_y$ as shown in the previous
appendix.  Time-symmetry implies that $\mathcal{H}_\mathbf{B}=\tauh\mathcal{H}_{-\mathbf{B}}\tauh^{-1}$, which  implies that the couplings are restricted to Eq. (\ref{TR_Coupling:Interpretation_Appendix}).

%Let $\tauh$ denote the antiunitary time-reversal operator acting\cite{Ludwig2016a} on $\Psi_{i}$. Weyl nodes related by time-reversal symmetry is the requirement $\tauh S_{2} \tauh^{-1} = S_{1}$ and  $\tauh \Psi_{2} \tauh^{-1} = \theta^{\dagger} \Psi_{1}$.
%Applying $\tauh^{2} \Psi_{1} \tauh^{-2}$ to a state $\lvert \alpha \rangle$ with $n_{\rm w}$ Weyl fermions  then $H_{0,1}(\bm{\nabla}_{\rb}) = \theta H_{0,2}^{*}(\bm{\nabla}_{\rb}) \theta^{\dagger}$. The choice of $\theta$ is arbitrary up to a unitary choice. Here we choose $\theta = \s_{y}$ as it brings both nodes to the isotropic form  Eq. (\ref{HamiltonEq2:PrincipalAxisTransformation_Appendix}).  The requirement $\mathcal{H}_{\Bb} = \tauh \mathcal{H}_{-\Bb} \tauh^{-1}$ implies that the couplings are restricted to Eq. (\ref{TR_Coupling:Interpretation_Appendix}).

\subsection{Inversion symmetric Weyl nodes}\label{Interaction_Appendix:InversionSymmetry}
Let $\hat{\rho}$ denote the unitary inversion operator acting on $\Psi_{i}$. 
Inversion symmetry transforms the spinors via
$\hat{\rho}\psi_2\hat{\rho}^{-1}=\theta^\dagger\psi_1$,$\hat{\rho}\psi_1\hat{\rho}^{-1}=\theta\psi_2$.  The standard isotropic form of
the Hamiltonian occurs if $\theta=\sigma_0$ as shown in the previous
appendix.  Inversion symmetry, i.e. $\mathcal{H}_\mathbf{B}=\hat{\rho}\mathcal{H}_{\mathbf{B}}\hat{\rho}^{-1}$, implies that the couplings are restricted to Eq. (\ref{I_Coupling:Interpretation_Appendix}).

%Let $\hat{\rho}$ denote the unitary inversion operator acting on $\Psi_{i}$. Weyl nodes related by inversion symmetry is the requirement $\hat{\rho} S_{2} \rhoh^{-1} = S_{1}$ and
%$\rhoh \Psi_{2} \rhoh^{-1} = \theta^{\dagger} \Psi_{1}$. Using  $\rhoh^{2} = 1$ and $\rhoh^{-1} = \rhoh^{\dagger}$  then $H_{0,1}(\bm{\nabla}_{\rb}) = \theta H_{0,2}(-\bm{\nabla}_{\rb}) \theta^{\dagger}$. 
%The choice of $\theta$ is arbitrary up to a unitary choice. Here we choose $\theta = \s_{0}$ as it brings both nodes to the isotropic form Eq. (\ref{HamiltonEq2:PrincipalAxisTransformation_Appendix}). 
%The requirement $\mathcal{H}_{\Bb} = \rhoh \mathcal{H}_{\Bb} \rhoh^{-1}$  implies that the couplings  are restricted to Eq. (\ref{I_Coupling:Interpretation_Appendix}).
%--------------- End: Interaction_Appendix -----------------------------------------------------------------------------

%%section: Susceptibility_Appendix
\section{Particle-hole Weyl pair\label{Susceptibility_Appendix}}
%--------------- Begin: Susceptibility_Appendix -----------------------------------------------------------------------------
The Weyl fermion correlator 
%\begin{subequations}
%\label{Susceptibility_Appendix_Eq1:ParticleHoleCorrelatior}
\begin{eqnarray}
\s^{(+)}_{\m\, \nu}(\rb,t) &=& V\left< \s^{(-)}_{\m}(\rb,t) \s^{(+)}_{\nu} \right>_{0} \label{Susceptibility_Appendix_Eq1:ParticleHoleCorrelatior}
\end{eqnarray}
%\end{subequations}
is an intermediate scattering function of non-hermitian operators $\s^{(+)}_{\m}(\rb,t) = \Psi_{2}^{\dagger}(\rb,t) \s_{\m}\Psi_{1}(\rb,t)$ and $\s^{(-)}_{\m}(\rb,t) = \s^{(+),\dagger}_{\m}(\rb,t)$. These excite an occupied state from the vicinity of one Weyl node to an empty state in the vicinity of the other Weyl node.  According to the fluctuation-dissipation theorem, the scattering function and the absorptive part of the generalized susceptibility are related by $\s^{(\pm)}_{\m\, \nu}(\qb,\omega) = \kappa(\omega,T) \chi''^{(\pm)}_{\m\, \nu}(\qb,\omega)$ where $ \kappa(\omega,T) = 2\hbar/[1-exp(-\beta \hbar\omega)]$ with $\beta = 1/k_{\rm B} T$. The susceptibility is decomposed into $\chi^{(\pm)}_{\m\, \nu}(\qb,\omega) = \chi'^{(\pm)}_{\m\, \nu}(\qb,\omega) + i \chi''^{(\pm)}_{\m\, \nu}(\qb,\omega)$. %, where the absorptive part ($z\in \mathds{C}$) is $\chi''^{(\pm)}_{\m\, \nu}(\qb,z) = K_{\mu\,\nu}^{(\pm)}(\qb,z)/2i$, in terms of $K_{\mu\,\nu}^{(\pm)}(\qb,z) = \frac{i V}{\hbar}\left< \left[\s_{\mu}^{(\mp)}(\qb,z) , \s_{\nu}^{(\pm)} \right] \right>_{0}.$
 By standard spectral decomposition\cite{Jensen1991a} at zero temperature and infinite volume limit, we get, for noninteracting Weyl fermions, that
\begin{widetext}
\begin{equation}
\chi''^{(\pm)}_{\mu\, \nu}(\qb,\omega) 
= \frac{\pi V}{(2\pi \hbar)^{3}} \int \dd \pbt_{i} \int \dd \pbt_{f} \, \d(\pbt - \Dtb)\, \d\left(\hbar\omega - \Delta\xi^{\rm w} \right)  \langle -\chi_{i};\pbt_{i} \rvert \s^{\mu} \lvert \pbt_{f} ; \chi_{f} \rangle \langle \chi_{f};\pbt_{f} \rvert \s^{\nu} \lvert \pbt_{i} ; -\chi_{i} \rangle, \label{Susceptibility_AppendixEq:AbsorpSusceptibility_InfiniteVolume} 
\end{equation} 
\end{widetext}
with change in internal energy $\Delta\xi^{\rm w} = \xi_{\pbt_{f}}^{+} - \xi_{\pbt_{i}}^{-}$. The energy of the occupied state $\xi_{\pbt_{i}}^{-}$ is negative, and the excited state $\xi_{\pbt_{f}}^{+}$ is positive. This is the Lindhard function weighted by a pseudospin correlation between the Weyl fermion ejected from the Fermi sea, and that scattered into the empty state. Now we
do a particle-hole transform, mainly for the reason that it makes the expressions more symmetric and the relativistic symmetry easier to see.
A neutron transfers energy $\hbar \omega$ to the WSM and creates a particle-hole Weyl pair.  The change in energy can be rewritten $\Delta \xi^{\rm w}  \to \xi_{\pbt_{f}}^{p} + \xi_{\pbt_{i}}^{h} = v_{\rm F} \left(\lvert \pbt_{f} \rvert + \lvert \pbt_{i} \rvert\right)$ by reinterpreting $- \xi_{\pbt_{i}}^{-}$ as the energy $\xi^h_{\pbt_{i}}$ of the created hole. In order to make this picture
consisten, we need to also redefine  $\pbt_{i}\rightarrow - \pbt_{i}$, i.e. a sign change on $\pbt_{i}$ with respect to the definition in Section \ref{Kinematics_Appendix}. In this particle-hole picture, Eq. (\ref{Susceptibility_AppendixEq:AbsorpSusceptibility_InfiniteVolume}) becomes
\begin{widetext}
\begin{equation}
\chi''^{(\pm)}_{\mu\, \nu}(\qb,\omega) 
= \frac{\pi V}{v_{\rm F}(2\pi\hbar)^3} \int  \frac{\dd \pbt_{i}}{2 \pt_{i}^{0}} \int \frac{\dd \pbt_{f}}{2 \pt^{0}_{f}}\, \d^{(4)}\left(\Q - \P\right) 2 \pt^{0}_{i} 2 \pt^{0}_{f}\, \langle \chi_{i}; \pbht_{i} \rvert \s^{\mu} \lvert \pbht_{f}; \chi_{f} \rangle \langle \chi_{f}; \pbht_{f} \rvert \s^{\nu} \lvert \pbht_{i};\chi_{i}  \rangle, \label{LorentzInvFormulation_Eq1b:AbsorptiveSusceptibility} 
\end{equation}  
\end{widetext}
where now the solutions at Weyl point $1$ have chirality $+\chi_i$\footnote{Note however that the hole actually \emph{does} have chirality $-\chi_i$; there is an additional complex conjugation involved in exchanging the created and annihilated states.}.
In this expression, the energy and $3$-momentum delta functions have been combined into an energy-momentum $4$-delta function. The neutron energy-momentum $4$-vector is $\Q^{\mu} = (\Q^{0},\Qb)$ with $\Q^{0} \equiv \hbar\omega/v_{\rm F}$ and $\Qb \equiv -\tilde{\Db} = \D \tilde{\kb}_{0} - \D \tilde{\qb}$, while the particle-hole Weyl pair energy-momentum $4$-vector is $\P^{\mu} = (\P^{0},\Pb)$ with $\P^{0} \equiv \D \xi^{\rm w}/v_{\rm F}$ and $\Pb \equiv  \D \pbt = \pbt_{1} +  \pbt_{2}$. The integration measure and $4$-delta are Lorentz invariant, but the integrand is not yet written in a relativistic form. In order to do this, we will transform from $2$-spinors  $\lvert \pbht; \chi \rangle$ to $4$-spinors $u^{\chi}_{\pbt}$, while simultaneously transforming Pauli matrices to gamma matrices whose Lorentz
transformation properties are more transparent.
We use the Weyl representation in which $\gamma^0=\sigma^x\otimes\sigma^0$,
$\gamma^i=i\sigma^y\otimes\sigma^i$,$\gamma^5=-\sigma^z\otimes\sigma^0$.
With these definitions, we find that a 4-spinor $\phi=(\phi_1,\phi_2)^T$ satisfies $\gamma^\mu\partial_\mu\phi=0$ if $\phi_1$ and $\phi_2$ satisfy the left- and right-handed Weyl equations respectively.  Equation (\ref{LorentzInvFormulation_Eq1b:AbsorptiveSusceptibility}) can now be written in terms
of 4-vectors by introducing special solutions $u_{\pbt} =  (\lvert \pbht; L \rangle, \lvert \pbht;  R \rangle)^{\rm T}$, $u_{\pbt}^{L}  = (\lvert \pbht; L \rangle, 0)^{\rm T}$, $u_{\pbt}^{R}  = (0, \lvert \pbht;  R \rangle)^{\rm T}$ and $\bar{u}_{\pbt}^{\chi} = u_{\pbt}^{\chi,\dagger}  \gamma^{0}$. 
We have to do further calculations to rewrite the $2\times 2$ Pauli matrices
in terms of $\gamma$'s, in order to determine the correct transformation rules.
We will temporarily rename $\chi''^{\mu\, \nu}_{(\pm)}(\qb,\omega)  \to T_{\chi\to\chi}^{\mu\,\nu}(Q)$ or $I_{\chi\to \bar{\chi}}^{\mu\,\nu}(Q) $ for time-reversal and inversion-symmetric nodes, although these are not necessarily tensors.
These are calculated in Section \ref{section:TRnodes_Appendix} and \ref{section:Inodes_Appendix}, respectively. 

Introducing projectors helps to carry out the calculations
and determine the Lorentz transformation properties.
 Because we have performed a particle-hole transformation, all
states have positive energy, and therefore, the only relevant projectors are
\begin{equation}
2 \pt^{0} u_{\pbt}^{\chi} \bar{u}_{\pbt}^{\chi}  = p_{\chi}\,p_{+}(\pt) \gamma^{0}, \label{Susceptibility_Appendix_Eq1:projector}
\end{equation}
which project into positive energy states with chirality $\chi = (L, R)$ by $p_{+}$ and $p_{\chi}$, respectively, given by $p_{+}(\pt) = \pt^{0} + \gamma^{0} \bm{\gamma}\cdot  \pbt$, $p_{L} = (\mathds{1} - \gamma^{5})/2$ and $p_{R}  =  (\mathds{1} + \gamma^{5})/2$.  Note that the projector is a Lorentz scalar
since $p_+(\tilde{p})\gamma^0=p^\mu\gamma_\mu$. (What we call a projector
is not technically a projector, but once adjusted slightly to give a Lorentz
invariant form.)

\subsection{Time-reversal symmetric Weyl nodes}\label{section:TRnodes_Appendix}
For time-reversal symmetric nodes, the matrix element Eq. (\ref{LorentzInvFormulation_Eq1b:AbsorptiveSusceptibility}) connects only nodes with same chirality, i.e. $\chi_{i} = \chi_{f} \equiv \chi$. One now seeks $4 \times 4$ operators
with the same properties and finds that $\gamma^0\gamma^\mu$ also connects
modes with the same chirality. The susceptibility can be rewritten
in terms of this operator. Transform
the susceptibility $T^{\mu\,\nu}_{\chi \rightarrow \chi}(\Q) \to (-1)^{\xi_{\chi}} \tilde{T}^{\mu\,\nu}_{\chi \rightarrow \chi}(\Q)$ with
\begin{equation}
\tilde{T}^{\mu\,\nu}_{\chi \rightarrow \chi}(\Q) = c \int\frac{\text{d}^3\,\pbt_{i}}{2 \pt_{i}^{0}}\int\frac{\text{d}^3\,\pbt_{f}}{2 \pt_{f}^{0}}\,\d^{(4)}\left(\Q-\P\right)\, \tilde{\tilde{T}}^{\mu\,\nu}_{\chi \rightarrow \chi}. \nonumber 
\end{equation}
Here $\tilde{\tilde{T}}^{\mu\,\nu}_{\chi \rightarrow \chi} = 2\pt_{i}^{0} 2\pt_{f}^{0}\, \bar{u}^{\chi}_{\pbt_{i}} \g^{\m} u^{\chi}_{\pbt_{f}} \bar{u}^{\chi}_{\pbt_{f}} \g^{\nu} u^{\chi}_{\pbt_{i}}$, constant $c = \pi V/[v_{\rm F}(2\pi\hbar)^3]$, $\xi_{R} = 0$ for any $\mu,\nu$, whereas $\xi_{L} = 0$ if $\mu = \nu =  0$ or $\mu,\nu \neq 0$, otherwise $1$.
The matrix element $\tilde{T}^{\mu\,\nu}_{\chi \rightarrow \chi}(\Q)$ is a Lorentz-invariant rank-$2$ tensor and by dimensional analysis
is quadratic in $Q$, thus the most general form it can have is $\tilde{T}^{\mu\,\nu}_{\chi \rightarrow \chi}(\Q) = a_{\chi}\left(\Q\cdot \Q\right) g^{\mu\,\nu} + b_{\chi}\Q^{\mu}
\Q^{\nu}.$ 
The scalars $a_{\chi}$ and $b_{\chi}$ can be determined from the two contractions $g_{\mu\,\nu} \tilde{T}^{\mu\,\nu}_{\chi \rightarrow \chi}(\Q) = \left(4 a_{\chi} + b_{\chi}\right)\Q\cdot \Q$ and $\Q_{\mu}\Q_{\nu} \tilde{T}^{\mu\,\nu}_{\chi \rightarrow \chi}(\Q) = \left(a_{\chi} + b_{\chi}\right)(\Q\cdot \Q)^2$,
evaluated in a frame where $Q$ is time-like $\tilde{\Q}^{\mu} = (\tilde{\Q}^{0},\bm{0})$--i.e. the center-of-momentum (COM) frame of the particle-hole pair. Using the projection operators Eq. (\ref{Susceptibility_Appendix_Eq1:projector}) gives $- a_{\chi}  = b_{\chi}=a$, the result Eq. (\ref{DifferentialCrossSection_Appendix_Eq1:TR-Susceptibility}), where
\begin{equation}
a = \frac{\pi^2}{3}\frac{V}{v_{\rm F}(2\pi\hbar)^3}.  \label{Susceptibility_Appendix_Eq1:a}
\end{equation}
In this frame, the conservation laws lead to a simple integral over the surface of a sphere.

\subsection{Inversion symmetric Weyl nodes\label{section:Inodes_Appendix}}
For inversion symmetric nodes the matrix element Eq. (\ref{LorentzInvFormulation_Eq1b:AbsorptiveSusceptibility}) connects only nodes with opposite chirality, i.e. $\chi_{i}  \equiv \chi$ and $\chi_{f} \equiv \bar{\chi}  = - \chi$.  As the amplitude corresponding to various $\mu,\nu = 0,1,2,3$ transforms differently, we will treat them case-by-case in the following.

\subsubsection{For $\mu = \nu = 0$}
The $I^{0\,0}$ component of the susceptibility in $2$-spinor space transforms into $4$-spinor space according to $I^{0\,0}_{\chi \rightarrow \bar{\chi}}(\Q) \to I_{\chi \rightarrow \bar{\chi}}(\Q)$, where
\begin{equation}
I_{\chi \rightarrow \bar{\chi}}(\Q) = c\int\frac{\text{d}^3\,\pbt_{i}}{2 \pt_{i}^{0}}\int\frac{\text{d}^3\,\pbt_{f}}{2 \pt_{f}^{0}}\,\d^{(4)}\left(\Q-\P\right)\, \tilde{I}_{\chi \rightarrow \bar{\chi}}, \nonumber 
\end{equation}
with $\tilde{I}_{\chi \rightarrow \bar{\chi}} = 2 \pt^{0}_{i} 2 \pt^{0}_{f}\, \bar{u}^{\chi}_{\pbt_{i}} u^{\bar{\chi}}_{\pbt_{f}}  \bar{u}^{\bar{\chi}}_{\pbt_{f}}  u^{\chi}_{\pbt_{i}}.$
Since $I_{\chi \rightarrow \bar{\chi}}(\Q)$ is a scalar, the most general form it can have is $I_{\chi \rightarrow \bar{\chi}}(Q) = f_{\chi}\left(Q\cdot Q\right)$.
The constant $f_{\chi}$ can be determined by evaluation in the neutron COM frame by using Eq. (\ref{Susceptibility_Appendix_Eq1:projector}). This gives $f \equiv f_{\chi}  =  (3/2)a$ with $a$ given by Eq. 
(\ref{Susceptibility_Appendix_Eq1:a}) and the result Eq. (\ref{DifferentialCrossSection_Appendix_Eq1:I-Susceptibility1}).  

\subsubsection{For $\mu \neq 0 , \nu \neq 0$}
The $I^{i\,j}$ component of susceptibility in the $2$-spinor space transforms into a rank-$4$ tensor according to $ I^{i\,j}_{\chi \rightarrow \bar{\chi}}(\Q)  \to I^{0\,i\,0\,j}_{\chi \rightarrow \bar{\chi}}(\Q)$,
where 
\begin{eqnarray}
I^{\alpha\,\beta\,\gamma\,\delta}_{\chi \rightarrow \bar{\chi}}(\Q) = c \int\frac{\text{d}^3\,\pbt_{i}}{2 \pt_{i}^{0}}\int\frac{\text{d}^3\,\pbt_{f}}{2 \pt_{f}^{0}}\,\d^{(4)}\left(\Q-\P\right)\,\tilde{I}^{\alpha\,\beta\,\gamma\,\delta}_{\chi \rightarrow \bar{\chi}} ,\nonumber 
\end{eqnarray}
with $\tilde{I}^{\alpha\,\beta\,\gamma\,\delta}_{\chi \rightarrow \bar{\chi}} = 2\pt_{i}^{0} 2\pt_{f}^{0}\, \bar{u}^{\chi}_{\pbt_{i}}  \s^{\alpha\,\beta}  u^{\bar{\chi}}_{\pbt_{f}} \,  \bar{u}^{\bar{\chi}}_{\pbt_{f}}  \s^{\gamma\,\delta}  u^{\chi}_{\pbt_{i}},$
which is a Lorentz rank-$4$ tensor. Here $\sigma^{\mu\,\nu}=\frac{i}{2} [\gamma^\mu,\gamma^\nu]$.
The expense of this transformation is that the tensor
has many extra components besides the ones we need.  However, the additional
entries are actually redundant due to the fact that $\sigma^{2\,3}=i\sigma^{0\,1}\gamma_5$ and $\gamma^5$ can be replaced by its eigenvalue.
Because this tensor is antisymmetric in $\alpha\beta$ and in $\gamma\delta$,
the most general form it can have is
\begin{eqnarray}
I^{\alpha\, \beta\, \gamma\, \delta}_{\chi \rightarrow \bar{\chi}}(\Q) = A^{\alpha\, \beta\, \gamma\, \delta}_{\chi\rightarrow \bar{\chi}}(\Q) &+&  B^{\alpha\, \beta\, \gamma\, \delta}_{\chi \rightarrow \bar{\chi}}(\Q)  + D^{\alpha\, \beta\, \gamma\, \delta}_{\chi \rightarrow \bar{\chi}}(\Q)  \nonumber \\ &+& E^{\alpha\, \beta\, \gamma\, \delta}_{\chi \rightarrow \bar{\chi}}(\Q), 
\end{eqnarray}
with
\begin{subequations}
\begin{eqnarray}
A^{\alpha\, \beta\, \gamma\, \delta}_{\chi \rightarrow \bar{\chi}}(\Q) &=&a_{\chi}(g^{\alpha\,\delta}\Q^{\beta}\Q^{\gamma} - g^{\alpha\,\gamma}\Q^{\beta}\Q^{\delta} + g^{\beta\,\gamma}\Q^{\alpha}\Q^{\delta} \nonumber \\
&& \hspace*{2.1cm} - g^{\beta\,\delta}\Q^{\alpha}\Q^{\gamma}),  \\
B^{\alpha\, \beta\, \gamma\, \delta}_{\chi \rightarrow \bar{\chi}}(\Q) &=& b_{\chi} (\Q\cdot \Q)(g^{\alpha\,\gamma}g^{\beta\,\delta} - g^{\alpha\,\delta}g^{\beta\,\gamma} ),   \\
D^{\alpha\, \beta\, \gamma\, \delta}_{\chi \rightarrow \bar{\chi}}(\Q) &=& d_{\chi} (\epsilon^{\alpha\,\beta\,\gamma\,\tau}\Q_{\tau}\Q^{\delta} - \epsilon^{\alpha\,\beta\,\delta\,\tau}\Q_{\tau}\Q^{\gamma}),  \\
E^{\alpha\, \beta\, \gamma\, \delta}_{\chi \rightarrow \bar{\chi}}(\Q) &=& e_{\chi} (\epsilon^{\alpha\,\gamma\,\delta\,\tau}\Q_{\tau}\Q^{\beta} - \epsilon^{\beta\,\gamma\,\delta\,\tau}\Q_{\tau}\Q^{\alpha}). 
\end{eqnarray}
\end{subequations}
Notice that a term $C^{\alpha\, \beta\, \gamma\, \delta}_{\chi\rightarrow \bar{\chi}}(\Q) = c_{\chi} \left(\Q\cdot \Q\right)\epsilon^{\alpha\,\beta\,\gamma\,\delta}$ is a linear combination of $D^{\alpha\, \beta\, \gamma\, \delta}_{\chi\rightarrow \bar{\chi}}(\Q)$ and $E^{\alpha\, \beta\, \gamma\, \delta}_{\chi\rightarrow \bar{\chi}}(\Q)$ and should therefore not be included. Now the coefficients
can be related with the help of the
redundancy, essentially,
\begin{equation}
\bar{u}^{\bar{\chi}}_{\pbt}\, \s_{\alpha\, \beta} \,u^{\chi}_{\pbt'}  =  \bar{u}^{\bar{\chi}}_{\pbt}\, i\frac{\chi}{2}\epsilon_{\alpha\, \beta\, \gamma\, \delta}\s^{\gamma\, \delta}  \,u^{\chi}_{\pbt'}. \label{Inodes_Appendix_Eq1:Redundancy} 
\end{equation} 
There turns out to be only one independent
scalar, $b_\chi=a_\chi/2,d_\chi=-e_\chi=i\chi(a_\chi/2)$.
This can be determined by the contraction $\Q_{\alpha}\Q_{\gamma}g_{\beta\,\delta}I^{\alpha\, \beta\, \gamma\, \delta}_{\chi \rightarrow \bar{\chi}}(\Q) =  -\frac{3}{2}a_{\chi}(\Q\cdot \Q)$ 
evaluated in the COM frame by using  Eq. (\ref{Susceptibility_Appendix_Eq1:projector}). This gives $a_{\chi}=a$ with Eq. 
(\ref{Susceptibility_Appendix_Eq1:a}), and the result Eq. (\ref{DifferentialCrossSection_Appendix_Eq1:I-Susceptibility3}).

Notice that Eq. (\ref{Inodes_Appendix_Eq1:Redundancy}) is an antisymmetric tensor $F$ with the extra symmetry property  $(\bar{\chi}/2) \epsilon^{\alpha\,\beta\,\gamma\,\delta}F_{\gamma\,\delta} = iF^{\alpha\,\beta}$,
which reduces the six independent components to $3$, since there are $F^{0\,1},F^{0\,2}, F^{0\,3}$, and the other components are all either $-1$ or $\pm  i$ times these.  
The amplitude is thus an \enquote{electromagnetic field tensor} 
\begin{equation}
F^{\alpha\, \beta}=\left(\begin{array}{cccc} 0& -E_x & -E_y & -E_z\\
E_x & 0&-B_z & B_y\\
E_y& B_z & 0 & -B_x\\E_z & -B_y & B_x & 0\end{array}\right),
\end{equation}
which transforms as a Lorentz rank-2 tensor, but with
an additional symmetry property $\mathbf{B}=i\mathbf{E}$, %=i\psi_{f;\chi}^\dagger \bm{\sigma}\psi_{i;\bar{\chi}}$, 
which is incidentally satisfied by the electromagnetic field of circularly polarized radiation.  This is called a self-dual tensor.

\subsubsection{For $\mu \neq 0 , \nu = 0$ and $\mu = 0 , \nu \neq 0$} 
This case starts in the same way as the previous one. This
tensor is found to be a component of an antisymmetric Lorentz rank-$2$ tensor quadratic in $\Q^{\mu}$, so without calculation (since this type of tensor
does not exist), we conclude the result Eq. (\ref{DifferentialCrossSection_Appendix_Eq1:I-Susceptibility2}).\\

%The fact that in the time-reversal symmetric case, all four values of $\mu$
%are united in a single 4-vector while in the inversion symmetric case, they 
%form two separate covariant tensors, can be understood with the help of representations of the Lorentz group\cite{Ramond1980a}. These are labelled by two spins $(s_1,s_2)$; in particular left- and right-handed Weyl spinors transform under $(\frac12,0)$ and $(0,\frac12)$.  Hence
%by the Wigner Eckart theorem it follows that the operators mapping
%between like spinors {\color{blue} transform under $(\frac12,0)\otimes \overline{(\frac12,0)}=(\frac12,\frac12)$, a 4-vector, while the operators mapping a left-handed to
%a right-handed point transform under $(\frac12,0)\otimes\overline{(0,\frac12)}$,
%.e., $(0,0)  \oplus (1,0)$.}

The fact that in the time-reversal symmetric case, all
four values of $\mu$ are united in a single 4-vector while in
the inversion symmetric case, they separate into a scalar and another covariant tensor, can be understood with the help of representations of the Lorentz group\cite{Ramond1980a}. These are labelled by
two spins $(s_1 , s_2 )$; in particular left- and right-handed
Weyl spinors transform under $(\frac12,0)$ and $(0,\frac12)$. The matrix elements
for transitions between two left-handed spinors (for example) $\psi_1,\psi_2$ are products of the components $\psi_{1\alpha}^*\psi_{2\beta}$,
which form the representation $\overline{(\frac12,0)} \otimes (\frac12,0)$ where the bar corresponds to the fact that the first spinor is complex-conjugated, and exchanges representations of types $(s_1,s_2)\rightarrow (s_2,s_1)$
(physically, an antiparticle of a left-handed particle is right-handed). This becomes $(\frac{1}{2},\frac{1}{2})$, a $4$-vector.
For a transition from a left-handed to a right-handed node,  the representation is $\overline{(0,\frac12)} \otimes (\frac12,0)=(0,0)\oplus (1,0)$, where $(1,0)$ is represented by the self-dual $2$-rank tensor. 

\section{Intranode scattering\label{IntraNode_Appendix}}
The argument in Section \ref{Kinematics_Appendix} shows that the neutron
speed must be high relative to the Weyl fermion speed if one wishes to 
measure intranode scattering, so we focused on scattering between
nodes at different momenta.  However,
for a material with a low Weyl fermion speed it \emph{would} be possible
to study intranode scattering without very high-energy neutrons.
In case intranode scattering is possible, some of the theory described above
applies to intranode scattering, but there are a few interesting differences.
  An important issue is the role played by minimal substitution
in finding the coupling of the Weyl fermions to the magnetic field of the neutrons, which we will begin by discussing.

In Section \ref{MagnetizationOperator_Appendix}, we suggested that the $\bm{\mathcal{J}}$ operator that
couples two distinct Weyl nodes should be found just by evaluating
the current operator matrix-elements, and interpreting the matrix elements
among the low-energy states as a $2\times2$ effective operator, which
can also be written in terms of a magnetization by using $\mathbf{J}=\mathrm{curl\ }\mathbf{M}$
to deduce $\bm{\mathcal{J}}=-\frac{2i}{\hbar}\mathbf{k}_0\times\bm{\mathcal{M}}$. The actual
parameters can be worked out only if one knows the detailed band
structure, where $\mathbf{J}(\mathbf{r})$, the current at $\mathbf{r}$ is represented in
first quantization by:
\begin{equation}
\mathbf{J}(\mathbf{r})=\frac{\hbar}{2mi}\{\delta(\mathbf{r}-\mathbf{R}), \bm{\nabla}_\mathbf{R}\}+\bm{\nabla}_\mathbf{r}\delta(\mathbf{r}-\mathbf{R})\times\bm{\mu} ,
\end{equation}
which is the Schrodinger current and the spin current.
Here $\mathbf{r}$ is the position where one is measuring the current (a c-number)
and $\mathbf{R}$ is the electron position operator.
If spin-orbit coupling is important, there is an additional contribution
that can be found using $\mathbf{J}=-\frac{\delta H}{\delta \mathbf{A}}$. Taking the
matrix element between two Bloch states and then taking Fourier transforms
with respect $\mathbf{r}$ gives the matrix elements connecting states
at certain momenta.

On the other hand, in the 4-band toy model, we began by 
introducing the vector potential into the \emph{effective} 4-band Hamiltonian via minimal substitution, and then differentiating with respect to $\mathbf{A}$
to find the current, which is not equivalent to starting
from a microscopic model of the system.
  The reason this
is approximately correct is the following:  The Hamiltonian in the presence
of a vector potential must be gauge invariant, and this is automatically
true when the vector potential is added by minimal substitution.  However, this does
not rule out other terms as long as they are gauge invariant,
like $\Bb\cdot \psi_2^\dagger\bm{\mathcal{M}}\psi_1+h.c.$  Now applying minimal
substitution to a single Weyl node as in Eq. (\ref{Kinematics_Appendix_Eq2:HamiltonianAisotropic}) or a single
 node as in the 4-band model, gives %a coupling of the form 
\begin{equation}
\mathcal{H}_{min} = -ev_{\rm F}\lambda_{lm}A^m\cdot\psi^\dagger\sigma_l\psi
\label{appendix_intra_singular}.
\end{equation}
This generates transitions within a single node only, so it 
does not generate the intranode scattering in Eq. (\ref{section:CurrentOperator_Appendix_Eq.1:MagnetizationOperator2}).
To understand such transitions, one has to just add the term mentioned
above explicitly.
%This is not applicable to understanding interactions
%with neutrons that cause the electron
%to jump from one Weyl node to another because it generates transitions
%only within a single node.  So to understand transitions like this,
%one has to just add some terms explicitly.

Now either for intranode scattering or for the 4-band model the
minimal substitution can be justified.  Although there can be other terms
present, the minimal substitution term has to be included for gauge invariance,
and it is larger than the others at low momenta. The vector potential of a neutron of spin $\bm{\tau}$ is given by $\mathbf{A}_{\rm n}(\mathbf{q}) =-i \mu_0 (\bm{\tau}\times\hat{\qb})/|\qb|$
%\begin{equation}
%\mathbf{A}_{\rm n}(\mathbf{q})=-\mu_0\frac{i}{q}(\bm{\mu}\times\hat{q})
%\%label{appendix_intra_magnet}
%\end{equation}
which diverges at small values of $q=|\qb|$, so this will cause stronger
scattering than a term like $\B_i \F_{ij}\psi^\dagger\bm{\sigma}^j\psi$, where
$\mathbf{B}$ is the magnetic field, whose Fourier transform is
$\mu_0(\bm{\tau}-(\bm{\tau}\cdot\hat{\qb})\hat{\qb})$,
as long as $q$ is low enough.  Now when $m$ and $\delta$ are nonzero, $q$
does not tend to zero for scattering between the nodes. But the coupling
to neutrons is still dominated by the minimal substitution
 contribution,
as long as the Weyl nodes are close, which happens when
$m$ and $\delta$ are small. 
%When $m$ and $\delta$ are nonzero, the coupling
%to neutrons is dominated by the minimal subtraction contribution,
%as long as the energies and momenta of the electrons are low enough.
%If $m$ and $\delta$ are small enough the Weyl points do not move
%far from the origin, so the effective theory based on minimal subtraction should still be correct.  

Now let us consider scattering between the Weyl fermions of a single node.
We have just seen that up to a certain energy we can focus on minimal
substitution.  Thus there are not all the free $\F$-parameters
that break Lorentz invariance.  However, Lorentz invariance is still broken by
the coupling to the neutrons.  %in an even more complicated way (although there \emph{are}
Equation \eqref{appendix_intra_singular} says that neutron
coupling will still be determined by the susceptibility $\chi''_{ij}(q,\omega)$, but it will be multiplied by $\mathbf{A}$,  more precisely,
\begin{equation}
\frac{\dd^2\sigma}{\dd\Omega \dd E_f}\propto \hspace*{-0.275cm}\sum_{l,l',m,m'} \hspace*{-0.275cm} \frac{(\bm{\tau}_{fi}\times \qb)_m\lambda_{lm}(\bm{\tau}_{if}\times \mathbf{q})_{m'}\lambda_{l'm'}\chi''_{mm'}(\tilde{\mathbf{q}},\omega)}{q^4},\label{fancytype}
%\frac{d^2\sigma}{d\Omega dE}\propto \frac{1}{q^4}\sum_{l,l',m,m'}(\bm{\tau}_{fi}\times \qb)_m\lambda_{lm}(\bm{\tau}_{if}\times \mathbf{q})_{m'}\lambda_{l'm'}\chi''_{mm'}(\tilde{\mathbf{q}},\omega),\label{fancytype}
\end{equation}
where $\bm{\tau}_{if}$ are the matrix elements of the neutron spin operators and $\tilde{q}_i=\sum_j\lambda_{ij}q_j$. \emph{Note that $\qb$ and $\qbt$ both appear} in this equation. The susceptibility is evaluated
at $\qbt$ because it was derived
above for Lorentz invariant coordinates, and it is in terms
of $\qbt$ that the velocity is isotropic.
Note that the ``kinetic momentum" appearing in the Weyl fermions'
cross-section is the same in this case as the momentum appearing in the
vector potential from the neutron since there is no offset
between the initial and final Weyl point.  
In Eq. (\ref{fancytype}) both the dipole field of the neutron and the Weyl fermion
dynamics contribute to the angular variation of the cross-section.
This leads to a more complicated breaking of Lorentz invariance than in
internode scattering, resulting from the fact that
the momentum respect to both coordinate systems
appears in the equation. Using the equal-chirality formula for the
susceptibility, Eq. (\ref{DifferentialCrossSection_Appendix_Eq1:TR-Susceptibility3}), and using the fact that $v_{\rm F}^2\lambda^T\lambda=v^2$, the squared velocity matrix of the
Weyl fermions, we obtain
\begin{equation}
\frac{\dd^2\sigma}{\dd\Omega \dd E_f}\propto \frac{1}{q^4}\left(|\mathbf{l}^*\cdot \mathbf{h}|^2+
\mathbf{l}^*\cdot \mathbf{l}\left[(\hbar\omega)^2-\mathbf{h}^* \cdot\mathbf{h}\right]\right)
\end{equation}
where we have defined the two vectors $\mathbf{h}=v\mathbf{q}$ and $\mathbf{l}=v(\bm{\tau}_{fi}\times \mathbf{q}$).
This is a \emph{quartic} function of $\hat{\mathbf{q}}$, which (by introducing
polar coordinates) is seen to be the sum of spherical harmonics with $l=0,2$
and $4$.  

% rather than $B$, which is
%given by \eqref{appendix_intra_singular}.
%Thus the cross-section, apart from factors having to do with the neutron
%density of states, is 
%\begin{equation}
%\frac{d^2\sigma}{d\Omega dE_F}\propto\frac{(ev_F\mu_0)^2}{|\mathbf{q}|^4}(\bm{\mu}^{if}\times\mathbf{q})_l(\bm{\mu}^{fi}\times\mathbf{q})_m\left(\tilde{q}_l\tilde{q}_m+\delta_{ij}[\left(\frac{\hbar\omega}{v_F}\right)^2-|\tilde{\mathbf{q}}|^2]\right).
%\end{equation}

%%section: DecompositionCrossSection_Appendix
%%\input{./DecompositionCrossSection_Appendix}
%%section: FullyPolarized_Appendix
%%\input{./FullyPolarized_Appendix}
\end{appendix}

\bibliography{./References_Ed3}
\end{document}